\documentclass[preprint,5p,twocolumn,times]{elsarticle}
\usepackage[noend]{algpseudocode}
\usepackage[section]{placeins}
\usepackage{algorithm}
\usepackage{amsmath}
\usepackage{amssymb}
\usepackage{booktabs}
\usepackage{caption}
\usepackage{color}
\usepackage{graphicx}
\usepackage{hyperref}
\usepackage{leftidx}
\usepackage{makecell}
\usepackage{mathrsfs}
\usepackage{multirow}
\usepackage{subfigure}
\def\0{\mathbf{0}}
\def\1{\mathbf{1}}
\def\2{\mathbf{2}}
\def\3{\mathbf{3}}
\def\4{\mathbf{4}}
\def\5{\mathbf{5}}
\def\6{\mathbf{6}}
\def\7{\mathbf{7}}
\def\8{\mathbf{8}}
\def\9{\mathbf{9}}

 \def\bB{\mathbf{B}}          
           
 \def\bD{\mathbf{D}}          
           
 \def\bF{\mathbf{F}}

 \def\bI{\mathbf{I}}          
           
 \def\bK{\mathbf{K}}      \def\bk{\mathbf{k}}    
           
 \def\bM{\mathbf{M}}          
 \def\bN{\mathbf{N}}          
           
       \def\bp{\mathbf{p}}    
 \def\bQ{\mathbf{Q}}      \def\bq{\mathbf{q}}    
\def\R{\mathbb{R}}

      \def\u{\mathbb{u}} \def\bu{\mathbf{u}}    
      \def\v{\mathbb{v}} \def\bv{\mathbf{v}}    
           
       \def\bx{\mathbf{x}}

\def\rdV{\mathrm{dV}}

\def\eb{\begin{equation}}
\def\ee{\end{equation}}

\graphicspath{figure/}
\graphicspath{figures/}

 \newcommand{\bveps}{\mbox{\boldmath$\varepsilon$}}

\newcommand{\bPhi}{\mbox{\boldmath$\Phi$}} 
\newcommand{\bPsi}{\mbox{\boldmath$\Psi$}} 

\newcommand{\bsig}{\mbox{\boldmath$\sigma$}}
\newcommand{\btau}{\mbox{\boldmath$\tau$}}

\graphicspath{{figure/}}

\def\bff{\mathbf{f}}
\def\u{\mathbf{u}}
\def\v{\mathbf{v}}

\usepackage{color}
\usepackage{ulem}
\newcommand{\rev}[2]{#2}

\begin{document}
\captionsetup[figure]{labelfont={bf},name={Fig.},labelsep=period}

\biboptions{sort&compress}

\begin{frontmatter}
\title{XVoxel-Based Parametric Design Optimization of Feature Models}
\author{Ming Li}
\author{Chengfeng Lin}
\author{Wei Chen}
\author{Yusheng Liu}
\author{Shuming Gao}
\author{Qiang Zou\corref{cor}}
\ead{qiangzou@cad.zju.edu.cn}
\cortext[cor]{Corresponding author.}
\address{State Key Laboratory of CAD$\&$CG, Zhejiang University, Hangzhou, 310027, China}

\begin{abstract}
Parametric optimization is an important product design technique, especially in the context of the modern parametric feature-based CAD paradigm. Realizing its full potential, however, requires a closed loop between CAD and CAE (i.e., CAD/CAE integration) with automatic design modifications and simulation updates. Conventionally the approach of model conversion is often employed to form the loop, but this way of working is hard to automate and requires manual inputs. As a result, the overall optimization process is too laborious to be acceptable. To address this issue, a new method for parametric optimization is introduced in this paper, based on a unified model representation scheme called eXtended Voxels (XVoxels). This scheme hybridizes feature models and voxel models into a new concept of semantic voxels, where the voxel part is responsible for FEM solving, and the semantic part responsible for high-level information to capture both design and simulation intents. As such, it can establish a direct mapping between design models and analysis models, which in turn enables automatic updates on simulation results for design modifications, and vice versa---effectively a closed loop between CAD and CAE. In addition, robust and efficient geometric algorithms for manipulating XVoxel models and efficient numerical methods (based on the recent finite cell method) for simulating XVoxel models are provided. The presented method has been validated by a series of case studies of increasing complexity to demonstrate its effectiveness. \rev{}{In particular, a computational efficiency improvement of up to 55.8 times the existing FCM method has been seen.}
\end{abstract}

\begin{keyword} Design optimization; Parametric modeling; Feature models; Extended voxels (XVoxels); Semantic voxels; Finite cell method (FCM); CAD/CAE integration
\end{keyword}
\end{frontmatter}

\section{Introduction}
\label{sec:intro}
Design optimization has been recognized as one of the dominant industrial practices for product design due to improved product quality, reduced cost, and shorter time to market \cite{shapiro2011geometric}. The optimization may be done in various ways, and parametric optimization is among the primary \cite{Sachin2017Parametric}. It optimizes engineering meaningful parameters that are embedded in feature-based CAD models with externally defined objective functions \cite{chen2007shape}. Design optimization of this sort has seen applications in many fields, including automotive, shipbuilding, and aerospace industries.

While parametric optimization is very relevant and beneficial in the context of modern feature-based CAD \cite{shah1995parametric}, its full realization is not trivial. Its working relies on a closed loop between CAD and CAE with automatic design modifications and simulation updates \cite{Boussuge2019,BOUSSUGE2022103372}. Forming such a loop is difficult because of the different information contents stored in CAD models and CAE models \cite{shapiro2011geometric}. Specifically, a CAD model is designated to have an accurate description of the design in order to automate any queries from manufacturing and assembling. It usually consists of feature history, geometric constraints, and parameter definitions (which, altogether, encompass design intent) \cite{shah1995parametric}. A CAE model contains data on the boundary conditions, material distribution, and volumetric meshes that are suitable for conducting the finite element method or the like, which encompass simulation intent \cite{bb-NOLAN201550,Boussuge2019}. 

\begin{figure}
    \centering
    \includegraphics[width=0.45\textwidth]{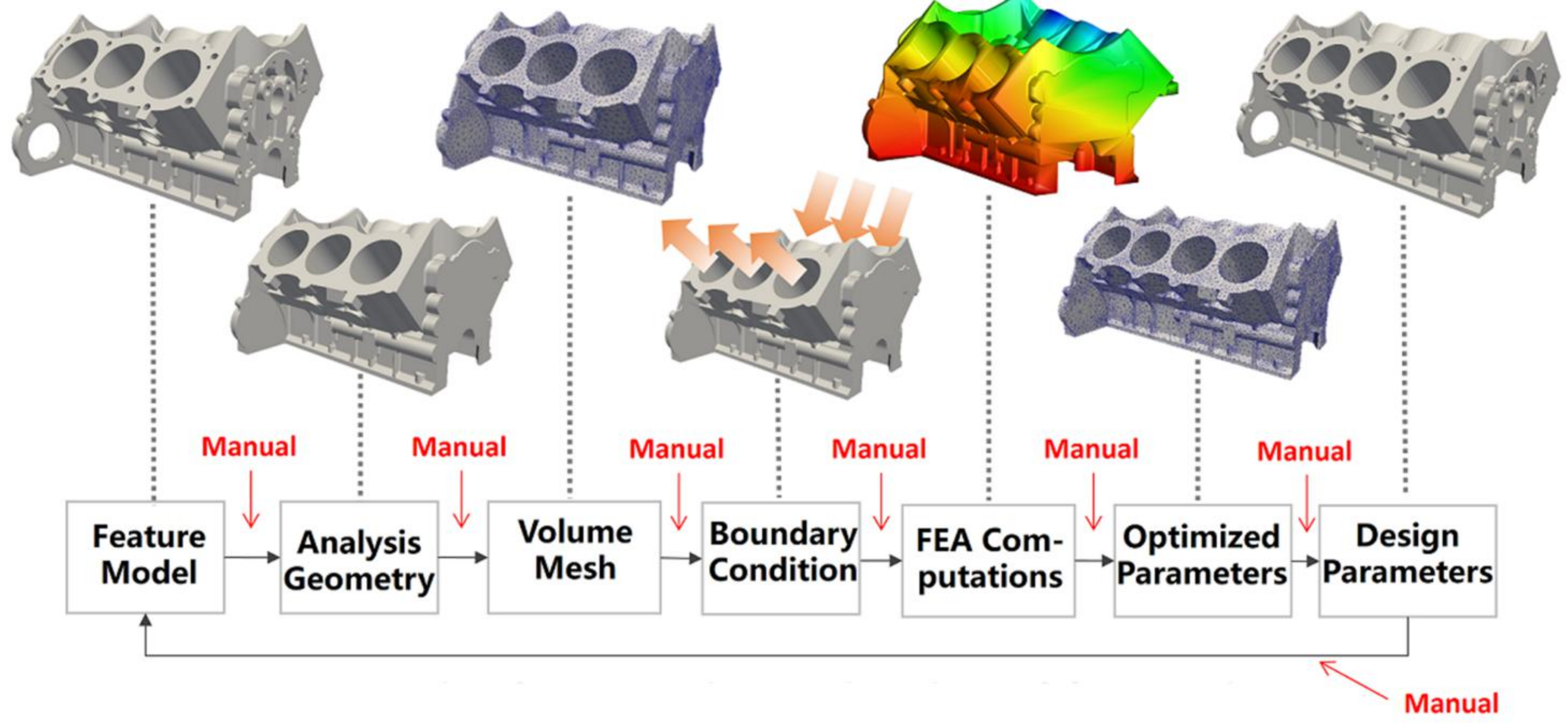}
    \caption{Traditional conversion-based process of parametric design optimization.}
    \label{fig-cadcae-td}
\end{figure}

To solve the discrepancy between CAD models and CAE models, the approach of model conversion is often employed. As illustrated in Fig.~\ref{fig-cadcae-td}, a typical conversion begins with a feature model, then goes through steps of boundary representation (B-rep) generation, model simplification, volumetric mesh generation, and boundary condition specification, cumulatively into a simulation model ready for FEM solving. The solving results will then be used to generate parametric modifications on the feature model for the next optimization iteration. Repeating these procedures will lead to an optimized design.

Despite its conceptual simplicity, there are several technical difficulties in the conversion. Typically the model simplification and mesh generation steps are hard to automate and require manual inputs~\cite{zhu2002b,Inna2009An,2000Automated,li2015A}, the design optimization and adjustment cannot be directly fed back to CAD modeling operations~\cite{bb-Liu2019CAD}, and the important design intent could be lost after conversion~\cite{bb-NOLAN201550,Boussuge2019}. In the context of iterative design optimization, such inefficiency will be much amplified and consequently, the overall process is too laborious to be acceptable. It has been reported that manual intervention accounts for about 80\% of the overall design time in the conversion-based process described above~\cite{Bazilevs2015Isogeometric,Boussuge2019}.

In view of the above issues, a unified representation scheme that can completely, compactly, and associatively represent the contents of both CAD and CAE models has been recognized as a much-desired method for parametric design optimization~\cite{Boussuge2019}. This paper follows this direction and proposes a new representation scheme called Extended Voxel (XVoxel) to address the problem. It essentially makes use of semantic voxels (as will be detailed in Section~\ref{sec:XVoxel}), where the voxel part is responsible for FEM solving, and the semantic part responsible for high-level information to capture both design and simulation intents. In a nutshell, XVoxel models provide the following advantages:

\begin{itemize}\setlength\itemsep{-0.25em}
    \item Design and simulation intents can be preserved in the loop of design, simulation, and optimization, which otherwise are lost in the conversion-based approach and have to be reconstructed. \rev{}{The relevant details can be found in Sections~\ref{sec:xvoxels} and \ref{sec:SimFeature}.}
    \item Generation of analysis geometries and volumetric meshes can be done virtually. As such, labor-intensive and non-robust model simplification and mesh generation can be possibly avoided, and boundary conditions can be well-retained over the course of optimization iterations. \rev{}{The relevant details can be found in Sections~\ref{sec:analysis-geometry} and \ref{sec:PMC}.}
    \item Modifications on design parameters and updates on simulation results can be associated automatically and locally, allowing automatic and efficient looping among design, simulation, and optimization. \rev{}{The relevant details can be found in Sections~\ref{sec:xvoxel-data-structure} and \ref{sec:SimFeature}.}
    \item Simulation of the feature models can be done efficiently on a coarse XVoxel model while attaining high accuracy through the combination of the fictitious domain technique and material-aware shape functions. \rev{}{The relevant details can be found in Section~\ref{sec:fcm-simulation} and \ref{sec:cbn-simulation}.}
\end{itemize}

The following sections begin with a review on existing parametric optimization methods in Section~\ref{sec:related-work}. A detailed description on XVoxel models is given in Section~\ref{sec:XVoxel}. The XVoxel-based simulation and design optimization are presented in Sections \ref{sec:analysis} and~\ref{sec:opt}, respectively. Application examples and comparisons with existing methods are provided in Section~\ref{sec:examples}, followed by conclusions in Section \ref{sec:conclusion}.


\section{Related work} 
\label{sec:related-work} 
Parametric optimization aims to find the optimal design parameters regarding certain performance metrics. Related approaches include conversion-based optimization, unified model-based optimization, and parameter-driven topology optimization. 
%

\subsection{Conversion-based parametric optimization}
\label{sec:conversion-cad-cae} 
The conversion-based parametric optimization is the de facto standard in practice, but it may require significant manual effort for complex CAD models or boundary conditions to ensure that all conversion steps can be carried out successively~\cite{shapiro2011geometric,BOUSSUGE2022103372}. A typical conversion procedure involves model simplification, volumetric mesh generation, boundary condition specification, and design modification. Despite the progress on simplifying geometries using methods like feature suppression \cite{zhu2002b}, direct modeling \cite{zou2019push}, virtual topology \cite{bb-Sheffer1997,bb-TIERNEY2017154}, etc., current methods either have restricted applicability or have robustness issues, thereby requiring considerable manual intervention. Generation of unstructured meshes, e.g., tetrahedron meshes, is an almost solved problem~\cite{si2015tetgen,bb-HuTetwild2020}. However, automatically generating structured meshes, which are preferable in applications requiring high computational accuracy and efficiency, still remains an open issue \cite{bb-hexEWC2001}. 

Boundary conditions are largely specified manually in practice, which thus requires huge human efforts in design optimization that loops even thousands of times between the feature model and its simulation. Clearly, this is unacceptable. A common way to address this issue is via assigning fixed boundary conditions or imposing simple varying loads, e.g., in topology optimization~\cite{bendsoe2003Topology}. These approaches however would restrict the range of the problem under study. The issue was addressed in a broad sense by defining simulation intent by incorporating concepts of cellular modeling and equivalencing~\cite{bb-NOLAN201550,Boussuge2019}. It shares similar spirits with the present work but does not involve standard voxels for performing the simulation. The involvement of manual intervention clearly decreases design efficiency and makes it hard, if not impossible, to automate design optimization. 

Note also that in the conversion-based parametric optimization, the underlying FE mesh is varied during each step of the design update, resulting in a varied design space. As a consequence, it usually tends to result in an unstable optimization convergence. 

\subsection{Unified model-based parametric optimization}
Existing unified model-based parametric optimization approaches mainly include isogeometric analysis (IGA), embedded domain, or their combinations. 

IGA, initialized by Hughes et al. \cite{Hughes2005Isogeometric}, uses a unified geometric representation scheme, i.e., NURBS (non-uniform rational B-spline), for both design and analysis. Basically, it discards the use of explicit meshes but employs the knot vector and spline basis of a NURBS surface to directly generate the elements and shape functions for FEM solving \cite{li2011isogeometric}. As the mesh generation step is eliminated (in principle), IGA provides a tighter integration between CAD and CAE for automatic design optimization, and the benefits extend beyond integration to higher simulation accuracy and efficiency \cite{QIAN20102059,SEO20103270,Sachin2017Parametric,XIE2020112696}. However, IGA only works well on the surface model having a regular parametric domain. For general shapes composed of trimmed NURBS surfaces or 3D volumetric models, quadrilateral meshing of its boundary, or hexahedral meshing of its volume are inevitable, which are challenging research topics in their own right. \rev{}{The XVoxel method to be presented does not have this issue because there are no B-rep models or meshing processes involved. This advantage manifests itself through situations where the B-rep model given to IGA is complex and introduces robustness issues in model simplification and difficulties in quad/hex meshing. It should, however, be noted that IGA has higher simulation accuracy and can directly take B-rep models as input, while XVoxel is not able to do so. IGA is thus preferred in such situations.}

Unlike IGA which revolves around the design model (i.e., NURBS), the embedded domain approach such as finite cell method (FCM)~\cite{schillinger2015finite} focuses on the other side, i.e., meshes. It uses the same regular background mesh (e.g., a grid) to carry out FEM solving regardless of the design model's variations. As such, no mesh generation is needed when the design model is modified during optimization. This is essentially achieved through high-order finite elements and weak enforcement of unfitted essential boundary conditions. FCM was also used together with IGA to utilize both of their advantages~\cite{bb-RANK2012104,Yingjun2016Isogeometric}, most of which did not discuss its work on feature models. Recently, Wassermann et al. \cite{bb-WASSERMANN20171703} studied the problem of conducting FCM on CSG model, which mainly studied the point membership classification problem for different primitives while the present study focuses on the overall integration flowchart for parametric optimization of feature models. 

The embedded domain approach is to be combined with the feature-based approach in this work to enable the embedding of design and simulation semantics within the background mesh (which is otherwise purely geometric), where the background mesh serves as a common data structure, and the embedded semantics provide automatic links between design modifications and simulation updates. 

\subsection{Parameter-driven topology optimization}
Research efforts have been devoted toward parametric optimization of CAD models, which mainly focus on finding the optimal shape parameters in describing a specific CAD part but seldom addressed the issue of integrating design semantics into the optimization process. 

Chen et al. considered using R-functions for design optimization with topological changes~\cite{chen2007shape,2008Shape}. Zhu et al. proposed a direct simulation approach for CAD models undergoing parametric modifications~\cite{zhu2016direct} using a model reduction technique called PGD (Proper Generalized Decomposition)~\cite{chinesta2010recent}. Schulz et al. developed an exploration tool for interactive exploration and optimization of parametric CAD models~\cite{Schulz2107} via pre-computations. More recently, Hafner et al. proposed a generic shape optimization method, called X-CAD, for CAD models based on the eXtended Finite Element Method (XFEM)~\cite{Chr2019X}. These approaches did not involve a complex model-conversion process but worked on an embedded background mesh so as to automate the overall process. 

To keep the design intent, the adjustment should be made on feature parameters or the feature history of CAD models \cite{zou2020decision}. Conducting topology optimization under constraints of specific CAD features has attracted research interests. Zhang and his colleagues have studied extensively the topic ~\cite{zhu2016topology,bb-CADJiu2020,zhu2020Areview_40,zhu2020Areview_41} for practical engineering design. Recently, Guo introduced a novel topology optimization method of MMC (Moving Morphable Components)~\cite{RN233,RN232,zhu2020Areview_43}, which uses deforming bars as primitive features in topology optimization process for ease of geometric control. 
However, most of the approaches only studied abstract and single parametric features without design history. Recently, Liu and To \cite{bb-Liu2019CAD} first included the feature modeling history of CAD models in the design optimization process. This work follows this direction but employs a more automatic and efficient method, i.e., XVoxel, to carry out the optimization of feature parameters by embedding design intent in the overall optimization process.

\section{XVoxel models}
\label{sec:XVoxel}
This section introduces features, voxels, and their combination into XVoxels, as well as the data structure and algorithms for constructing and manipulating XVoxel models.


\subsection{From features and voxels to XVoxels}\label{sec:xvoxels}
\begin{figure}[htbp]
    \centering
    \includegraphics[width=0.4\textwidth]{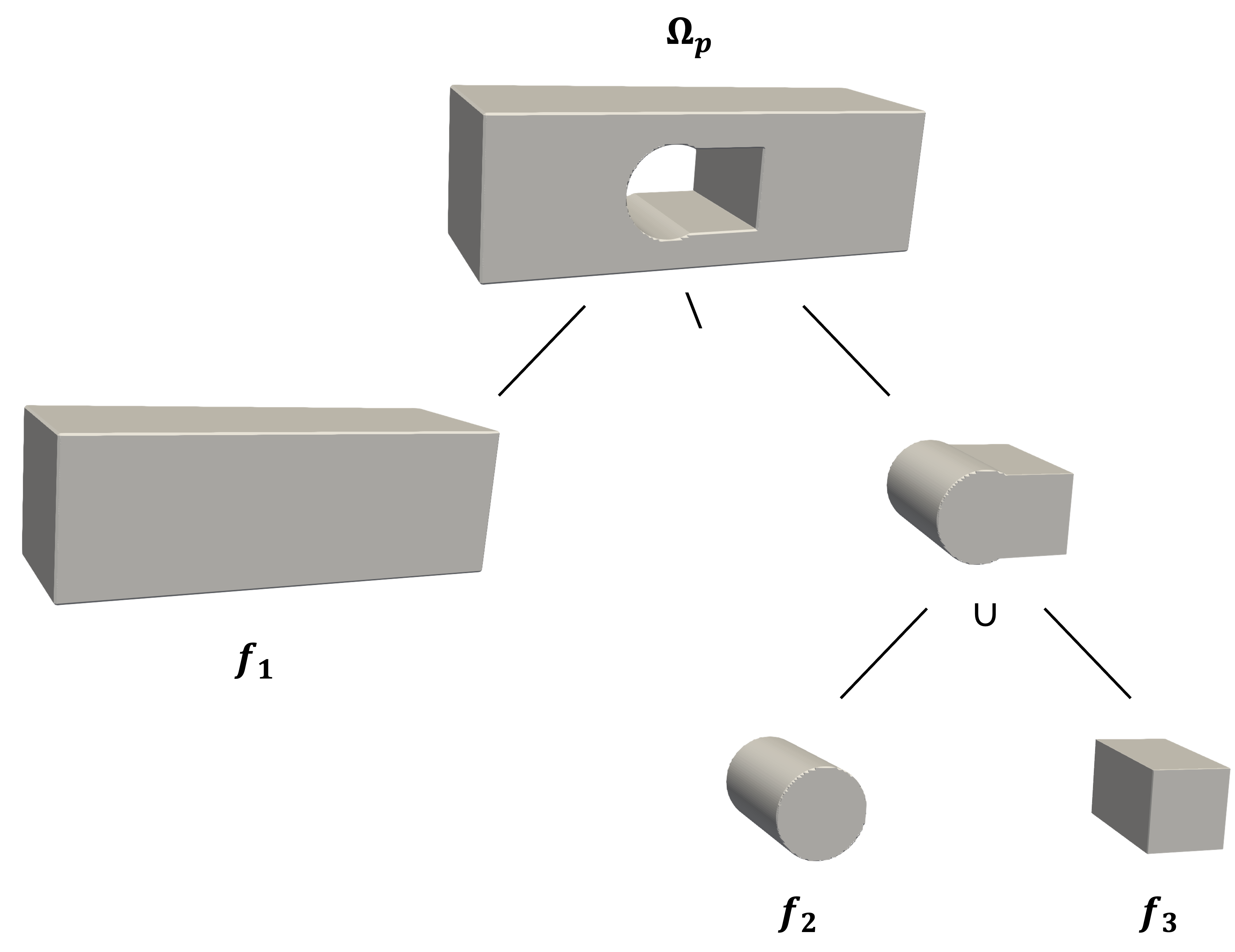}
    \caption{A CSG example.}
    \label{fig-csg}
\end{figure}

There is no widely accepted definition of features. The one this work employs is given by Shah \cite{shah1995parametric}: a feature is a generic portion of a model's shape that has certain engineering significance. Roughly speaking, features are clusters of geometric entities in a CAD model, which can be used as information containers to carry domain-specific attributes, e.g., materials and boundary conditions. A feature model is a set of features, combined in a way similar to traditional constructive solid geometry, as shown in Fig.~\ref{fig-csg}. Practically almost all of today's commercial CAD systems use features as an internal representation for constructing and/or editing their CAD models \cite{zou2022robust}. The user designs a feature by first defining a topology of geometric entities then specifying geometric constraints relating them. A feature can be positioned anywhere in space, or relatively to existing features (through, again, geometric constraints). As such, geometric entities of a CAD model are stored associatively and hierarchically. Changes to the parameters of those features can then be propagated automatically in a pre-defined fashion \cite{shah1998designing}. This is the basis upon which parametric design optimization becomes possible.

A voxel is a cube-like element in space, and a voxel model is a collection of voxels comprising a three-dimensional geometry of interest. A voxel model can be stored as an array of voxels occupied by the geometry or a grid with binary labels indicating the occupancy relationship between each voxel and the geometry; see also Fig.~\ref{fig-xvoxel}. The former storage scheme is often used to represent static geometries, and the latter used to represent dynamic geometries (and therefore the chosen one in this work).

This work proposes to combine features with voxels, i.e., embedding features into voxels. Traditional CAD/CAE integration methods consistently use features as information containers to store design intent (e.g., shape parameterization) and simulation intent (e.g., meshing procedures and boundary conditions) \cite{lee2005cad}. In this work, voxels are information containers where design intent and simulation intent reside. This shift leads to the notion of semantic voxels (named XVoxels in this work). The primary benefit of doing so is that explicit generation of analysis geometry and meshes can be mostly avoided, and then an automatic, closed loop between CAD and CAE can be achieved. This will be demonstrated in the next few subsections. We begin with the specific data structure used to represent XVoxel models and some primitive operations used to manipulate them.

\subsection{XVoxel representation and operations}
\label{sec:xvoxel-data-structure}
\begin{figure*}[htbp]
    \centering
    \includegraphics[width=\textwidth]{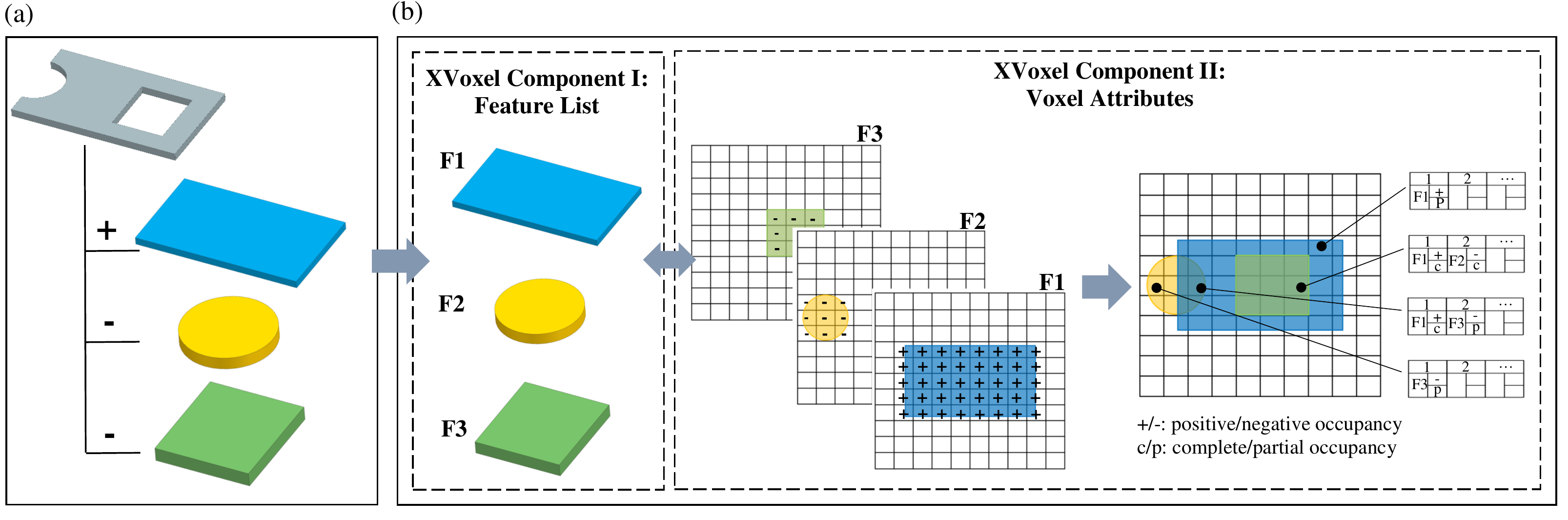}
    \caption{Illustration of the XVoxel data structure: (a) a feature model (the plus sign means Boolean addition, and the minus sign Boolean subtraction); and (b) its corresponding XVoxel model.}
    \label{fig-xvoxel}
\end{figure*}

An XVoxel model consists of two components: a list of features and an array of voxel attributes, as shown in Fig.~\ref{fig-xvoxel}. The feature list is nothing but an unordered set of features (with boundary conditions, material properties, etc. already associated). The voxel attributes associate each voxel with the features occupying it. Three feature attributes are stored: feature occupancy, feature nature, and feature history. For a voxel, feature occupancy describes whether it is completely or partially occupied by a feature; feature nature indicates whether an occupying feature is adding material or subtracting material; feature history refers to the precedence of all occupying features of the voxel.

Consider, for example, the model in Fig.~\ref{fig-xvoxel}, and focus on feature F1. It occupies the voxels colored blue. Voxels at its boundary have partial occupancy, while those in its interior have complete occupancy. The plus signs in Fig.~\ref{fig-xvoxel}b indicate that the occupied voxels are positive (the same as F1's nature). Following the same principle, two additional arrays of voxel attributes can be generated for features F2 and F3, as shown in Fig.~\ref{fig-xvoxel}b. Combining these three arrays of voxel attributes in their chronological order (i.e., F1 $\rightarrow$ F2 $\rightarrow$ F3) results in an XVoxel model, where each voxel maintains an ordered list of 3-tuples $(feature\ index, feature\ nature, occupancy\ completeness)$, as shown by the rightmost four lists in Fig.~\ref{fig-xvoxel}b.

In XVoxel models, determining a model's actual shape relies merely on XVoxel nature, which refers to the nature of the last feature in the attributes list of individual XVoxels. This is because whether the last feature adds or subtracts material, it will override any preceding operations. One exceptional situation is when the last feature partially occupies an XVoxel; this XVoxel's nature is a compound result of the last few features in the attribute list, from the last feature with a complete occupancy to the end. In the following, an XVoxel of this kind is referred to as compound nature. Special algorithms will be developed in the next subsection to handle this situation when using XVoxel to conduct its property simulation.

The above statements seemingly imply that there is no need for storing all historical feature natures of an XVoxel, but only the last one (or ones). They are actually saved for providing easy ways to carry out XVoxel operations, as detailed below. In particular, the novel idea of constantly storing negative feature nature, rather than immediately discarding it after feature Booleans as in conventional feature modeling approaches, allows all operations to work locally, efficiently, and robustly.

\textbf{Feature Addition} \quad This operation creates a new feature by instantiating a chosen feature class with user-specified feature parameter values. After instantiation, the feature's shape extent is used to determine which voxels it occupies, then append the feature's attributes (i.e., the 3-tuple described above) to the end of those voxels' attribute lists. This addition operation is the basis of constructing an XVoxel model from a given feature model. We simply repeat this operation over all features of the model in their chronological order.

\textbf{Feature Deletion} \quad The selected feature is simply removed from the XVoxel model's feature list, with feature dependencies updated accordingly and its attributes removed from relevant XVoxels' attribute lists. To facilitate the retrieval of relevant XVoxels, we further associate each feature with a list of XVoxel indices it occupies in the XVoxel data structure (which can be easily recorded during feature addition). For every single relevant XVoxel, we linearly search the corresponding feature entry in its attribute list and, once found, simply remove it from its current position. (Note that in practice, because attributes in each XVoxel are stored as a linked list, a postprocessing step to correct the linking pointers of remaining entries in the list is needed.) If parallel computing is enabled, we can search and do the removal for all XVoxels simultaneously, without the need for the associativity from features to relevant voxels. Multiple features can also be deleted in parallel. It should, however, be noted that to avoid race conditions when deleting feature attributes at the same XVoxel, we lock the list when the entry removal operation is being carried out for a feature.

\textbf{Parameter Editing} \quad This operation modifies features' parameter values. In the background, we first delete it from the XVoxel model, then re-add its modified version to the XVoxel model according to its original precedence in the feature history. As such, no additional algorithms are needed. Considering that features are often interdependent \cite{zou2019variational}, the above two procedures are modified to include the dependent features of the feature being edited.

\textbf{Feature Rearrangement} \quad This operation modifies the order of features (under the condition that feature dependencies will not be broken). What we need to do is simply updating the orders in individual XVoxels' attribute lists to accommodate the rearrangement.

As can be seen, there is no time-consuming and non-robust geometric computing involved in the above operations, except for the determination of voxels occupied by a feature to be added in the addition operation. All operations boil down to manipulating entries in a certain linked list, which is easy to implement, robust, and efficient. For the determination of occupied voxels, the essential task involved is to voxelize the shape of a given feature, using the same resolution as the XVoxel model. Note that voxelization is done on individual feature shapes here, which are usually primitives like cuboids or spheres, not on the overall combined shape of all features, which is otherwise complex. Many algorithms exist to voxelize a feature's B-rep model, and the method developed by Young and Krishnamurthy \cite{young2018gpu} is employed in this work due to its high efficiency. The B-rep model of a feature is often made readily available during feature instantiation, a function provided by almost all modern commercial CAD modelers.

\subsection{Virtual model simplification}
\label{sec:analysis-geometry}
Model simplification\footnote{It should be noted that a more general concept than model simplification is model idealization, which includes an additional dimension reduction task~\cite{bb-ARMSTRONG1994573,2000Automated,chong2004automatic}. As XVoxels models are three-dimensional, they are not able to handle dimension-reduced geometries in their current form. The authors wish to extend XVoxels to representing low-dimension geometries in our future studies.} is to remove some design features (e.g., small drilled holes) that are of little significance to simulation. This task becomes straightforward if XVoxel models are used. What we need to do is applying the delete operation described in the previous subsection. The only issue is that, similar to traditional feature-based model simplification approaches, directly removing a feature may cause the persistent naming problem, ultimately breaking the design-analysis cycle. To be more specific, features are made interdependent in feature modeling to enable automatic propagation of parameter changes \cite{shah1998designing}. Removing a feature makes any inter-dependencies related to it undefined, and then the whole model becomes invalid, which is the so-called persistent naming problem \cite{shapiro1995parametric}.

To solve this issue, we customize the delete operation slightly. The delete operation in Section~\ref{sec:xvoxel-data-structure} directly removes a feature from the XVoxel model's feature list. Instead, we retain it but make it transparent to voxel attributes by associating the feature list with a bitmask whose 0-elements indicate that features at their positions have been removed, virtually. As such, the difficult persistent naming problem is avoided and meanwhile, there are no real geometric operations involved in model simplification. Another benefit of doing so is that boundary conditions, once associated with certain features, can retain over the course of optimization iterations regardless of design modifications because those features are completely stored in XVoxel models.

\subsection{Point membership classification}
\label{sec:PMC}
Traditionally, what comes next after model simplification is generating boundary-conformed meshes for downstream task of simulations. This is, however, a field not all major questions have been answered \cite{Hughes2005Isogeometric}. This work reformulates the problem as an underlying problem of point membership classification (PMC) for Gaussian integral point selection. It is to be further combined with the recently developed method of FCM for physical simulation, which embeds the physical domain of computation (i.e., the geometry of the feature model) in a larger, regular mesh like a grid, and then transforms the FE computation onto the embedding meshes \cite{schillinger2015finite}. (A detailed introduction to FCM will be given in the next section.) 
This way of working is a perfect match for XVoxel models.

If an XVoxel has positive nature and complete occupancy, it is completely within the model shape. Then the stiffness matrix for this XVoxel can be computed in the exact same way as conventional FEM does. If an XVoxel has compound nature, the XVoxel crosses the boundary of the model shape. According to FCM, voxel subdivision is needed to generate stiffness matrices for such boundary XVoxels, which in turn relies on the operator of point membership classification (PMC) to determine if a sample integration point within a boundary XVoxel is IN/ON/OUT the model shape.

Because XVoxel models have prepared the history of feature occupancy for every XVoxel, the problem of PMC against the overall model shape can be converted to a sequence of much simpler PMCs against individual features \cite{rossignac2022ibnc}. The conversion consists of three major steps: (1) screening relevant features; (2) evaluating PMC against each screened feature; and (3) compiling evaluation results to the final IN/ON/OUT decision. Clearly, not all features occupying an XVoxel contribute to its final shape (i.e., which portion of the XVoxel is solid or void). According to the XVoxel's attribute list, candidate features include those ranging from the last feature with a complete occupancy to the end feature (see Section~\ref{sec:xvoxel-data-structure}). For this reason, the screening step can be simply done by tracing from the back of the attribute list up to the first entry having the complete occupancy attribute.

Having relevant features in place, we next determine the IN/ON/OUT relationship between a query integration point and each of the features. Let the relevant features be denoted by $f_1, f_2, \cdots, f_n$, and their corresponding implicit representation denoted by $\phi_1, \phi_2, \cdots, \phi_n$. In this work, feature implicitization is done by first triangulating its B-rep model with a sufficient high accuracy, then building a KD-tree for the triangles to allow fast query of the (approximated) signed distance between a given point and the feature, similar to the method presented in \cite{Wang2013Thickening}. Note that alternative methods surely exist \cite{jones20063d}, and we choose this one for its simplicity and efficiency. Whether a given point $\bx$ is IN/ON/OUT feature $f_i$ is determined by the sign of $\phi_i(\bx)$:
\begin{equation}
    \left\{
    \begin{array}{lll}
         & \phi_i(\bx) > 0 & \rightarrow \quad \bx \mbox{ IN }\ f_i,  \\
         & \phi_i(\bx) = 0 & \rightarrow \quad \bx \mbox{ ON }\ f_i,  \\
         & \phi_i(\bx) < 0 & \rightarrow \quad \bx \mbox{ OUT }\ f_i.
    \end{array}
    \right.
\end{equation}

To compile individual classification results to the final IN/ON/OUT decision, we again make use of the feature history stored in each XVoxel. First, the features classified as OUT are filtered out from the relevant feature set because they contribute nothing to the process of adding/removing material. Then, the final IN/ON/OUT decision is the same as the nature of the last remaining relevant features: if the nature is positive, the material is added to the query point, and the final decision is IN/ON; otherwise, the final decision is OUT. This is because the last material removing/adding operation overrides all the preceding operations.

Altogether, they yield a method to generate a ``mesh'' suitable for FCM solving from an XVoxel model. The mesh is not explicitly generated but through combining the fixed grid carrying the XVoxel model and an implicit PMC operator developed specifically for XVoxel models. The method is thus easy to implement. It should, however, be noted that, the use of triangulation in feature implicitization will introduce errors, and therefore possible misclassifications in PMC. In fact, this is generally acceptable since we can triangulate at a high accuracy. Also, due to the integral nature of FCM, it is not very sensitive to such misclassifications.


\section{XVoxel-based simulation}
\label{sec:analysis}
In this work, simulation is to be carried out using a fictitious domain approach following a FCM-like framework~\cite{schillinger2015finite}, which can work directly on voxel models. This approach's low computational efficiency is improved in two aspects: (1) by introducing material-aware piecewise matrix-valued shape functions, called CBN (Curved Bridge Node) shape functions following the previous study in~\cite{li2022analysis}; (2) by utilizing the local computation of XVoxel models. 

\subsection{Finite cell method (FCM) for XVoxel-based simulation}\label{sec:fcm-simulation}
The basic idea of FCM is to use a simple regular structured mesh to approximate the solution fields. This is achieved by combining the fictitious domain idea with the benefits of high-order finite elements, thus avoiding the costly and even labour-intensive meshing process. 

\begin{figure}[]
    \centering
    \includegraphics[width=0.45\textwidth]{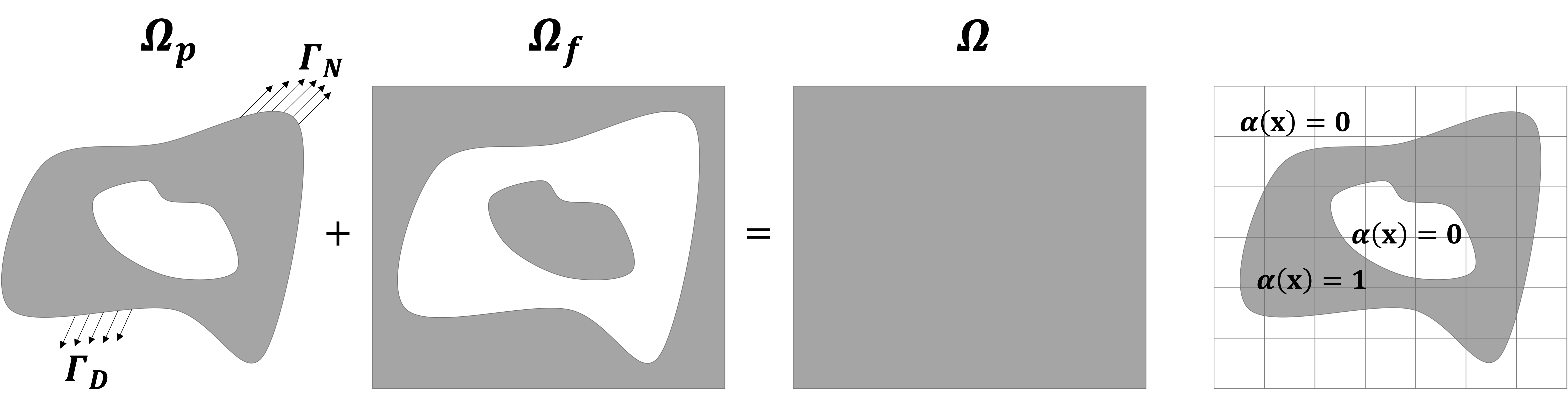}
    \caption{The embedded domain $\Omega$ consists of the physical domain $\Omega_{p}$ and the fictitious domain $\Omega_{f}$, and the influence of $\Omega_{f}$ is penalized by material parameter $10^{-q}$.}
    \label{fig-fcm_concept}
\end{figure}

The FCM concept is interpreted by a 2D linear elasticity problem in Fig. \ref{fig-fcm_concept}. Let $\Omega_{p} \in \R^2$ be the physical domain, $\Gamma_D$ the Dirichlet boundary, and $\Gamma_N$ the Neumman boundary under external loading $\btau$. A linear elasticity analysis problem on $\Omega_{p}$ is studied to find the displacement $\bu$ satisfying
\begin{equation}\label{eq-weak}
    a(\u,\v)=l(\v),\quad \forall~ \v\in H_0^1(\Omega),
\end{equation}
where
\eb\label{eq-a}
a(\u,\v)=\int_{\Omega_p} \bveps(\bu)^T \bD \bveps(\bv)~\rdV= \int_{\Omega}  H(\bx) \bveps(\bu)^T\bD \bveps(\bv)~\rdV,
\ee
and
\eb\label{eq-l}
l(\v)=\int_{\Omega_p} \bff\cdot \bv~\rdV+\int_{\Gamma_N} \btau\cdot \bv~d\Gamma= \int_{\Omega} H(\bx)\bff\cdot \bv~\rdV+\int_{\Gamma_N} \btau\cdot \bv~d\Gamma,
\ee
where $H^{1}(\Omega)$ and $H_{0}^{1}(\Omega)$ are the usual Sobolev vector spaces, $\bff$ is the body force, $\bsig(\bu)$ is the second-order \emph{stress tensor} defined via Hooke's law,
\eb
\bsig(\bu) = \bD:\bveps(\bu),\quad \bveps(\bu)=\frac{1}{2}(\nabla \bu + \nabla \bu^T)
\ee
for a fourth-order elasticity tensor $\bD$.

Here in Eqs.~\eqref{eq-a} and~\eqref{eq-l}, the computation domain is converted from $\Omega_p$ to the embedded domain $\Omega$ by incorporating the fictitious domain material which is defined via a Heaviside function $H(\Phi(\bx))$ \rev{at a material control parameter $\epsilon$}{},
\eb\label{eq:H}
\small
\rev{}{
H(\Phi(\bx))= \begin{cases}
    1,                                                                                                              & \text { if } \Phi(\bx) > 0,                  \\
    \alpha,                                                                                                        & \text { otherwise },   \\
    \end{cases}
}
\ee
where $\Phi$ is the SDF (Signed Distance Function) of the feature model $\Omega_p$\rev{, $\bp$ is the vector of design parameters, $\epsilon$ controls the magnitude of regulation,}{} and $\alpha$
\rev{$=10^{-q}$ by default}{ is a small positive coefficient, say $10^{-8}$, }
to avoid ill-conditionedness on the stiffness matrix. The Nitsche's method~\cite{nitsche1971variationsprinzip} was usually adopted to weakly impose Dirchlet boundary conditions in FCM; we are not going into details here.

Following a classical Galerkin FE method, the solution $\bu(\bx)$ to Eq.~\eqref{eq-weak} is approximated as a linear combination of \emph{higher-order} shape (base) functions $\bN^{\alpha}(\bx)$ for each regular grid (or voxel) $\Omega^{\alpha}\subset \Omega$.   Specifically, the overall displacement on any point of $\bx\in \Omega^p$ can be interpolated from an assembly sum 
\eb\label{eq-uNQ}
\bu(\bx)\approx \bN(\bx) \bQ=
\sum_{\alpha=1}^M\bN^{\alpha}(\bx)~\bQ^{\alpha}, \quad \bx\in \Omega,
\ee
where $\bN(\bx)$ is the collection of bases $\bN^{\alpha}(\bx)$, $\bQ$ is the collection of $\bQ^{\alpha}$, a displacement vector per voxel $\Omega_{\alpha}$.

Accordingly, the displacement $\bQ$ to Eq.~\eqref{eq-weak} is computed as the solution to a linear system
\eb\label{eq-kuf}
\bK \bQ=\bF,
\ee
where the stiffness matrix and load vector are assembly from their element stiffness matrix $\bK_{\alpha}$ and element load vector on a regular grid (or XVoxels)
\eb\label{eq-localk}
\bK=\sum_{\alpha} \bK^{\alpha},\quad   \bF=\sum_{\alpha} \bF^{\alpha}.
\ee

\begin{figure}[]
    \centering
    \includegraphics[width=0.45\textwidth]{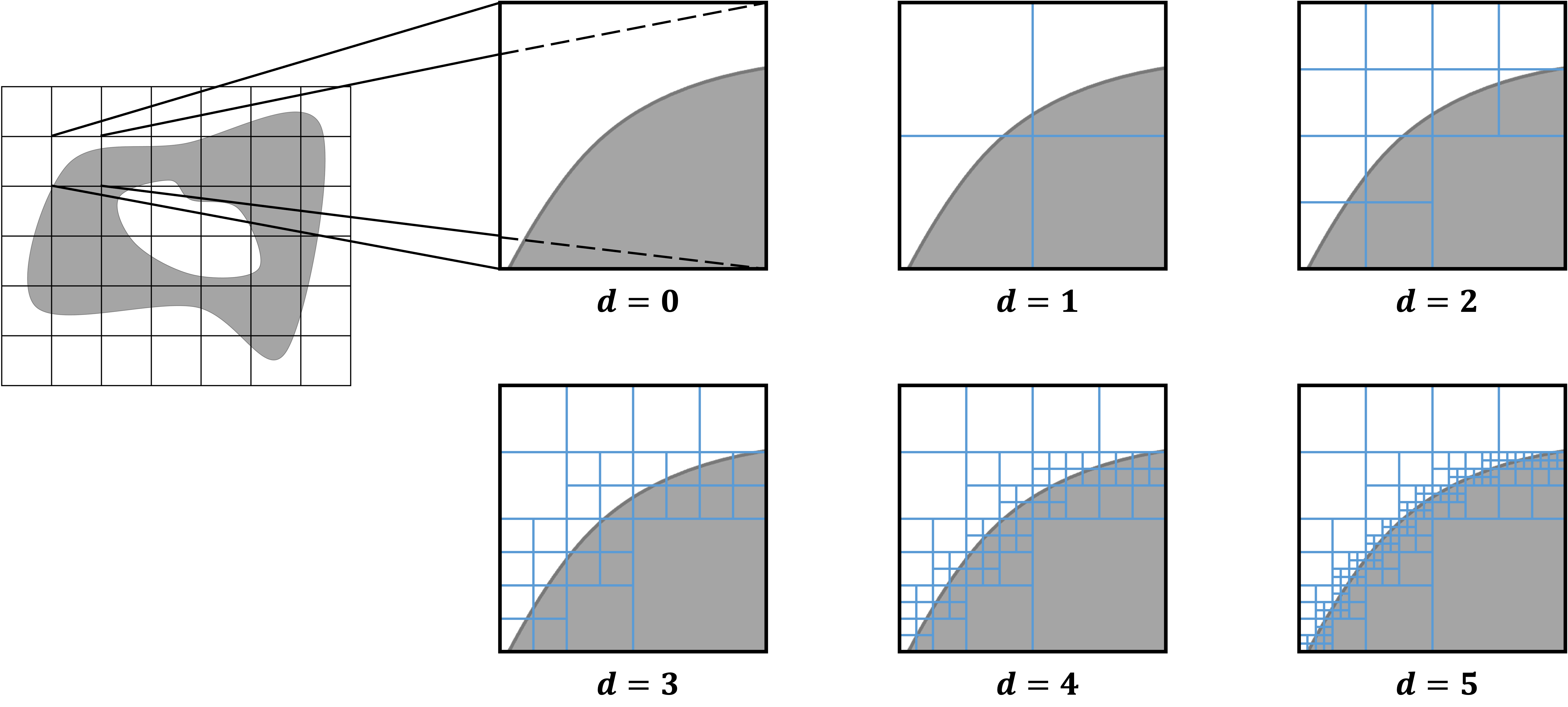}
    \caption{In 2D, adaptive Gaussian integration is recursively refined towards boundary of physical domain (gray domain), yielding a quadtree structure (thin blue grids) of finite cells (bold black grids); Quadtrees depth ranging from 0 to 5 are shown here. The voxels along the boundary are called cut voxels. }
    \label{fig-fcm_refinement}
\end{figure}

FCM transfers the challenges of mesh generation to the numerical integration of discontinuous integrands in Equation \eqref{eq-a} and \eqref{eq-l}. An adaptive Gauss integration is usually applied to improve its accuracy; see Fig. \ref{fig-fcm_refinement}. The high-order shape functions $\bN^{\alpha}(\bx)$ in FCM requires a huge number of Gaussian points, which involves huge computational costs as compared with FEA and occupies the dominant computational costs of FCM.

\subsection{CBN shape functions for efficient FCM computation}\label{sec:cbn-simulation}
Following previous study~\cite{li2022analysis}, material-aware CBN shape functions are introduced in this work to replace the higher-order shape functions in FCM to accelerate the computations.

In our adopted version of CBN, a $4\times 4$ grid is formed by introducing twelve additional virtual nodes on each face of a voxel element besides its original $4$ nodes (all together $80$ nodes for a 3D  XVoxel). Displacement on these CBN nodes are collected into a vector $\bQ$ and taken as DOFs for solution computation. 

The higher order shape function $\bN^\alpha$ in Eq.~\eqref{eq-uNQ} is replaced by the following one composed of linear shape functions on fine mesh via a transformation matrix
$\tilde{\bPhi}^\alpha$:
\eb\label{eq-gshapef}
\bN^{\alpha}(\bx)= \bN^{\alpha,h}(\bx)~\tilde{\bPhi}^\alpha, \quad \bx \in \Omega^{\alpha}
\ee
where $\bN^{\alpha,h}(\bx)$ is an assembly of the nodal shape functions on the fine mesh of $\Omega^{\alpha}$. 

The CBN transformation matrix $\tilde{\bPhi}^{\alpha}$ aims to map the CBN nodal values to the interior values in $\Omega^{\alpha}$. It is derived as a product of \emph{boundary interpolation matrix} $\bPsi$ and \emph{boundary--interior transformation matrix} $\tilde{\bM}^{\alpha}$, as follows,
\eb\label{eq:NH}
\tilde{\bPhi}^{\alpha} = \tilde{\bM}^{\alpha} ~ \bPsi,
\ee
where $\bPsi$ and $\tilde{\bM}^{\alpha}$ maps the displacements from the CBNs to the boundary nodes and then to the full fine nodes in $\Omega^{\alpha}$.

The boundary interpolation matrix $\bPsi$ maps the CBN nodal values to the fine mesh boundary nodal values of $\Omega^{\alpha}$. It is derived by constructing a bi-cubic B\'ezier interpolation surface over the face of interest, taking the CBN as control points. The matrix $\bPsi$ is derived by evaluating the surfaces at fine mesh nodes within the face, and collecting them in a matrix form all the values row by row for the six faces of the mesh $\alpha$.

The transformation matrix $\tilde{\bM}^{\alpha}$ maps the boundary node values to those of the fine mesh in $\Omega^{\alpha}$. It is derived from the local simulation on the fine mesh of $\Omega^{\alpha}$ with the equilibrium equation
\begin{align}\label{eq:micros}
    \begin{bmatrix}
        \bk_{b}  & \bk_{bi} \\
        \bk_{ib} & \bk_{i}
    \end{bmatrix}\begin{bmatrix}
                     \bq_b \\
                     \bq_i
                 \end{bmatrix} = \begin{bmatrix}
                                     \bff_b \\
                                     0
                                 \end{bmatrix},
\end{align}
where $\bk_{b},\ \bk_{i},\ \bk_{bi},\ \bk_{ib}$ are the sub-matrices of the local stiffness matrix $\bk^{\alpha}$ on $\Omega^{\alpha}$, $\bq_b, \ \bq_i$ is respectively vector of the boundary, interior nodes, and $\bff_b$ is vector of exposed forces on the boundary nodes formed by harmonic analysis~\cite{li2022analysis}.

We have from the second-row the relation of $\bq_i=\bM^{\alpha}\bq_b$, for ${\bM^{\alpha}} = {-\bk_{i}^{-1}}\bk_{ib}$. Accordingly, assembling $\bq_i$ and $\bq_b$ as ${\bq}= [\bq_b,\bq_i]^T$, we have the form of  $\tilde{\bM}^{\alpha}$,
\eb\label{eq:bitrans}
\tilde{\bM}^{\alpha}=[\bI_{2b}, -\bk_{i}^{-1}\bk_{ib}]^T,
\ee
where $\bI_{2b}$ is the $2b \times 2b$ identity matrix.

Once the CBN shape functions are derived, the solution to the linear elasticity problem in Eq.~\eqref{eq-weak} can be similarly attained, following a classical Galerkin FE method. More technical details are referred to~\cite{li2022analysis}.

\subsection{XVoxel-based local simulation of feature models}\label{sec:SimFeature}
Based on the approach of FCM for simulation, in combination with CBN, the XVoxel-based approach for simulation of modified feature model is developed below.

The feature model is generally modified via updating feature parameters, which may change the topology and geometry of the final B-rep model. As long as the feature model are updated, element stiffness matrix $\bK^{\alpha}$ of each voxel need to be re-computed, which accounts most for the computation costs. FCM equipped with local voxel updates can accelerate the computations in two ways: (a) the element stiffness matrix of each full-voxel or void-voxel is identical; the voxels along the boundary are called \emph{cut voxels}; (b) the element stiffness matrices of voxels not affected by updated features remain unchanged. Case (a) can be easily resolved via a pre-computation strategy. For case (b), the element stiffness matrix can be incrementally updated by updating and querying voxel-feature membership table via the PMC algorithm described in Section~\ref{sec:PMC}, where voxels affected by the updated features can be quickly located, called \emph{active voxels}, and consequently only their stiffness matrices are re-computed. This can significantly reduce computation costs.

\section{XVoxel-based parametric design optimization}\label{sec:opt}
Using XVoxels, the parametric design optimization works over a fixed regular grid under controlled simulation accuracy, and on direct updates of feature parameters. During the process, the sensitivities with respect to the design parameters is derived for parameter updates. The locality information of XVoxel provides an efficient sensitivity computation either via finite difference or via a derived analytical expressions. 

Let $\Omega_p$ be a CAD model with features $f_1, f_2, \cdots, f_n$. For ease of explanation, each feature $f_i$ is assumed to take only one parameter $p_i$. The classical compliance minimization problem is studied to find the optimized design parameters $\bp = (p_1, p_2, \cdots, p_n)$:
\begin{equation}
    \begin{aligned}
         & \min\limits_{\bp} C(\bu, \bp)=\bu^{T} \bK \bu,                                         \\
         & \text { s.t. }\left\{\begin{array}{l}
                                    \bK \bu=\bF,                                                      \\
                                    V=\int_{\Omega} H(\Phi(\bx, \bp)) \mathrm{d} \Omega \leq \bar{V}, \\
                                    \underline{p}_i \leq p_{i} \leq \bar{p}_i, i=1,2 \ldots, n,
                                \end{array}\right.
    \end{aligned} \label{eq-compliance minimization}
\end{equation}
in which $V$ and $\bar{V}$ are the total structural volume and maximum volume constraint, the Heaviside function $H(\cdot)$ is used to indicate structural boundary, $\underline{p}_i$ and $\bar{p}_i$ are lower and upper bounds of the design variable $p_i$.

The optimization problem Eq.~\eqref{eq-compliance minimization} is to be solved following a numerical gradient-based approach Globally Convergent Method of Moving Asymptotes
(GCMMA)~\cite{Zillober1993} for its robust convergence in design optimization. It approximates the original nonconvex problem through a set of convex sub-problems by using the gradients of the optimization objective and constraints with respect to the design variables $\bp$ derived below.

The gradient computation follows the chain rule. First consider the sensitivities of stiffness matrix $\bK$ with respect to design parameter $p_i$. Rewritting $\psi(\bx) = \bB^T \bD \bB$ for conciseness, we have
\begin{equation}
    \begin{aligned}
        \frac{\partial \bK}{\partial p_{i}} & = \frac{\partial}{\partial p_i} \int_{\Omega} \bB^T\bD\bB H(\Phi(\bx, \bp)) \mathrm{d} \Omega                                \\
                                            & = \int_{\Omega} \bB^T \bD \bB \frac{\partial H(\Phi)}{\partial \Phi} \frac{\partial \Phi}{\partial p_{i}} \mathrm{~d} \Omega \\
                                            & = \int_{\Omega} \psi(\bx) \frac{\partial H(\Phi)}{\partial \Phi} \frac{\partial \Phi}{\partial p_{i}} \mathrm{~d} \Omega.
    \end{aligned}
    \label{eq-dK_volume}
\end{equation}

The key point of above equation is to compute derivative of Heaviside function. We bring in Dirac delta function $\hat{\delta}(\Phi)$
\begin{equation}
    \hat{\delta}(\Phi)=\nabla H(\Phi) \cdot \frac{\nabla \Phi}{\|\nabla \Phi\|}=\frac{\mathrm{d} H(\Phi)}{\mathrm{d} \Phi} \nabla \Phi \cdot \frac{\nabla \Phi}{\|\nabla \Phi\|}=\frac{\mathrm{d} H(\Phi)}{\mathrm{d} \Phi}\|\nabla \Phi\|,
\end{equation}
where
\begin{equation}
    \|\nabla \Phi\|=\sqrt{\left(\frac{\partial \Phi}{\partial x}\right)^{2}+\left(\frac{\partial \Phi}{\partial y}\right)^{2} + \left(\frac{\partial \Phi}{\partial z}\right)^{2}}.
\end{equation}

Consequently, Eq. \eqref{eq-dK_volume} is rewritten as
\begin{equation}
    \begin{aligned}
        \frac{\partial \bK}{\partial p_{i}} & =\int_{\Omega} \psi(\bx) \frac{\partial H(\Phi)}{\partial \Phi} \frac{\partial \Phi}{\partial p_{i}} \mathrm{~d} \Omega                                                     \\
                                            & =\int_{\Omega} \psi(\bx) \frac{\partial \Phi}{\partial p_{i}} \frac{1}{\|\nabla \Phi\|}\left(\frac{\partial H(\Phi)}{\partial \Phi}\|\nabla \Phi\|\right) \mathrm{d} \Omega \\
                                            & =\int_{\Omega} \psi(\bx) \frac{\partial \Phi}{\partial p_{i}} \frac{1}{\|\nabla \Phi\|} \hat{\delta}(\Phi) \mathrm{d} \Omega                                                \\
                                            & =\int_{\partial \Omega_{p}} \psi(\bx) \frac{\partial \Phi}{\partial p_{i}} \frac{1}{\|\nabla \Phi\|} \mathrm{d} \Gamma,
    \end{aligned}
    \label{eq-dK_globalBoundary}
\end{equation}
where $\partial \Omega_{p}$ denotes boundary of feature model $\Omega_p$. This way, the volume integral of sensitivities is transformed into a boundary integral.

According to the expression of $\Phi(\bx, \bp)$ in Eq.~\eqref{eq:H}, we further have for Eq.~\eqref{eq-dK_globalBoundary},
\begin{equation}
    \frac{\partial \Phi(\bx, \bp)}{\partial p_{i}}=\sum_{j=1}^{n}\frac{\partial \Phi(\bx, \bp)}{\partial \phi_{j}} \cdot \frac{\partial \phi_{j}}{\partial p_{i}},
\end{equation}
\begin{equation}
    \left\|\nabla \Phi(\bx, \bp)\right\|=\left\|\sum_{j=1}^{n} \frac{\partial \Phi(\bx, \bp)}{\partial \phi_{j}} \nabla \phi_{j}\right\|.
\end{equation}

Noting that design variable $p_i$ is only associated to one feature $f_i$, we have
\begin{equation}
    \begin{aligned}
        \frac{\partial \phi_j}{\partial p_i} = 0, \quad j \neq i.
    \end{aligned}
\end{equation}

Let $S_{i} = \frac{\partial \Phi (\bx, \bp)}{\partial \phi_i}$ be the logical operationde defined by parent bifurcation nodes of $f_i$ in CSG tree, and $S_{i} \in \{-1, 1\}$. Accordingly, the integral domain of Eq. \eqref{eq-dK_globalBoundary} can be reduced from boundary $\partial \Omega_p$ of whole feature model $\Omega_p$ to boundary $\partial f_i$ of a feature $f_i$, that is, computations.
\begin{equation}
    \frac{\partial \Phi(\bx, \bp)}{\partial p_{i}}=\frac{\partial \Phi(\bx,\bp)}{\partial \phi_{i}} \cdot \frac{\partial \phi_{i}}{\partial p_{i}}= S_i\frac{\partial \phi_{i}}{\partial p_{i}},
\end{equation}
which avoids redundant integration.

Similarly, we have
\begin{equation}
    \left\|\nabla \Phi(\bx, \bp)\right\|=\left\| S_{i} \nabla \phi_{i}\right\| = \left\|\nabla \phi_{i}\right\|
\end{equation}
for quadrature points along boundary of feature $f_i$.

Accordingly, the sensitivities in Eq. \eqref{eq-dK_globalBoundary} is reduced to
\begin{equation}
    \begin{aligned}
        \frac{\partial \bK}{\partial p_i} & = \int_{\partial \Omega_{p}} \psi(\bx) \frac{\partial \Phi}{\partial p_{i}} \frac{1}{\|\nabla \Phi\|} \mathrm{d} \Gamma     \\
                                          & = \int_{\partial f_i} S_{i} \psi(\bx) \frac{\partial \phi_i}{\partial p_{i}} \frac{1}{\|\nabla \phi_i\|} \mathrm{d} \Gamma.
    \end{aligned}
    \label{eq-dK_localBoundary}
\end{equation}

Afterwards, we consider sensitivities of structural compliance $C$,
\begin{equation}
    \begin{aligned}
        \frac{\partial C}{\partial p_{i}} & =\frac{\partial \bF^{\mathrm{T}}}{\partial p_{i}} \bu+\bF^{\mathrm{T}} \frac{\partial \bu}{\partial p_{i}}                                                              \\
                                          & =\frac{\partial \bF^{\mathrm{T}}}{\partial p_{i}} \bu+\bF^{\mathrm{T}} \bK^{-1}\left(\frac{\partial \bF}{\partial p_{i}}-\frac{\partial \bK}{\partial p_{i}} \bu\right) \\
                                          & =2 \frac{\partial \bF^{\mathrm{T}}}{\partial p_{i}} \bu-\bu^{\mathrm{T}} \frac{\partial \bK}{\partial p_{i}} \bu.
    \end{aligned}
\end{equation}
Assuming for simplicity the independence of load $\bF$ and feature design variables, the first term in the above equation is zero. Based on sensitivities of stiffness matrix in Eq. \eqref{eq-dK_localBoundary}, we have
\begin{equation}
    \begin{aligned}
        \frac{\partial C}{\partial p_{i}} & = - \bu^T \frac{\partial \bK}{\partial p_i} \bu                                                                                                          \\
                                          & = - \bu^T (\int_{\partial f_i}S_{i} \psi(\bx) \frac{\partial \phi_i}{\partial p_{i}} \frac{1}{\left\|\nabla \phi_i\right\|} \mathrm{d} \Gamma) \bu       \\
                                          & = - \bu^T (\int_{\partial f_i} S_{i} \bB^T \bD \bB \frac{\partial \phi_i}{\partial p_{i}} \frac{1}{\left\|\nabla \phi_i\right\|} \mathrm{d} \Gamma) \bu.
    \end{aligned}
\end{equation}

\begin{figure}[htbp]
    \centering
    \includegraphics[width=0.3\textwidth]{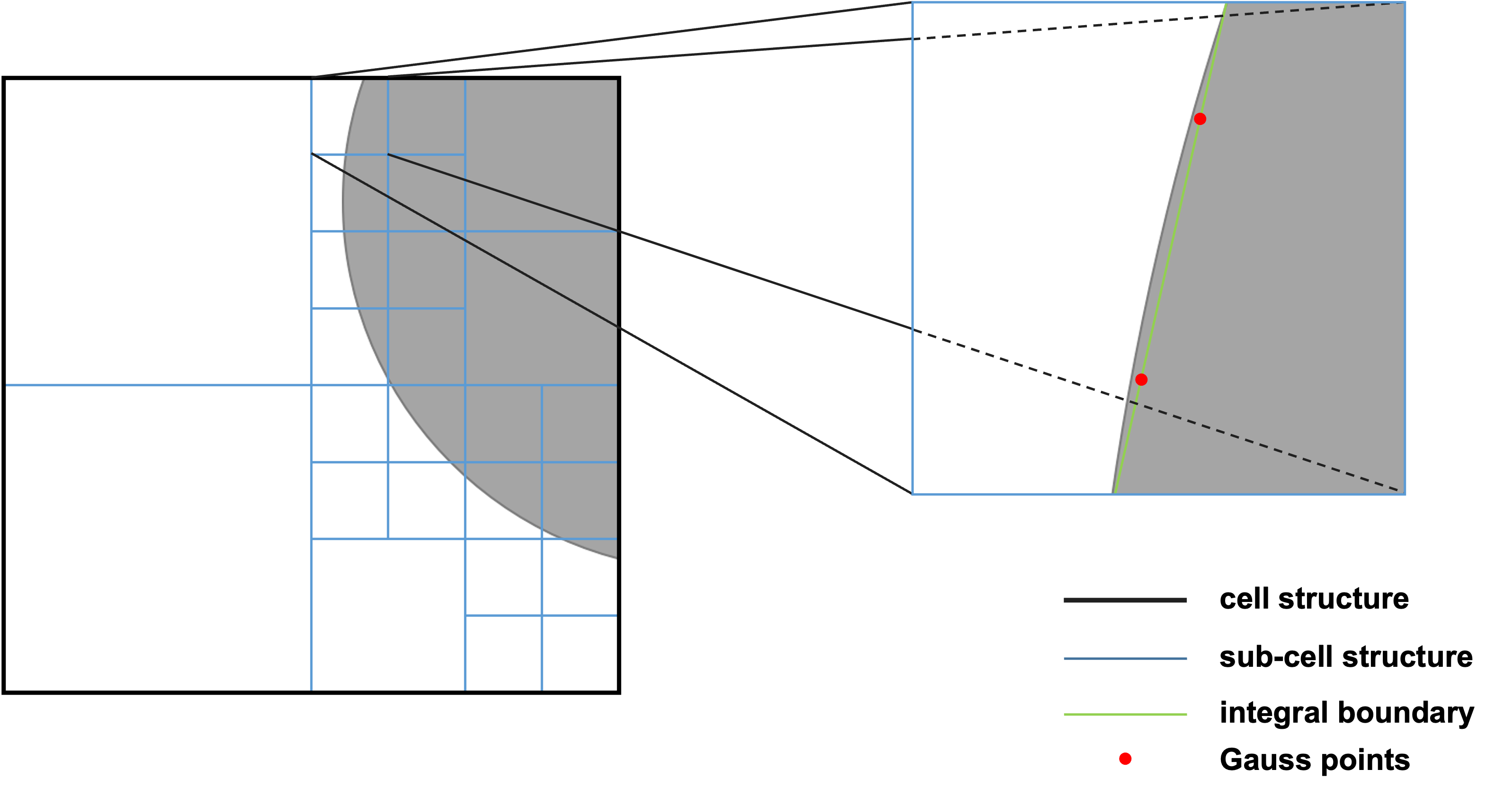}
    \caption{Sensitivity computation (in 2D) as adaptive boundary integration of finite cell (bold black grid): In each sub-cell (thin blue grid), the boundary is approximated with line segments (green), along which Gauss quadrature points (red) are taken.}
    \label{fig-boundary_integral}
\end{figure}

In our numerical implementation, we use triangles in 3D (lines in 2D) to approximate structural boundary of model for adaptive integration; see Fig. \ref{fig-boundary_integral} for an illustration.

\section{Numerical examples and \rev{}{discussions}}\label{sec:examples}
The proposed XVoxel-based method for parametric design optimization of feature models has been implemented in Matlab on a computer with Intel Core i7-12700 3.6 GHz CPU, 64GB RAM. Five different examples are shown to demonstrate its effectiveness: the first three on simulation during interactive editing to test its computational accuracy and efficiency, and the last two on its usage for feature-based design optimization. The CAD model sizes are all measured in micrometer, and the material has a Young's modulus $E = 2e^{11} Pa$ and Poisson's ratio $\nu = 0.3$. In FCM computing, all examples have the octree refinement depth $d=3$ and the shape function order $p=2$, except for the first example where $d=4$ and $p=3$.

The locality characteristics of XVoxel are measured in terms of the number of active voxels (i.e. voxels affected by local feature updates) against that of FCM. The fidelity of XVoxel or FCM is measured via its displacement residual (in terms of top $10 \%$) against the benchmark:
\begin{equation}
    r_u=\frac{\|\bu_1 - \bu_0\|}{\|\bu_0\|},
\end{equation}
where $\bu_1$, $\bu_0$ are the computed and the benchmark displacements, respectively.


The experimental settings and results are summarized in Table \ref{tb-performance}, including mesh size, DOFs, timing (per step/iter), the number of active voxels and relative error $r_{\bu}$. In all these examples, FEA simulation results on tetrahedral meshes (in Ansys Workbench 22R1) were taken as the benchmark. Three other approaches were tested to show the method's simulation accuracy and efficiency: standard FCM approach~\cite{schillinger2015finite}, XVoxel-FM combining FCM with XVoxel, XVoxel-CBN combining CBN-based FCM~\cite{li2022analysis}; the last two are our approaches. The DOFs of FEA and XVoxel were set approximately same for the comparisons to be fair.

As can be seen from Table~\ref{tb-performance}, the error of FCM (XVoxel) is as low as $0.005\%$, demonstrating its high accuracy. FCM and XVoxel-FCM always have the same simulation accuracy, and CBN-FCM is very close to them. Other examples may have higher error due to the need for balancing accuracy and effiency. Note that FCM (XVoxel) can reach a prescribed accuracy voxel refinement or degree elevation~\cite{schillinger2015finite}. In all examples, FEA is much more efficient than FCM while XVoxel-FCM improves the efficiency, resulting in a similar computation time to FEA, due to its local computations. XVoxel-CBN greatly improves the efficiency of XVoxel-FCM due to its usage of piecewise linear shape functions. We have to mention again that in comparison with FEA FCM or XVoxel (either FCM or CBN version) has a prominent advantage in its much easier and more robust voxolization than FEA's tetrahedral meshing. 

\begin{table*}[htbp]
    \centering
    \caption{Summary of the performance of XVoxel on the tested numerical examples in comparison with those using FEA and FCM (standard FCM, XVoxel-FCM combining FCM with XVoxel, XVoxel-CBN combining CBN-based FCM and XVoxel). Here, $d$ is the depth of octree refinement for element integration and $p$ is the order of shape functions.
    }
    \resizebox{\textwidth}{!}{
        \begin{tabular}{cccccccccccccc}
            \toprule
            \multirow{2}{*}{Example}                    &
            \multirow{2}{*}{Step  (Iter)}               &
            \multicolumn{2}{c}{Mesh Size}               &
            \multicolumn{3}{c}{DOFs}                    &
            \multicolumn{4}{c}{Timings (per Step/Iter)} &
            \multicolumn{2}{c}{Active Voxel Number}     &
            \multirow{2}{*}{$r_{\mathbf{u}}$ (\%)}                                                                                                                                                                                                                                      \\
                                                        &     & FEA               & FCM(XVoxel)              & FEA               & FCM (XVoxel-FCM)          & XVoxel-CBN                  & FEA               & FCM     & XVoxel-FCM & XVoxel-CBN & FCM   & XVoxel &                   \\
            \midrule
            \multirow{5}{*}{\#1}                        & 1   & 13,801            & \multirow{5}{*}{675}     & 63,372            & \multirow{5}{*}{63,480}   & \multirow{5}{*}{47,280}     & 5.3               & 35.1    & 35.1       & 6.0        & 33    & 33     & 0.0061            \\
                                                        & 2   & 14,118            &                          & 64,833            &                           &                             & 5.2               & 28.8    & 28.9       & 2.9        & 21    & 21     & 0.0169            \\
                                                        & 3   & 13,807            &                          & 63,600            &                           &                             & 5.2               & 23.2    & 23.7       & 1.8        & 21    & 21     & 0.0263            \\
                                                        & 4   & 13,844            &                          & 63,714            &                           &                             & 6.2               & 17.3    & 17.7       & 1.3        & 15    & 15     & 0.0049            \\
                                                        & 5   & 13,908            &                          & 64,029            &                           &                             & 5.2               & 11.6    & 11.8       & 0.8        & 9     & 9      & 0.0047            \\
            \midrule
            \multirow{6}{*}{\#2}                        & 1   & 22,789            & \multirow{6}{*}{4,004}   & 109,632           & \multirow{6}{*}{108,135}  & \multirow{6}{*}{255,024}    & 8.5               & 44.9    & 44.9       & 13.8       & 1,182 & 1,182  & 0.0328            \\
                                                        & 2   & 22,579            &                          & 108,777           &                           &                             & 7.3               & 44.2    & 6.3        & 1.1        & 1,182 & 121    & 0.0825            \\
                                                        & 3   & 22,326            &                          & 108,006           &                           &                             & 7.2               & 42.6    & 9.4        & 1.8        & 1,182 & 183    & 0.0812            \\
                                                        & 4   & 22,465            &                          & 109,350           &                           &                             & 7.3               & 44.3    & 9.8        & 2.1        & 1,185 & 189    & 0.1555            \\
                                                        & 5   & 22,611            &                          & 109,569           &                           &                             & 7.3               & 43.7    & 5.6        & 1.4        & 1,191 & 128    & 0.3221            \\
                                                        & 6   & 22,326            &                          & 108,336           &                           &                             & 8.5               & 40.2    & 8.5        & 1.4        & 1,141 & 227    & 0.0263            \\

            \midrule
            \multirow{12}{*}{\#3}                       & 1   & 222,816           & \multirow{12}{*}{39,775} & 1,019,364         & \multirow{12}{*}{998,325} & \multirow{12}{*}{2,365,080} & 22.1              & 140.1   & 143.1      & 80.9       & 6,040 & 6,040  & 0.0928            \\
                                                        & 2   & 217,244           &                          & 1,001,976         &                           &                             & 19.7              & 151.5   & 33.9       & 5.2        & 6,048 & 397    & 1.0381            \\
                                                        & 3   & 218,659           &                          & 1,006,104         &                           &                             & 21.6              & 152.0   & 35.0       & 6.0        & 6,156 & 491    & 1.1859            \\
                                                        & 4   & 215,383           &                          & 996,989           &                           &                             & 23.1              & 166.2   & 35.3       & 5.4        & 6,356 & 412    & 1.3995            \\
                                                        & 5   & 215,185           &                          & 996,513           &                           &                             & 20.8              & 161.3   & 28.0       & 1.1        & 6,369 & 97     & 1.4215            \\
                                                        & 6   & 214,206           &                          & 991,353           &                           &                             & 23.5              & 167.1   & 35.8       & 4.5        & 6,545 & 458    & 1.4720            \\
                                                        & 7   & 216,983           &                          & 1,002,681         &                           &                             & 21.0              & 168.0   & 31.0       & 3.3        & 6,561 & 216    & 1.4834            \\
                                                        & 8   & 215,132           &                          & 997,344           &                           &                             & 20.9              & 174.4   & 34.8       & 3.9        & 6,681 & 325    & 2.5318            \\
                                                        & 9   & 214,288           &                          & 993,975           &                           &                             & 18.0              & 181.3   & 30.3       & 2.0        & 6,789 & 190    & 2.3235            \\
                                                        & 10  & 216,411           &                          & 1,004,442         &                           &                             & 21.3              & 210.0   & 29.5       & 2.2        & 6,633 & 102    & 2.4980            \\
                                                        & 15  & 214,690           &                          & 992,787           &                           &                             & 23.7              & 201.5   & 48.2       & 9.0        & 5,837 & 677    & 2.7177            \\
                                                        & 20  & 213,749           &                          & 992,295           &                           &                             & 23.5              & 223.4   & 42.6       & 5.5        & 6,485 & 459    & 3.0647            \\
            \midrule
            \multirow{4}{*}{\#4}                        & 1   & \multirow{4}{*}{} & \multirow{4}{*}{9,100}   & \multirow{4}{*}{} & \multirow{4}{*}{240,975}  & \multirow{4}{*}{250,638}    & \multirow{4}{*}{} & 1,490.8 & 130.3      & 55.2       & 2,256 & 2,256  & \multirow{4}{*}{} \\
                                                        & 10  &                   &                          &                   &                           &                             &                   & 1,411.0 & 109.6      & 25.2       & 2,174 & 308    &                   \\
                                                        & 30  &                   &                          &                   &                           &                             &                   & 1,401.8 & 101.3      & 24.8       & 2,166 & 274    &                   \\
                                                        & 100 &                   &                          &                   &                           &                             &                   & 1,467.5 & 102.4      & 25.3       & 2,172 & 255    &                   \\

            \midrule
            \multirow{4}{*}{\#5}                        & 1   & \multirow{4}{*}{} & \multirow{4}{*}{33,150}  & \multirow{4}{*}{} & \multirow{4}{*}{279,265}  & \multirow{4}{*}{282,006}    & \multirow{4}{*}{} & 146.0   & 148.0      & 60.4       & 4,592 & 4,592  & \multirow{4}{*}{} \\
                                                        & 10  &                   &                          &                   &                           &                             &                   & 141.5   & 85.8       & 30.7       & 3,904 & 1,644  &                   \\
                                                        & 30  &                   &                          &                   &                           &                             &                   & 156.0   & 92.7       & 33.2       & 3,952 & 1,611  &                   \\
                                                        & 100 &                   &                          &                   &                           &                             &                   & 166.6   & 97.2       & 32.6       & 3,962 & 1,486  &                   \\
            \bottomrule
        \end{tabular}
    }
    \label{tb-performance}
\end{table*}

\subsection{Example \#1: an L-shaped model for simulation accuracy testing}

\begin{figure*}[htbp]
    \centering
    \subfigure[]{\includegraphics[width=0.2\textwidth,trim=0 0 0 0,clip]{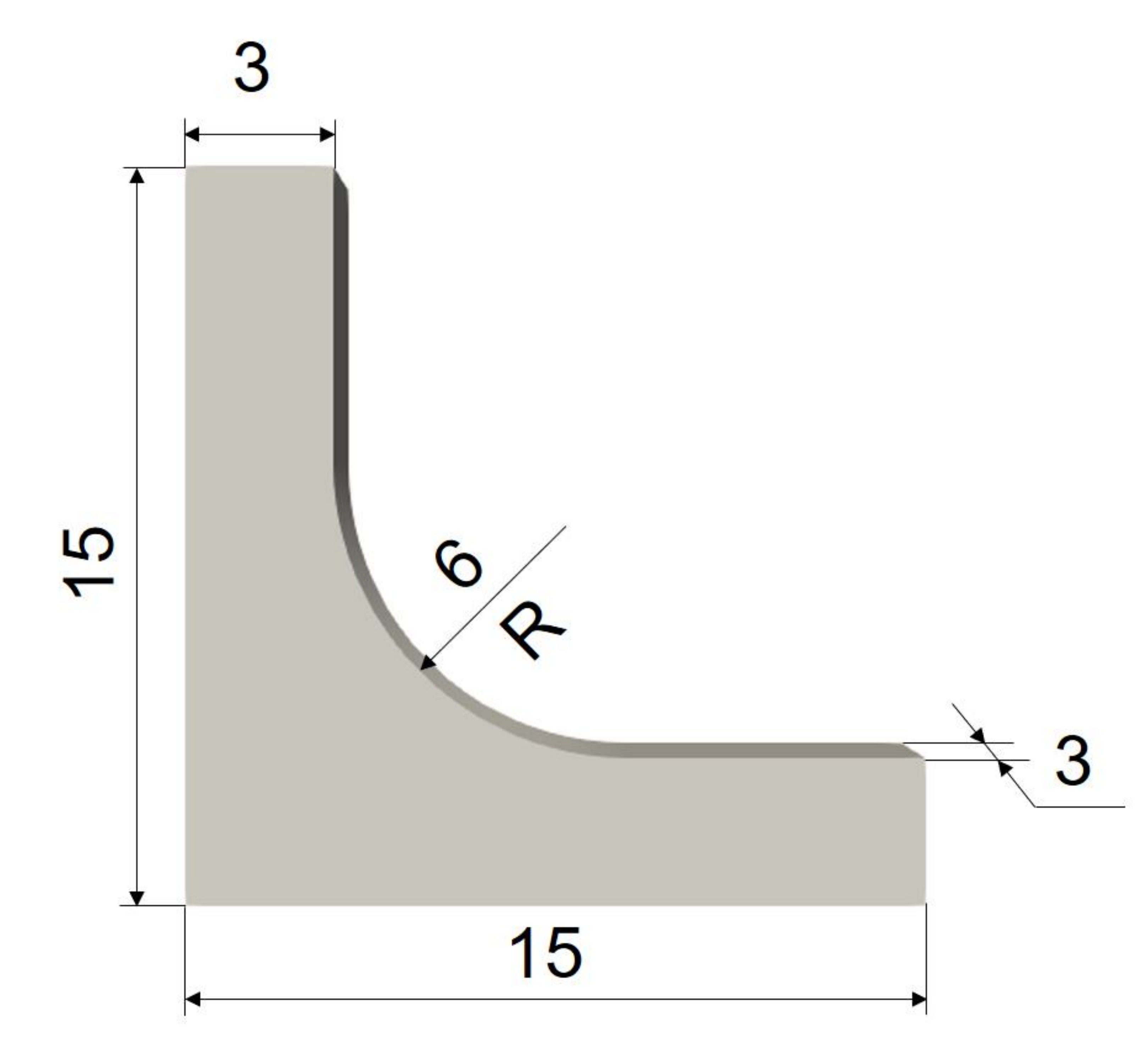}}\hspace{0.4cm}
    \subfigure[]{\includegraphics[width=0.18\textwidth,trim=0 0 0 0,clip]{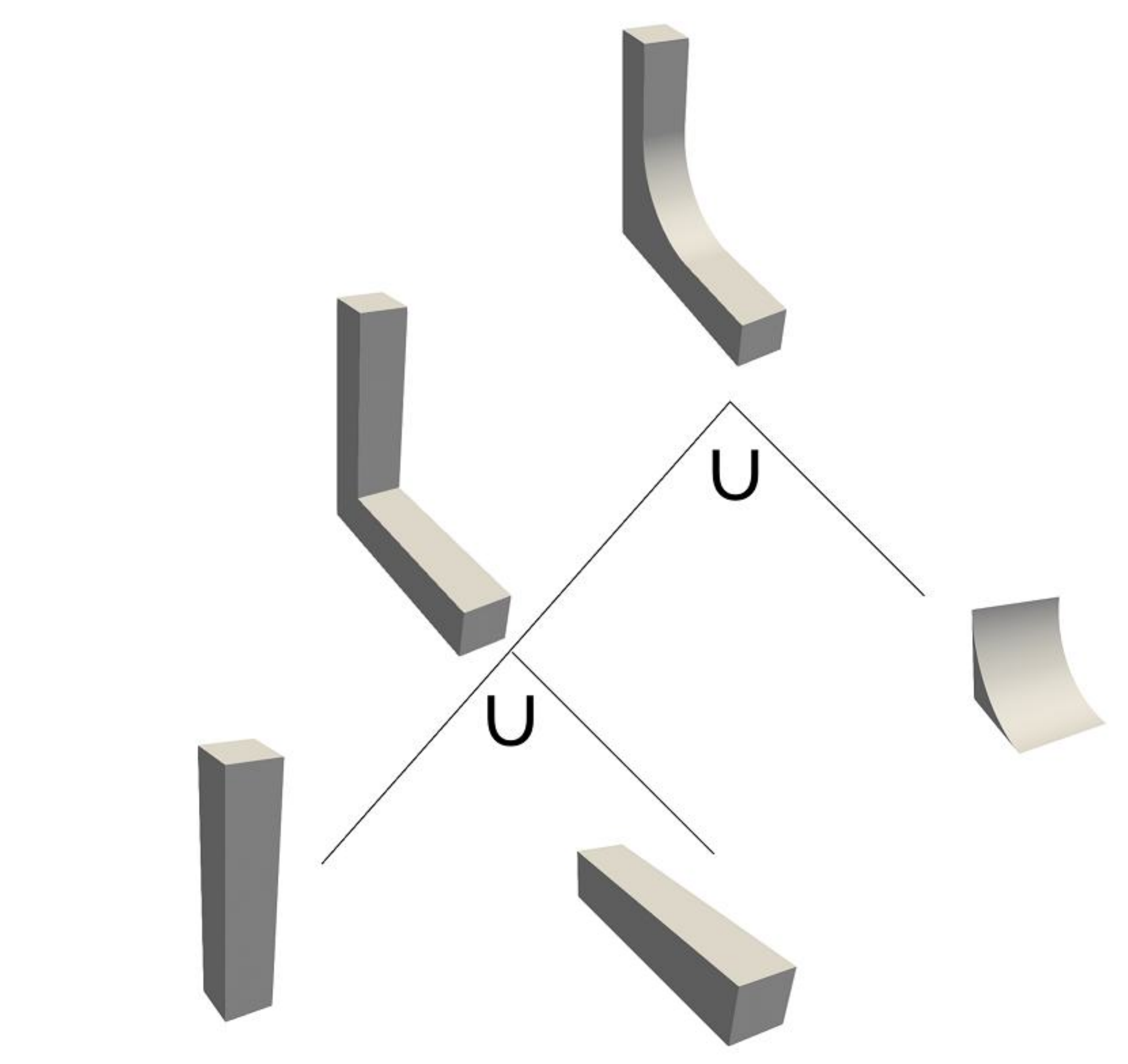}}
    \subfigure[]{\includegraphics[width=0.16\textwidth,trim=0 0 0 0,clip]{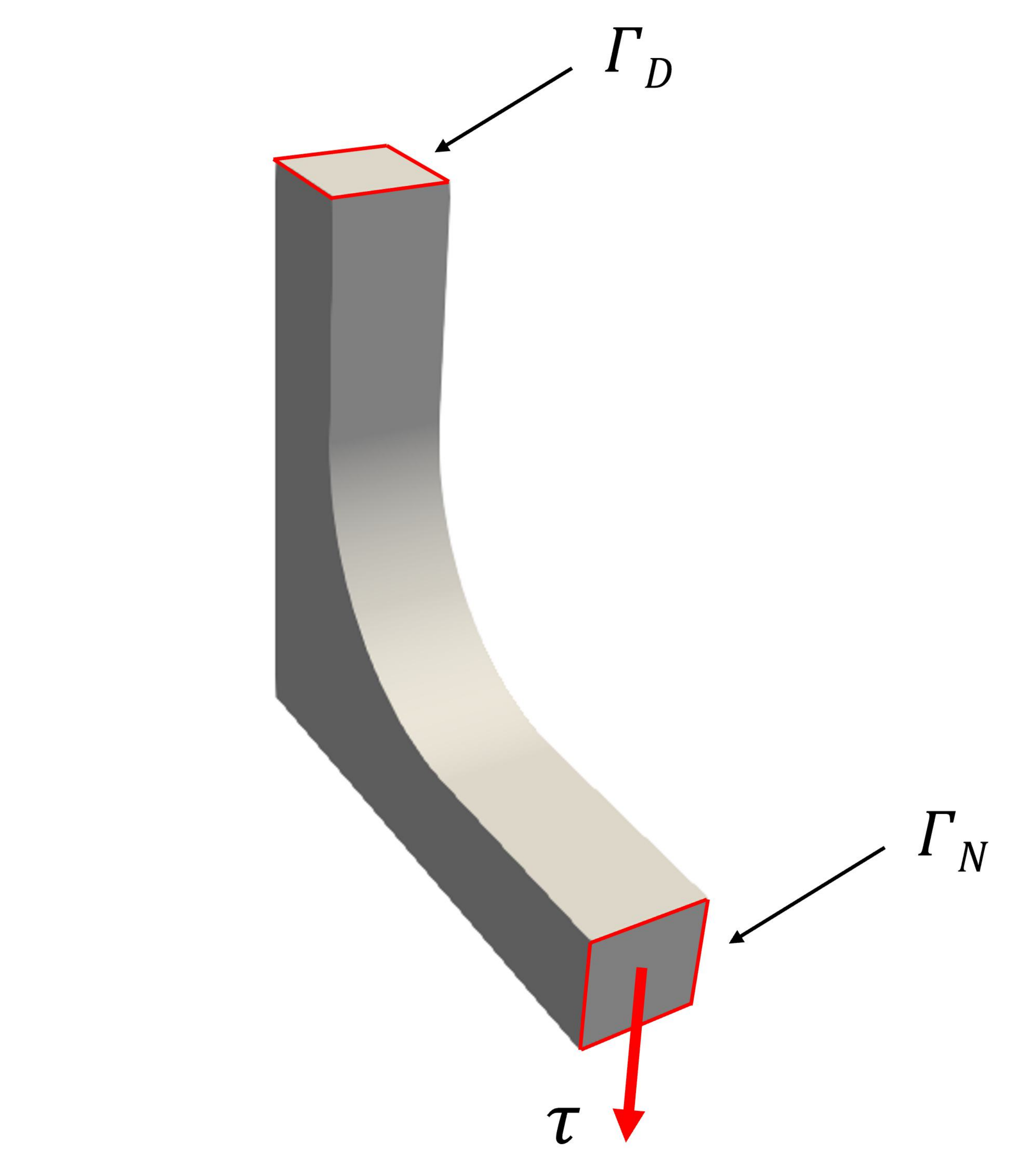}}
    \subfigure[]{\includegraphics[width=0.18\textwidth,trim=20 20 20 20,clip]{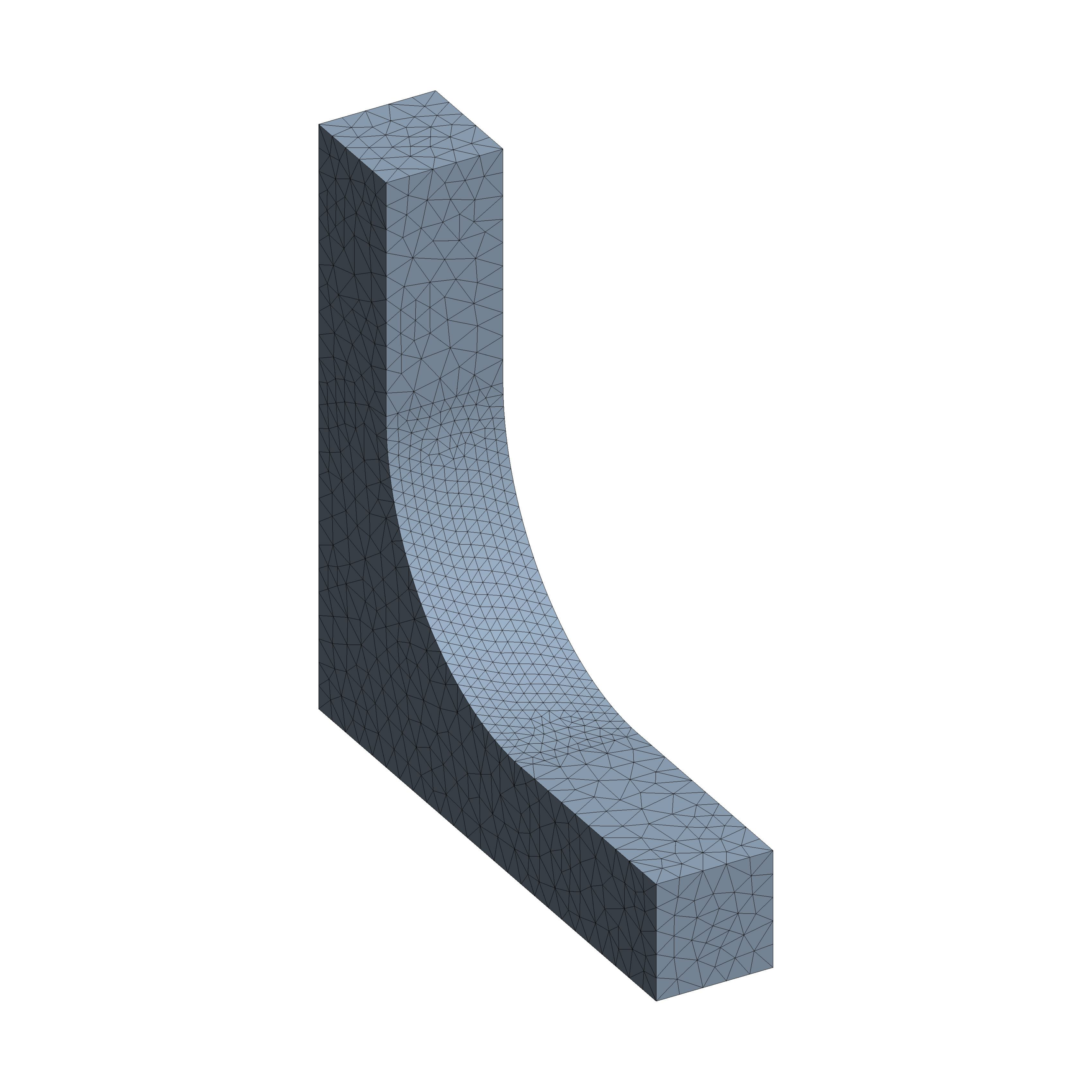}}
    \subfigure[]{\includegraphics[width=0.19\textwidth,trim=0 0 0 0,clip]{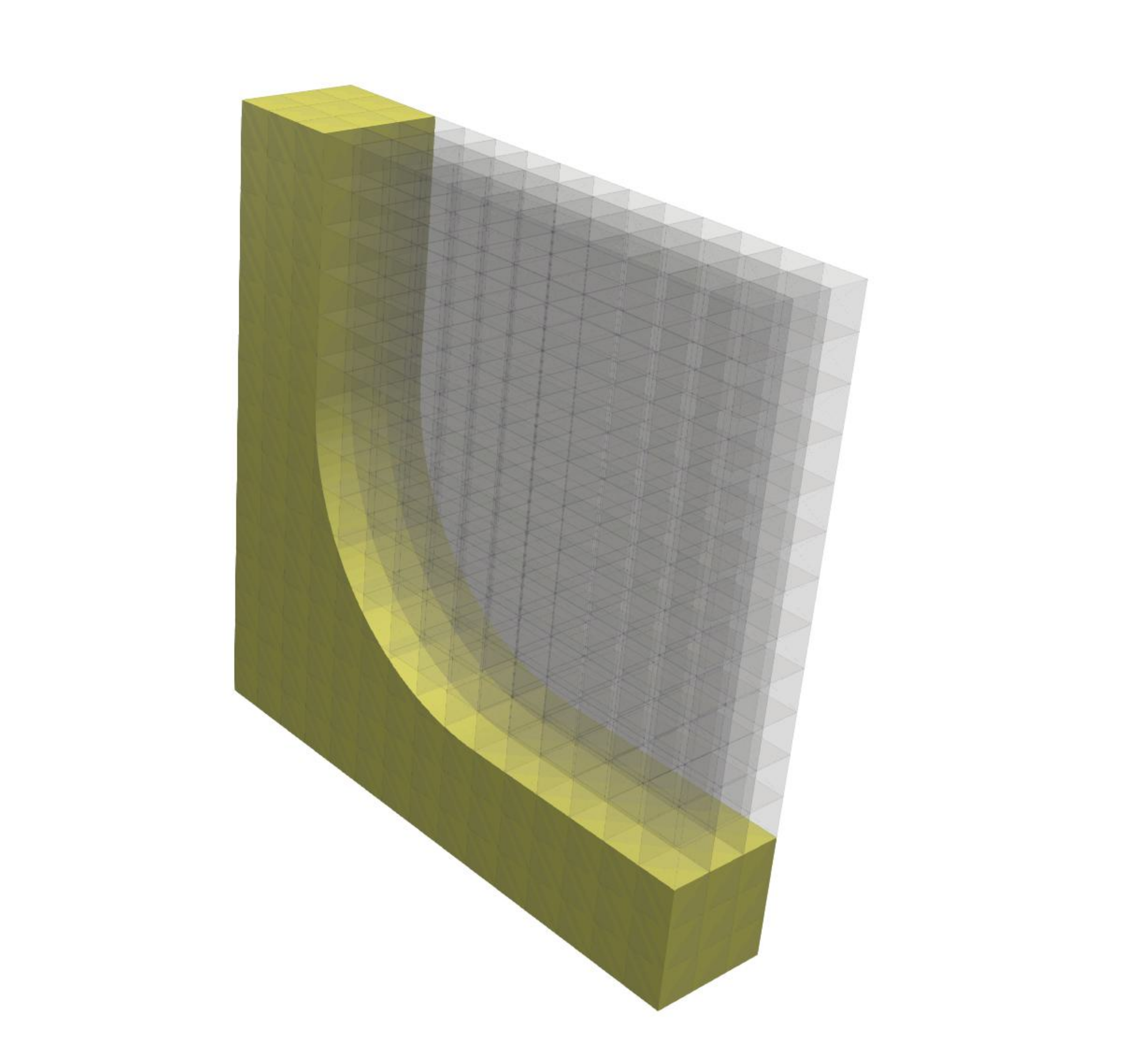}}
    \caption{
        Example \#1.
        (a) Parameters (mm) of the L-shaped model, where the radius R gradually varies from $6$ to $2$ with a
        step size of $-1$;
        (b) The CSG of the model;
        (c) Boundary conditions, where $\gamma_D$ is fixed and $\tau=100N/mm^2$ in $\gamma_N$;
        (d) FEA mesh with 6,325 tetrahedral elements;
        (e) FCM (XVoxel) mesh with $3\times15\times15$ voxels.
    }
    \label{fig-Lshape_model_info}
\end{figure*}

\begin{figure*}[htbp]
    \centering
    \subfigure[Relative error $r_{\bu}$]{\includegraphics[width=0.3\textwidth]{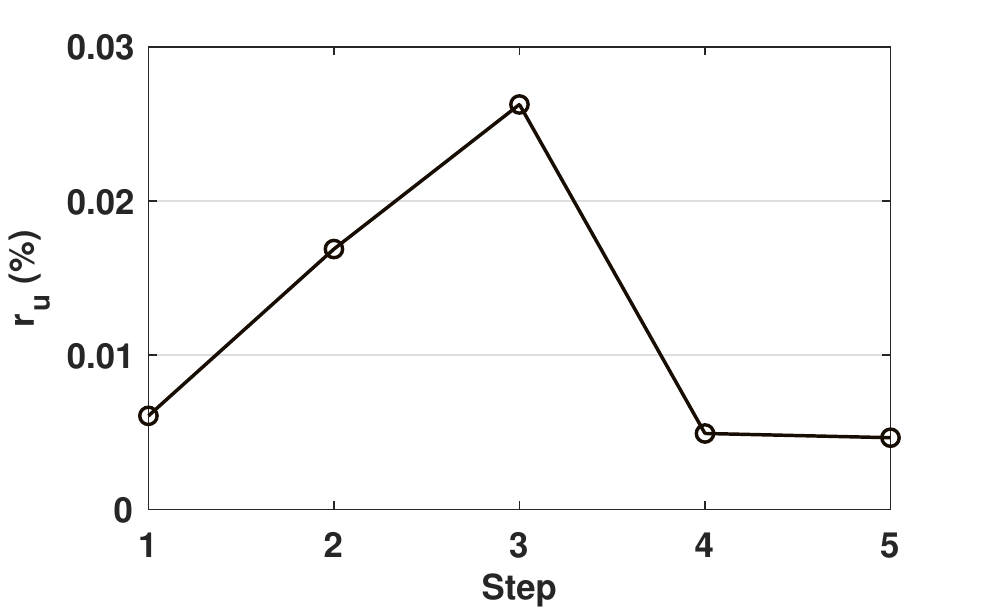}}
    \hspace{0.5cm}
    \subfigure[The number of active voxels]{\includegraphics[width=0.3\textwidth]{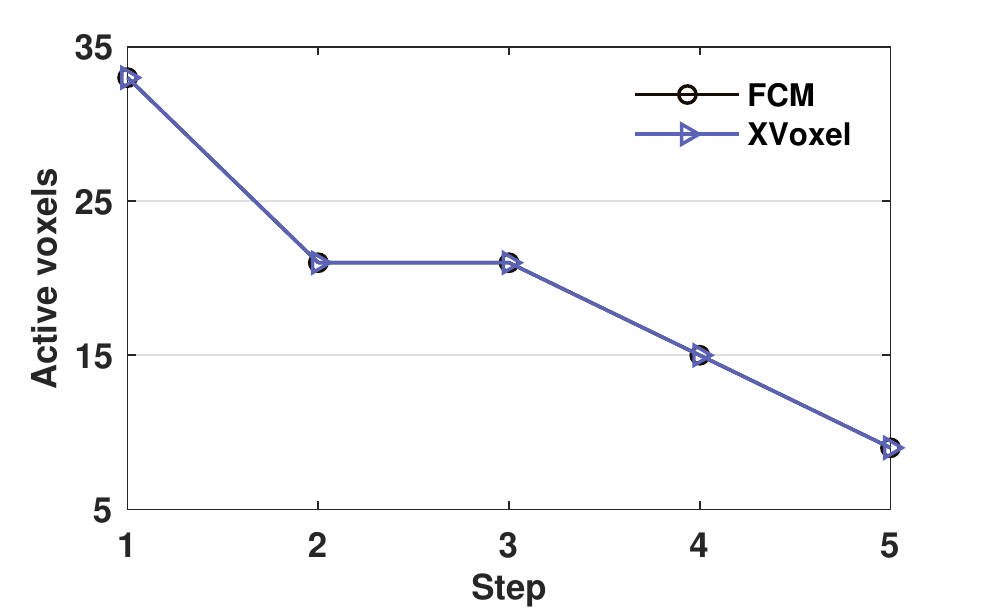}}
    \hspace{0.5cm}
    \subfigure[Timing]{\includegraphics[width=0.3\textwidth]{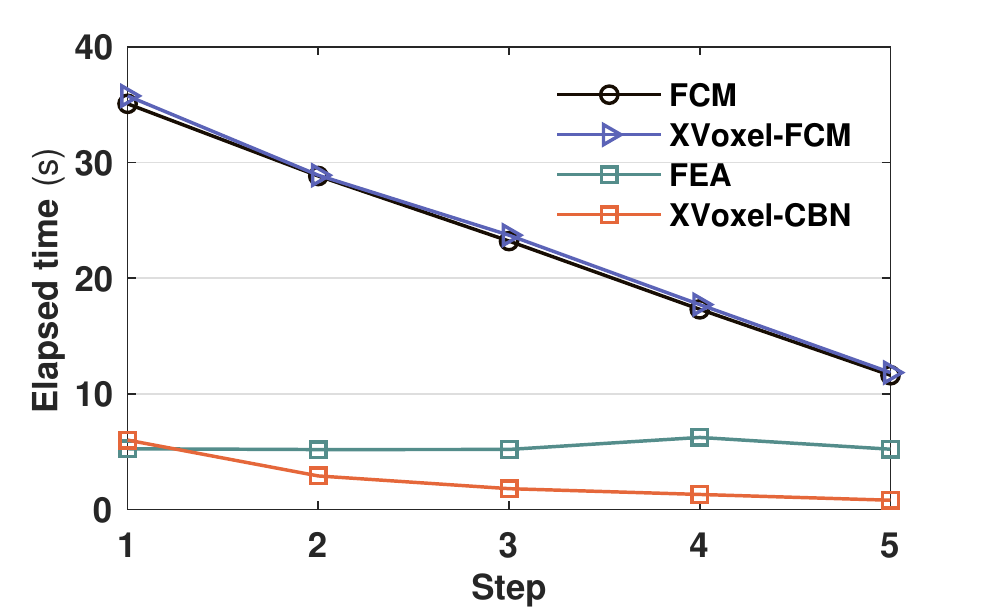}}
    \caption{Performance statistics of Example \#1 in Fig.~\ref{fig-Lshape_model_info}
    \rev{}{: (a) Displacement residual $r_u$ between XVoxel and FCM;
    (b) The number of active voxels by FCM and XVoxel;
    (c) Timing of four methods FEA, FCM, XVoxel-FCM (XVoxel based on FCM) and XVoxel-CBN (XVoxel based on CBN).}
    }
    \label{fig-Lshape_statistics}
\end{figure*}

\begin{figure*}[htbp]
    \centering
    \includegraphics[width=0.75\textwidth]{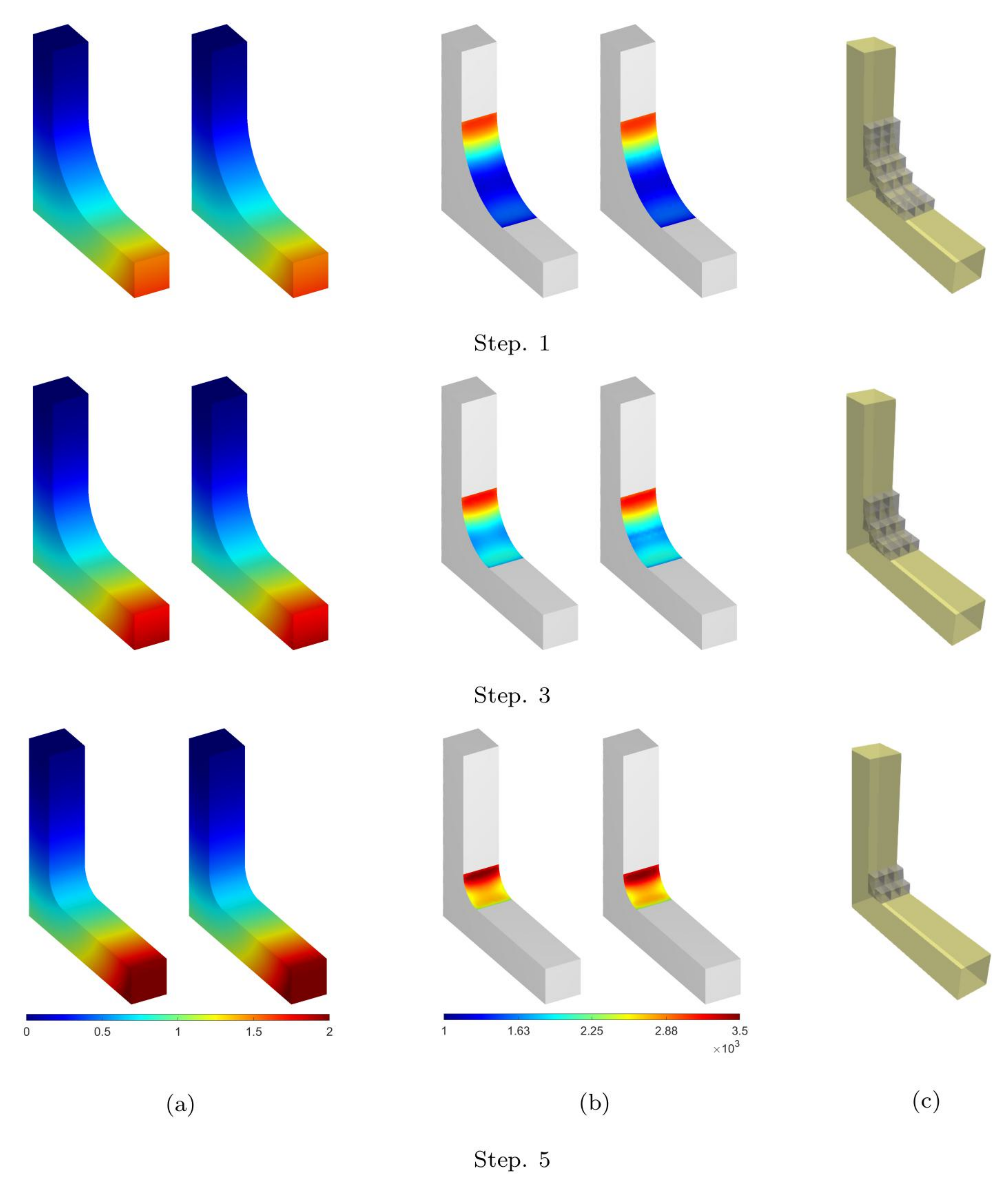}
    \caption{Results for an L-shaped model in editing step 1, 3 and 5. (a): Displacement norm (mm) from FEA (left) and FCM/XVoxel (right);
        (b) Von Mise stress (MPa) from FEA (left) and FCM/XVoxel (right);
        (c) Active voxels (in grey) of XVoxel. }
    \label{fig-Lshape_results}
\end{figure*}

%
The accuracy of the proposed method was first tested on a classic L-shaped model as shown in Fig. \ref{fig-Lshape_model_info}, constructed by combining two cubes and one rounded corner. The model is fixed on its upper face and subject to a downward traction of $\tau=100N/mm^2$ on its right face. The tetrahedral mesh of FEA has 6.3K elements, and the FCM has 768 voxels.

The rounded corner radii was varied from 6mm to 2mm at a step of $-1$mm. The simulation error, the number of active voxels, and timings for each step were respectively plotted in Figs.~\ref{fig-Lshape_statistics}(a),(b),(c). FEA and XVoxel-FCM has a very close approximation at an error $r_{\mathbf{u}} = 0.03\%$, as can also be observed from distributions of their displacement norm and von Mises stress in Figs.~\ref{fig-Lshape_results}(a),(b). We also notice from Figs.~\ref{fig-Lshape_results}(c) and~\ref{fig-Lshape_statistics}(b),(c) that FCM and XVoxel-FCM have exactly the same number of active voxels and computational timings in this example. This is because the cut cells in FCM are just the active voxels due to the regular shape of the L-shaped model.


\subsection{Example \#2: a connector model for simulation efficiency testing}

\begin{figure*}[htbp]
    \centering
    \subfigure[]{\includegraphics[width=0.2\textwidth]{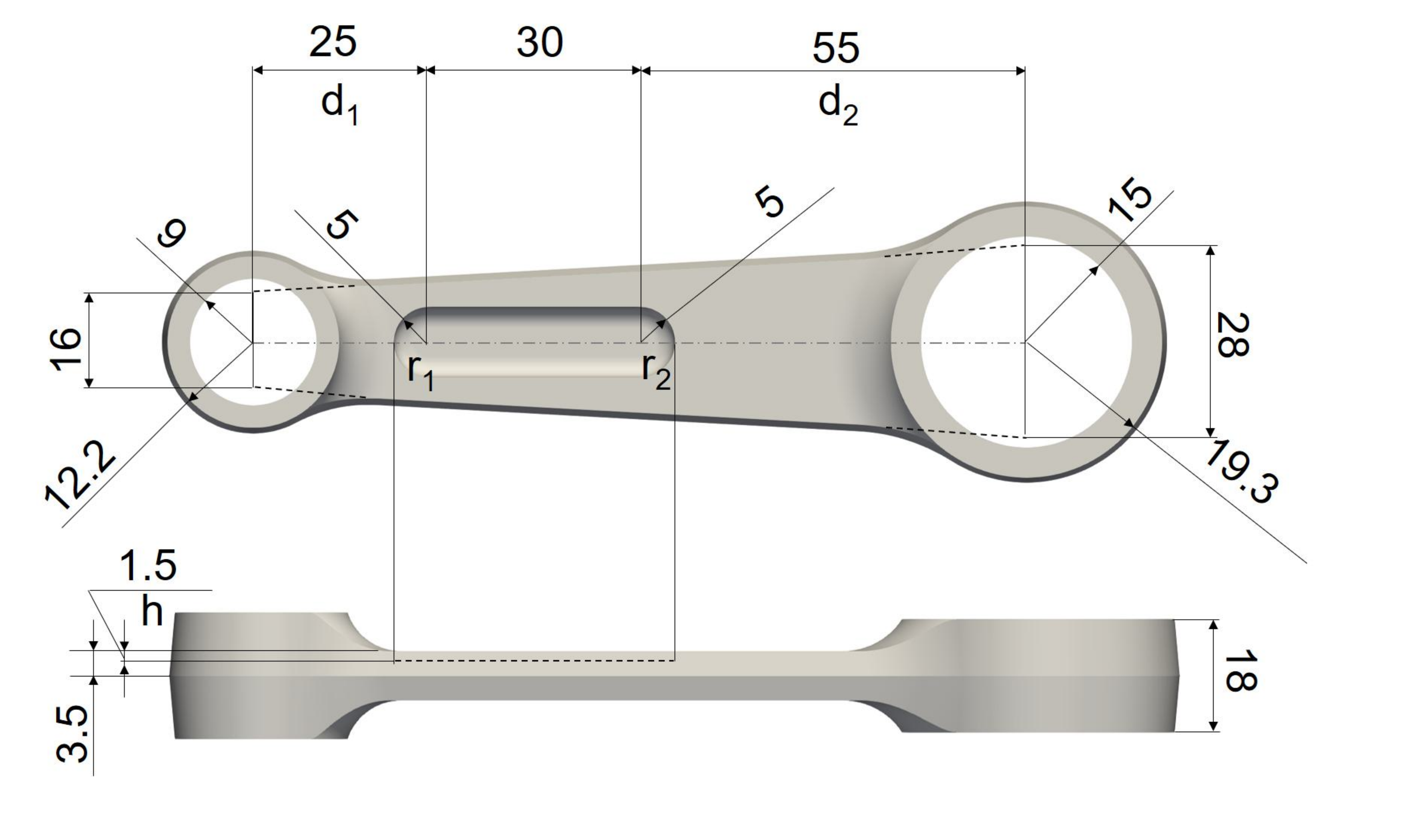}}
    \subfigure[]{\includegraphics[width=0.18\textwidth]{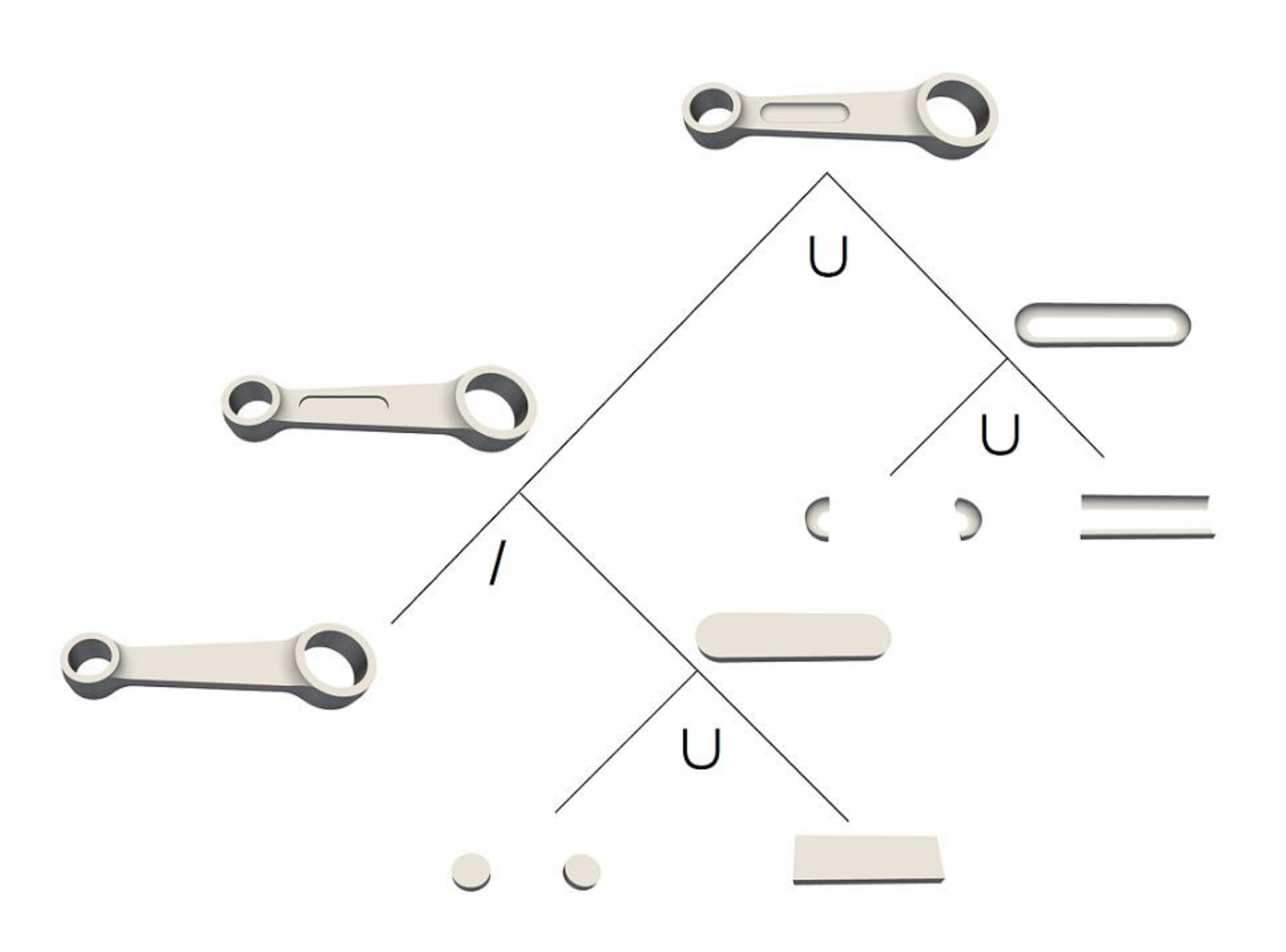}}
    \subfigure[]{\includegraphics[width=0.17\textwidth,trim=0 45 0 0,clip]{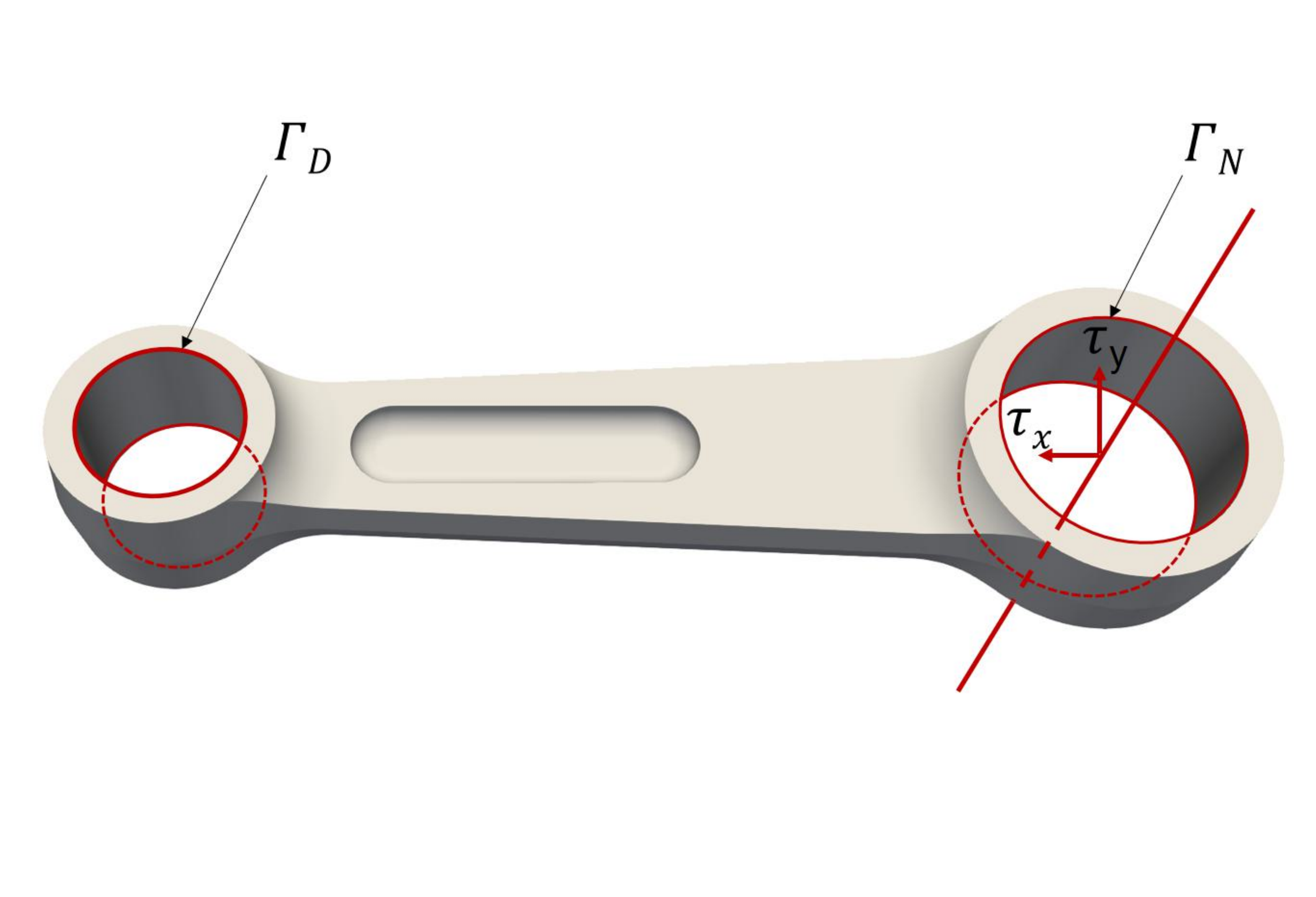}}
    \subfigure[]{\includegraphics[width=0.18\textwidth]{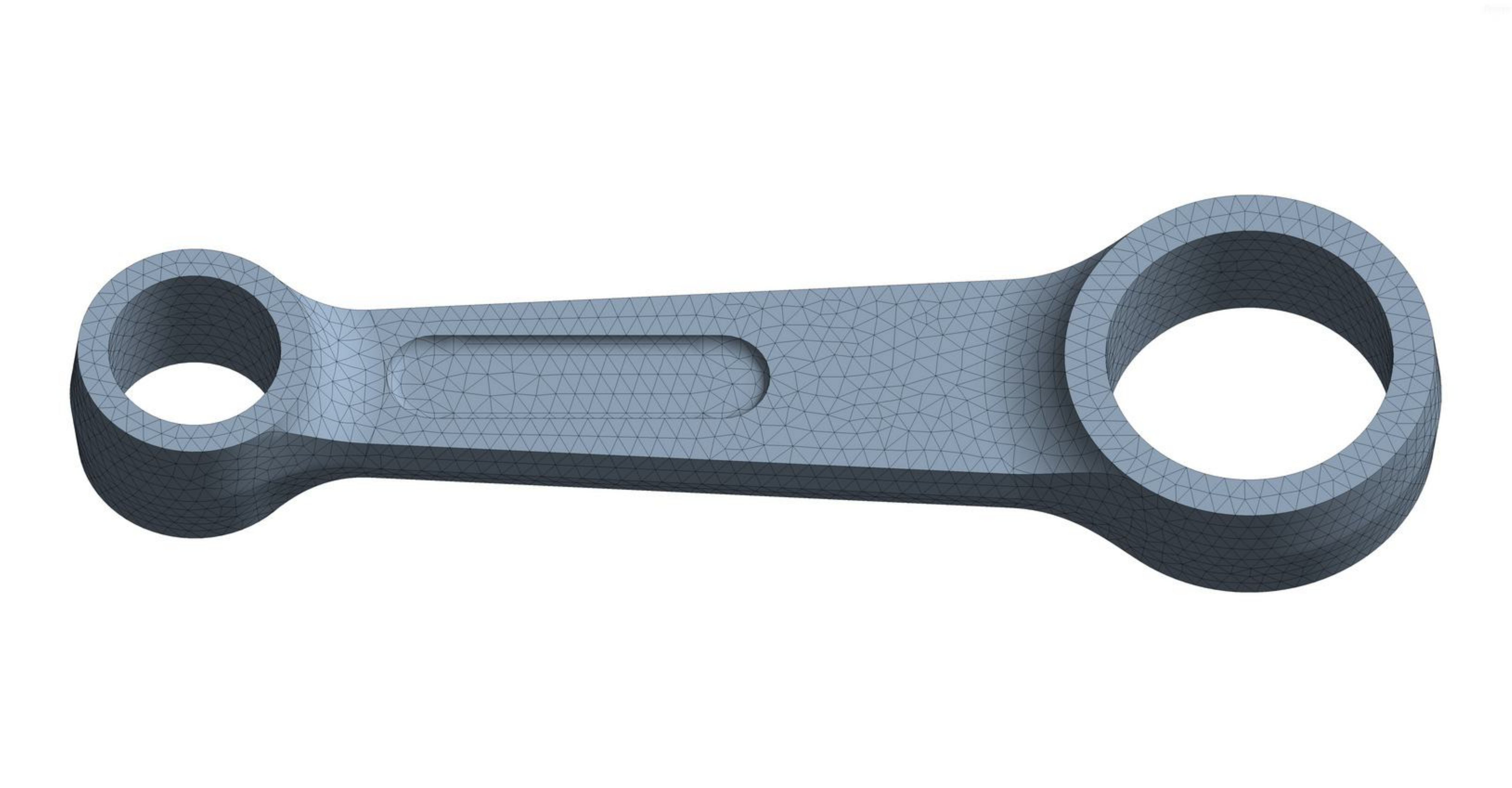}}
    \subfigure[]{\includegraphics[width=0.18\textwidth,trim=200 100 200 0,clip]{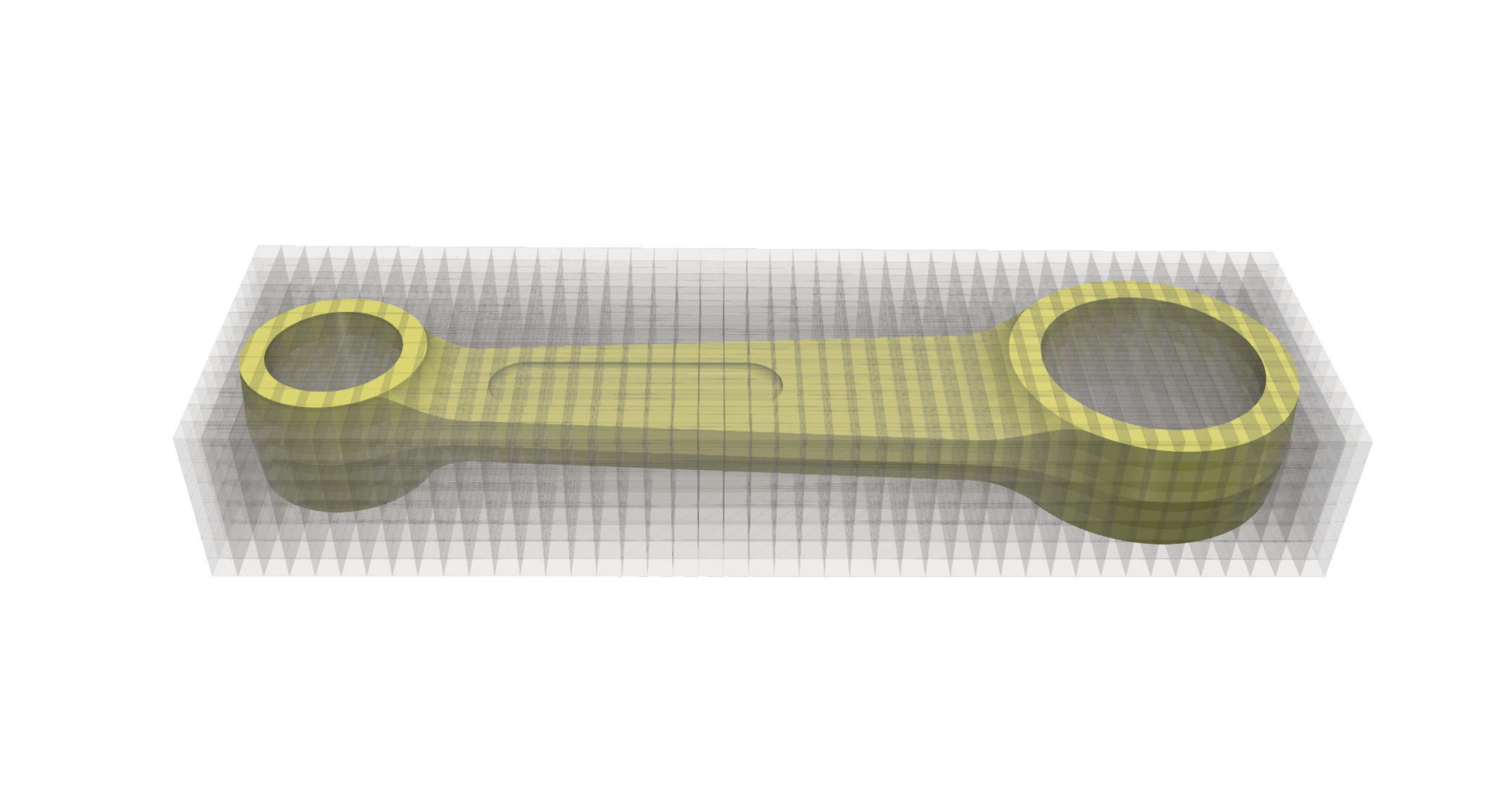}}

    \caption{
        Example \#2.
        (a) Parameters of the piston model\rev{}{, where modified variables are $d_1, d_2, r_1, r_2$ and $h$};
        (b) The CSG of the model;
        (c) Boundary conditions where $\Gamma_D$ is fixed and $\tau_x=100N/m^2$ and $\tau_y=200N/m^2$ were applied to $\Gamma_N$ as (sinusoidal) bearing loads;
        (d) FEA mesh of 22,789 tetrahedral elements in step 1;
        (e) FCM (XVoxel) mesh of $55\times16\times9$ voxels.
    }
    \label{fig-Piston_model_info}
\end{figure*}

\begin{figure*}[htbp]
    \centering
    \subfigure[Relative error $r_{\mathbf{u}}$]
    {\includegraphics[width=0.3\textwidth]{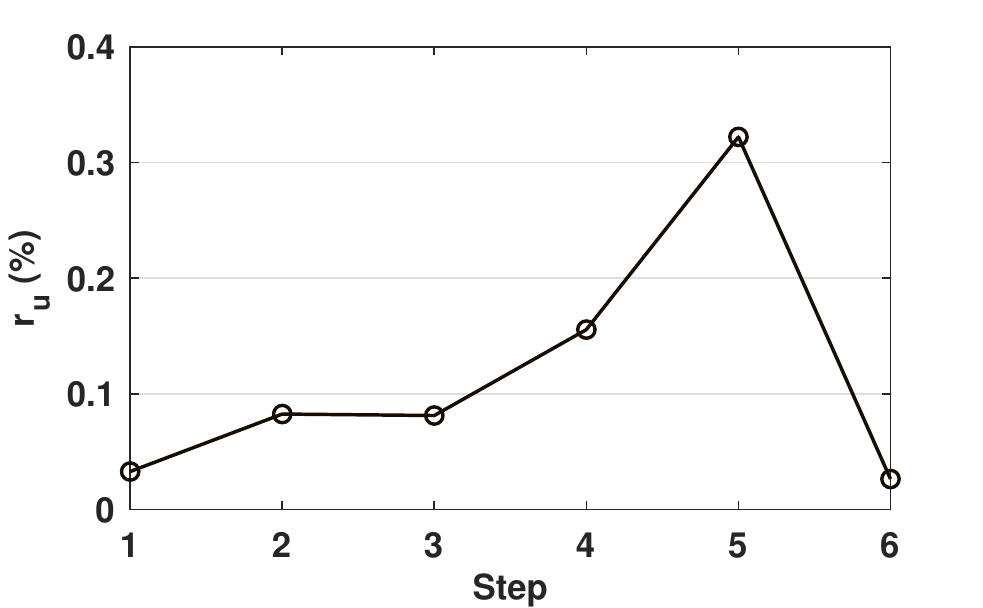}}
    \hspace{0.5cm}
    \subfigure[The number of active voxels]{\includegraphics[width=0.3\textwidth]{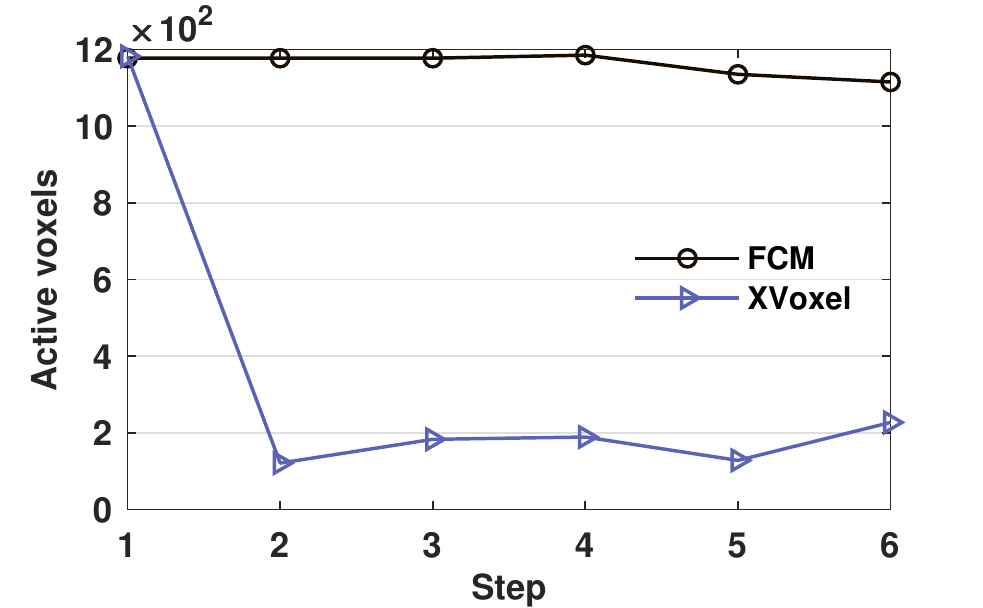}}
    \hspace{0.5cm}
    \subfigure[Timing]{\includegraphics[width=0.3\textwidth]{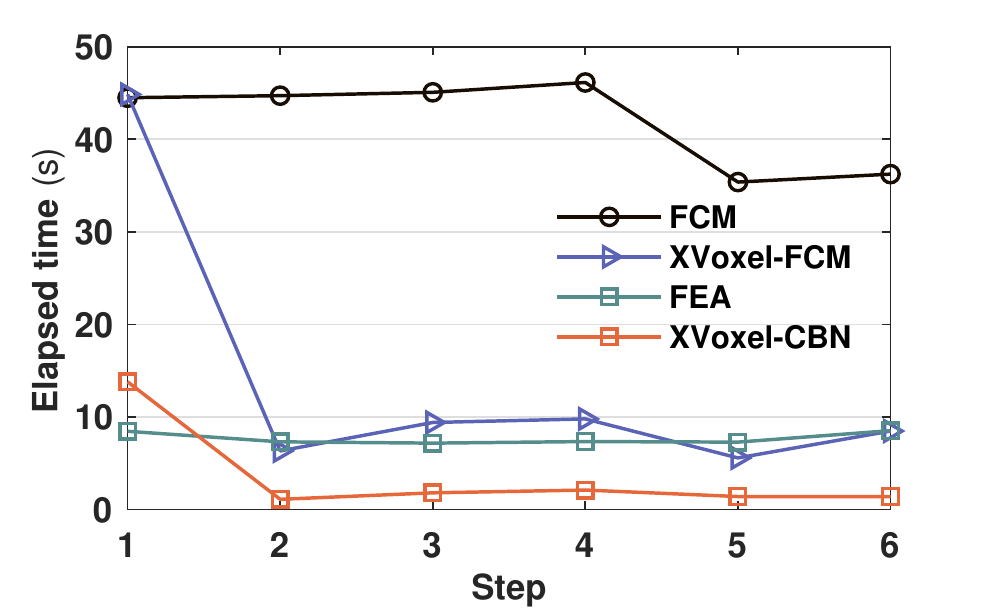}}
    \caption{Performance statistics of Example \#2 in Fig.~\ref{fig-Piston_model_info}
    \rev{}{: (a) Displacement residual $r_u$ between XVoxel and FCM;
    (b) The number of active voxels by FCM and XVoxel;
    (c) Timing of four methods FEA, FCM, XVoxel-FCM  and XVoxel-CBN.}
    }
    \label{fig-Piston_statistics}
\end{figure*}

\begin{figure*}[htbp]
    \centering
    \includegraphics[width=1\textwidth]{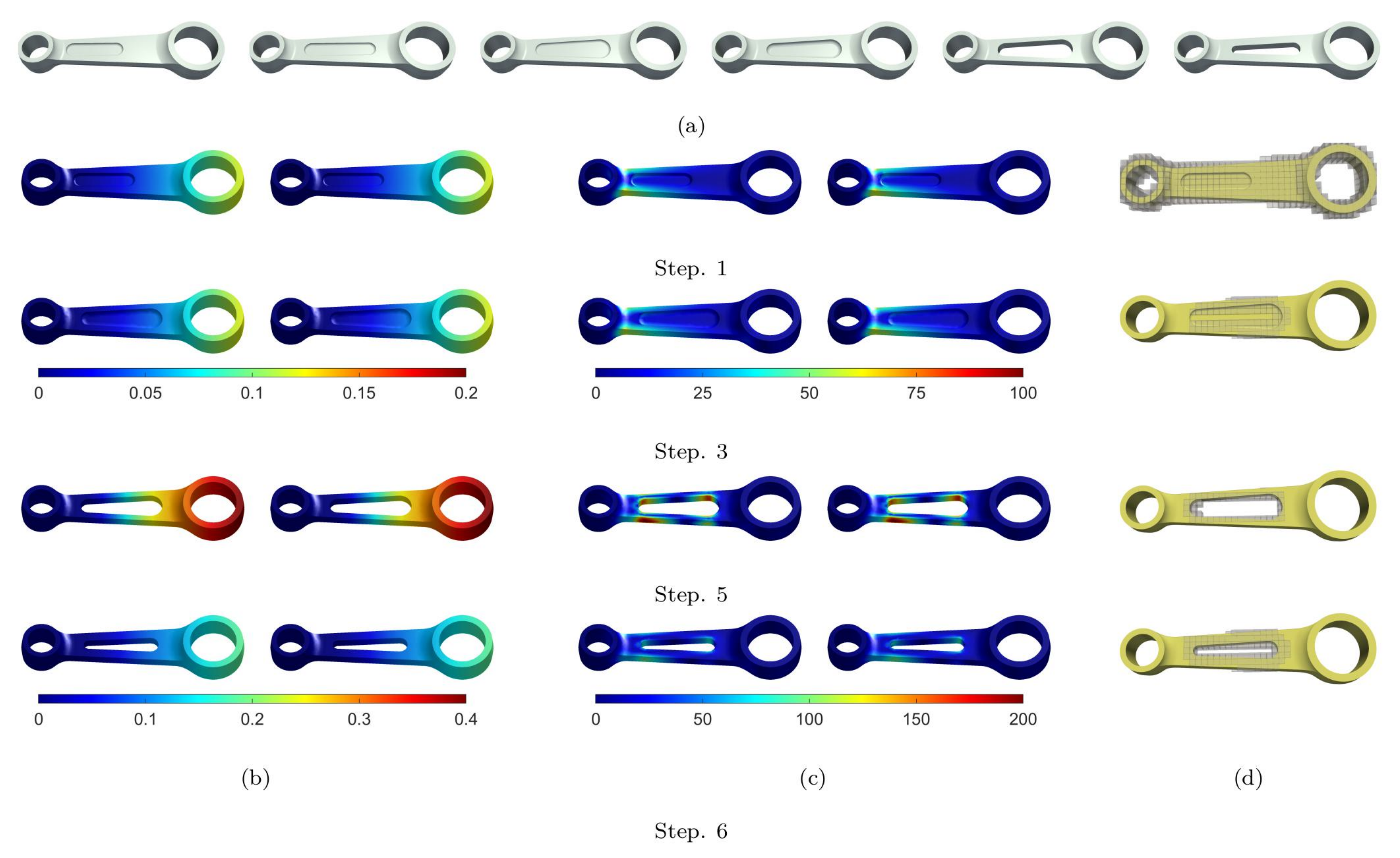}
    \caption{(a)The models for steps 1 to 6\rev{}{: Initial model with a pair of symmetrical grooves;
    Modify $d1$ and $d2$ to translate grooves;
    Increase the right cylinder’s radius $r2$;
    Increase grooves’ depth $h$;
    Increase groove' depth $h$ to run through the model;
    Decrease radii $r_1$ and $r_2$ to narrow the groove
    };
        Results for a connector model in editing steps 1, 3, 5 and 6 
        (b) Displacement norm (mm) by FEA (left) and FCM/XVoxel (right),
        (c) Von Mise stress (MPa) from FEA (left) and FCM/XVoxel (right),
        (d) Active voxels (in grey) of XVoxel. }
             
    \label{fig-Piston_results}
\end{figure*}

The second test was conducted on the engine connector as shown in Fig. \ref{fig-Piston_model_info} to test XVoxel's ability in handling more complex models. The model is fixed on its left hole, subject to horizontal and vertical tractions on its right hole. The FEA tetrahedral mesh has 23K elements while the XVoxel has 7.9K voxels. The connector model was modified in Fig. \ref{fig-Piston_results}(a) by the following six steps of feature operations:
\begin{enumerate}
    \item Add a pair of inner groove made of two cylinders and their tangents.
          Note that the two sides of connector are symmetrical, and we only consider design parameters on one side.
    \item Translate the two cylinders by changing parameters $d_1$ from 25 to 30, $d_2$ from 55 to 40.
    \item Modify the right cylinder's radius $r_2$ from 5 to 7.5.
    \item Modify the inner groove's depth $h$ from 1.5 to 2.5.
    \item Modify design parameter $h$ from 2.5 to 3.5 and remove the round corner so that the inner groove goes through the whole model.
    \item Modify design parameters $r_1$ from 5 to 3, $r_2$ from 7.5 to 5.
\end{enumerate}

XVoxel has a close approximation to FEA with a maximal $r_{\bu} = 0.33\%$ as observed from plot in Fig. \ref{fig-Piston_statistics}(a), or the comparison of simulation results in Figs.~\ref{fig-Piston_results}(b) and (c). We also noticed from Figs. \ref{fig-Piston_results}(d) and \ref{fig-Piston_statistics}(b) and (c) that the XVoxel (XVoxel-FCM and XVoxel-CBN) has much less computational time than FCM during the model modifications (after the first step) as its local voxel update immensely decreases the number of active voxels.
%


\subsection{Example \#3: a pump model under drastic topology variation and varying loads}

\begin{figure*}[]
    \centering
    \subfigure[]{\includegraphics[width=0.46\textwidth]{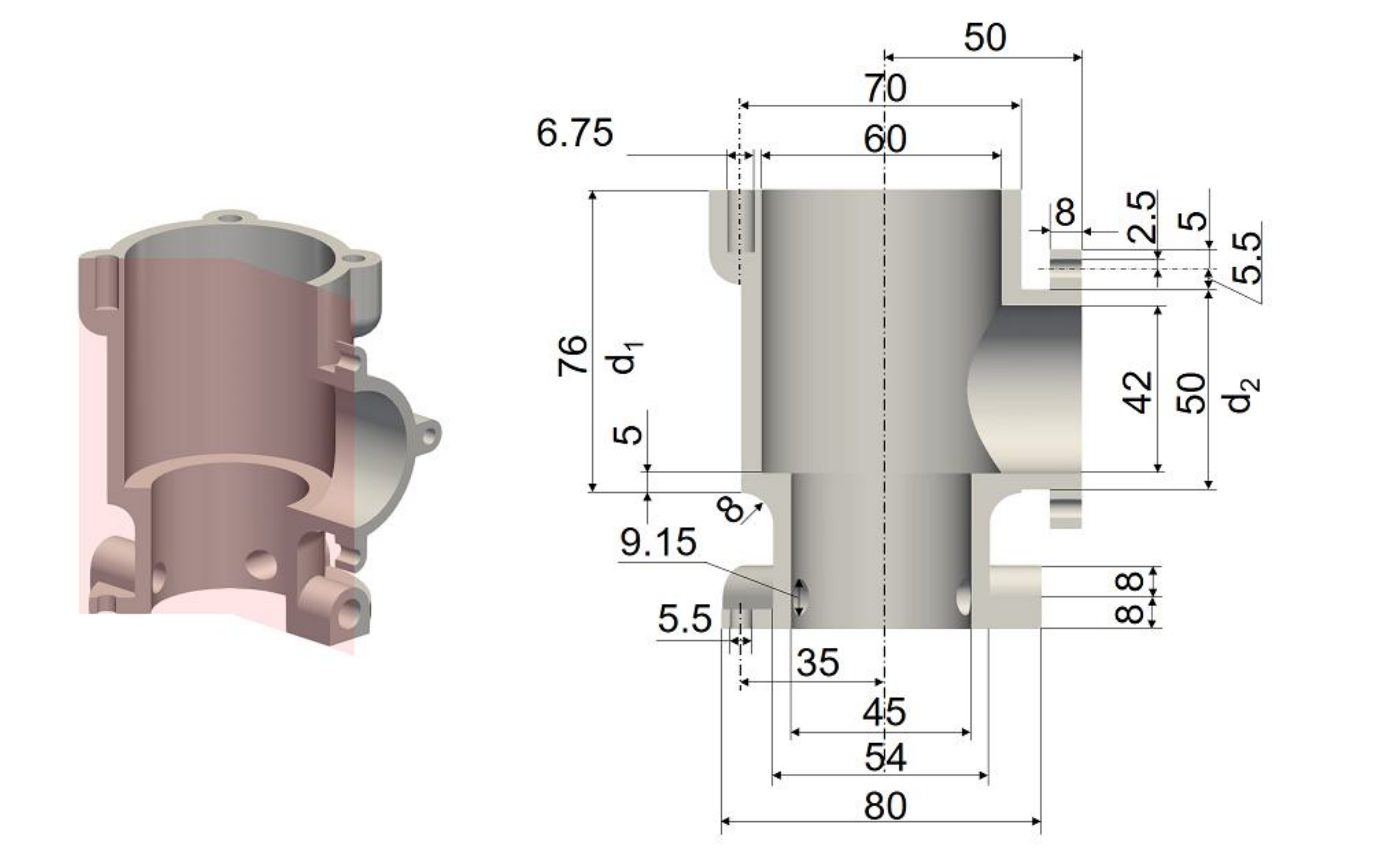}}
    \subfigure[]{\includegraphics[width=0.5\textwidth]{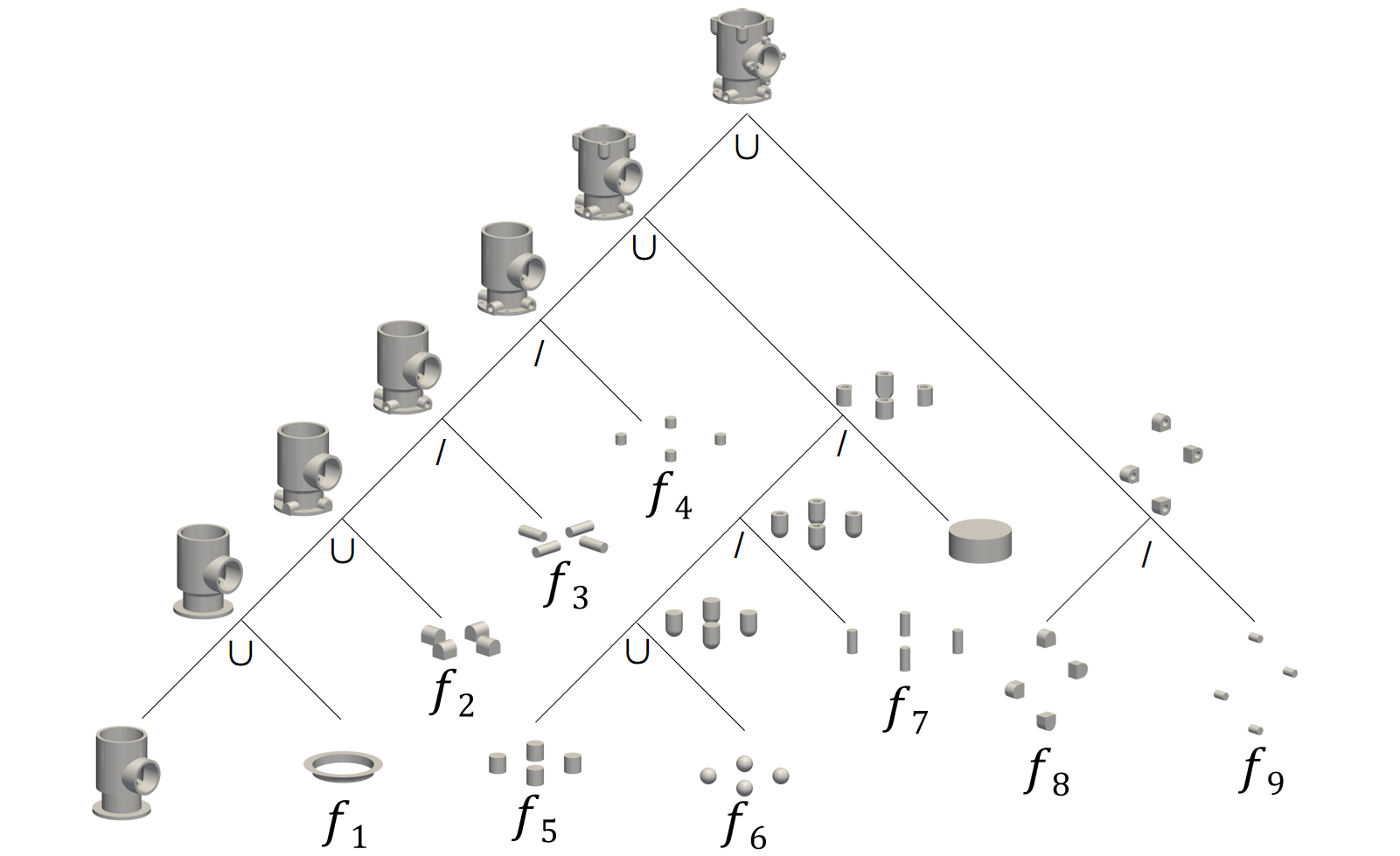}}
    \subfigure[]{\includegraphics[width=0.29\textwidth,trim=0 40 0 0,clip]{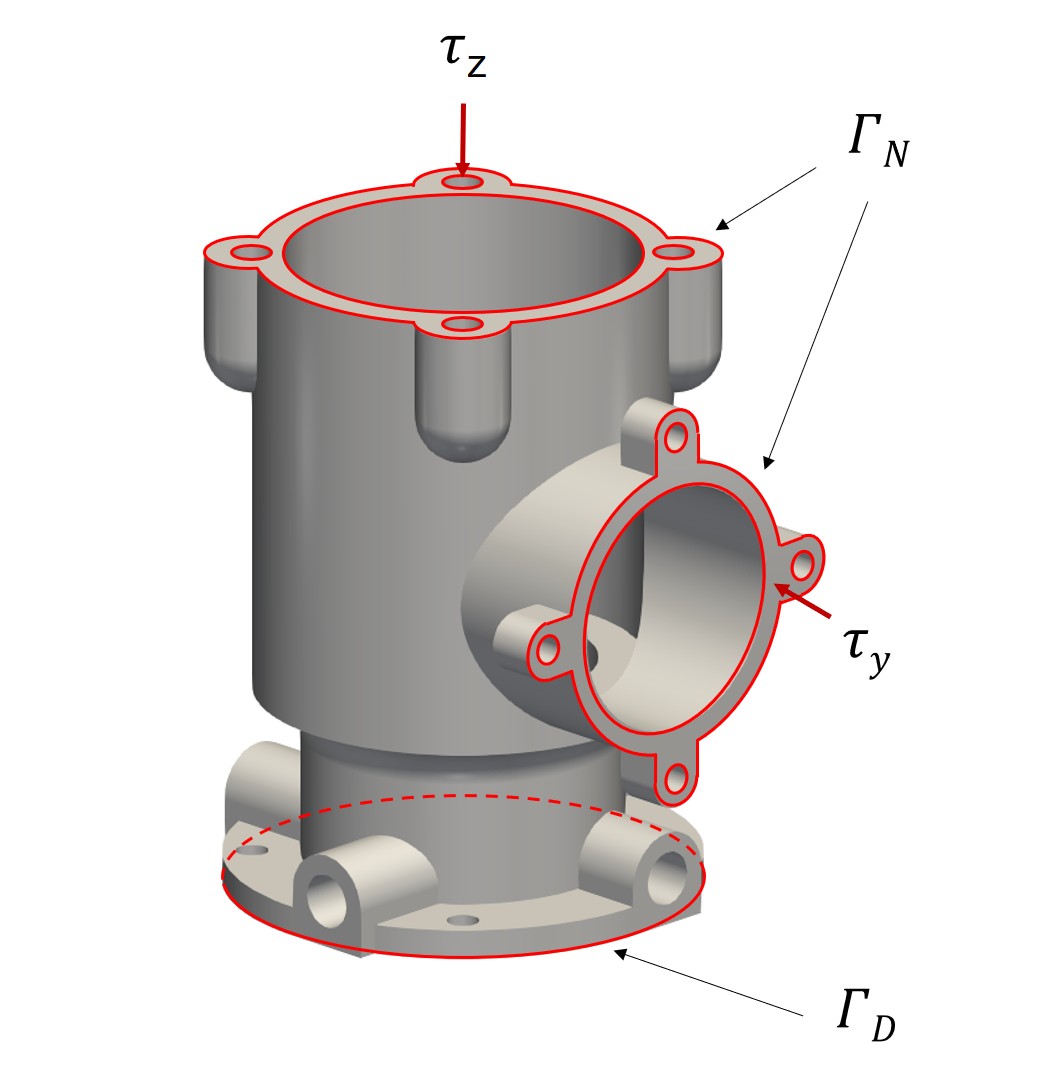}}
    \subfigure[]{\includegraphics[width=0.32\textwidth,trim=1500 300 1500 300,clip]{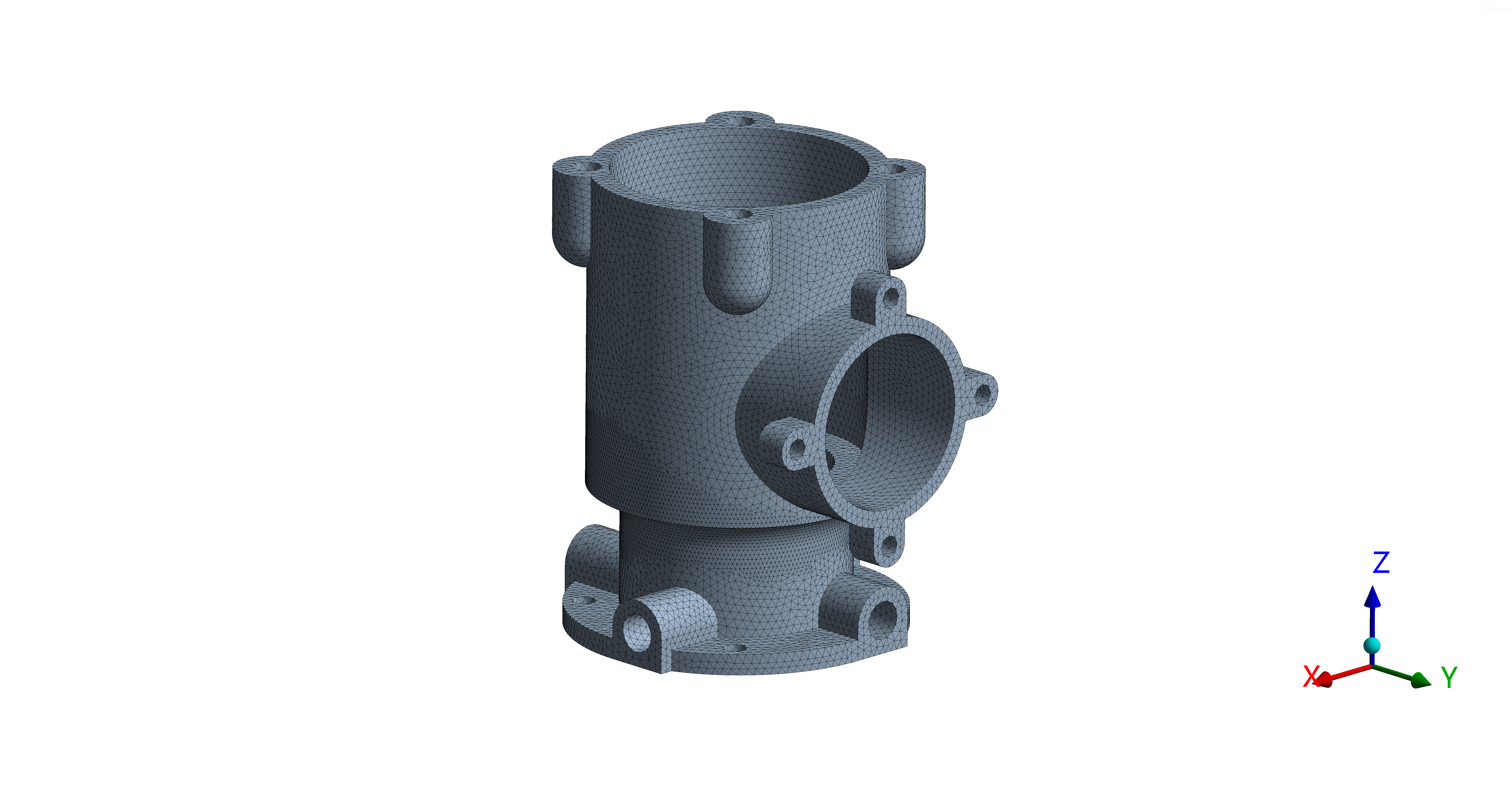}}
    \subfigure[]{\includegraphics[width=0.32\textwidth,trim=550 90 450 0,clip]{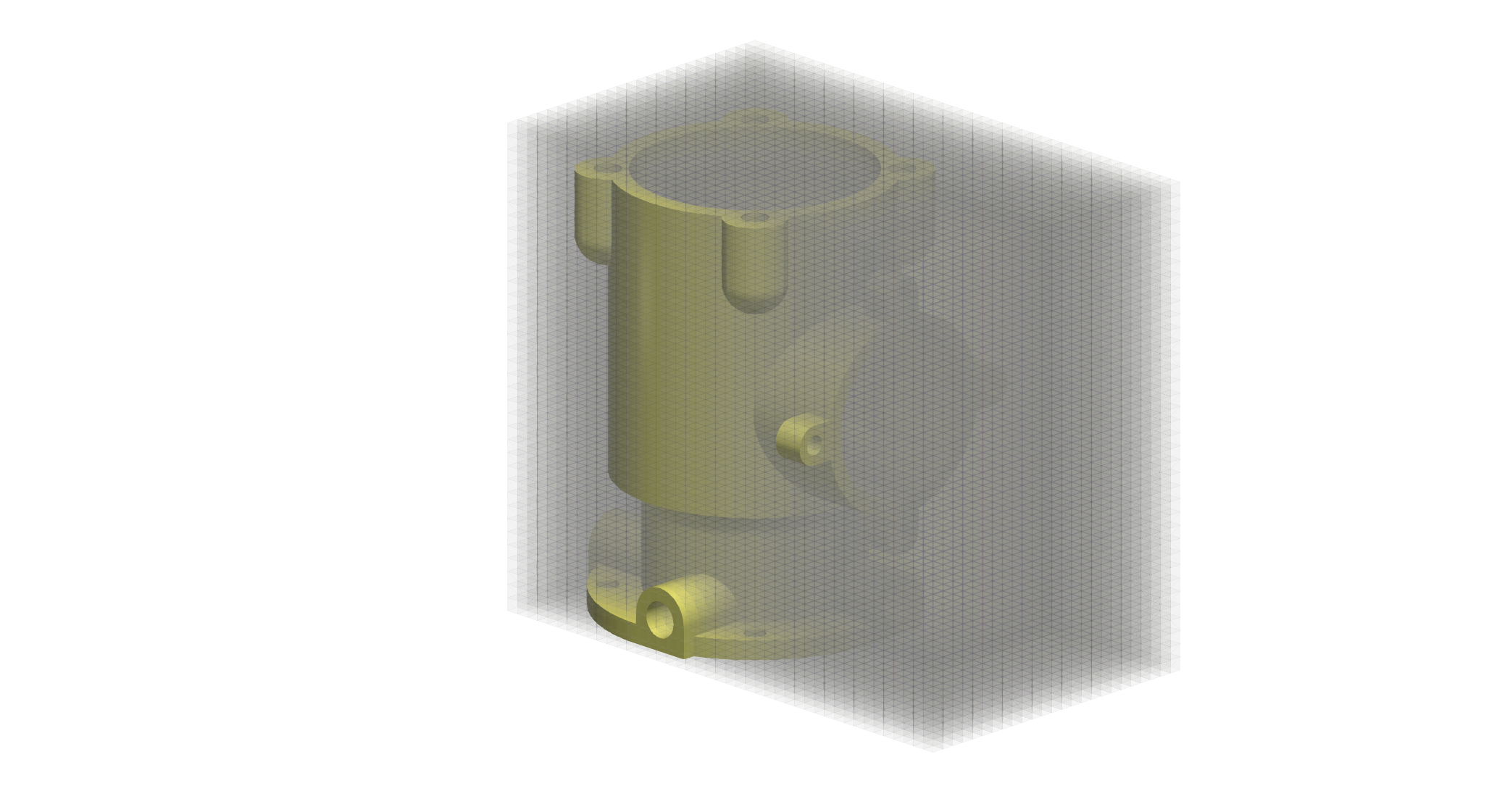}}
    \caption{
        Example \#3.
        (a) Parameters (mm) of the pump model at the 10th step\rev{}{, where $d_1$ and $d_2$ are the variables to modify in step $11-20$};
        (b) The CSG of the model at the 10th step\rev{}{, where $f_i$ are the features to add in step $1-10$, $i=1, ..., 9$};
        (c) Boundary condition, where $\gamma_D$ is fixed and $\tau_z=200N$ and $\tau_y=100N$;
        (d) FEA mesh with 216,411 tetrahedron at the 10th step;
        (e) FCM (XVoxel) mesh with $25\times43\times37$ voxels.
    }
    \label{fig-Cylinders_model_info}
\end{figure*}

\begin{figure*}[]
    \centering
    \includegraphics[width=1\textwidth]{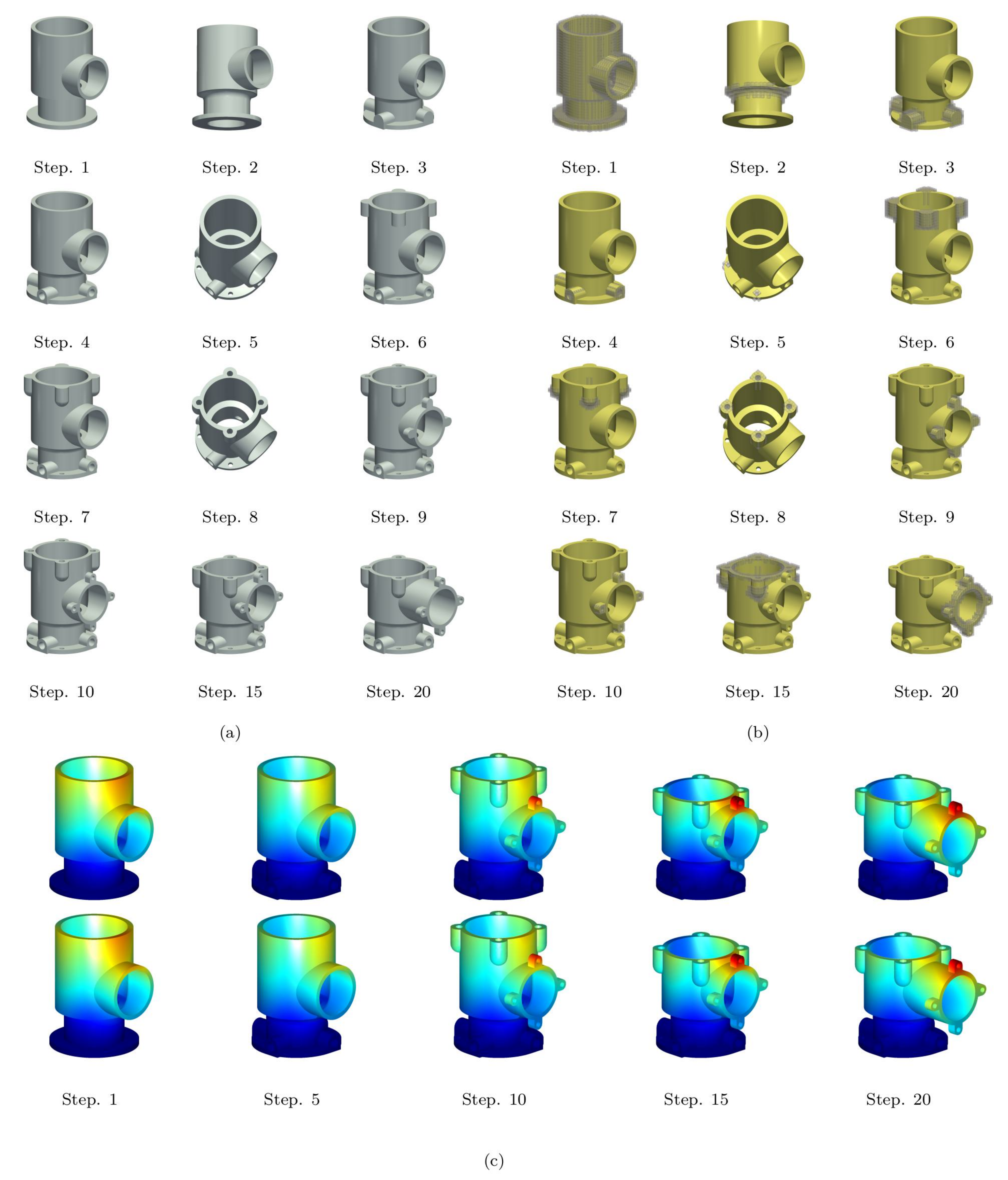}
    \caption{
        (a)The models in editing steps 1-10, 15 and 20,
        (b)Active voxels (in grey) of XVoxel in editing steps 1-10, 15 and 20,
        (c)The simulation results for a pump model in editing steps 1, 5, 10, 15 and 20. Displacement norm (mm) of FEA (up) and of FCM/XVoxel (down).
    }
    \label{fig-Cylinders_results}
\end{figure*}

\begin{figure*}[]
    \centering
    \subfigure[Relative error $r_{\mathbf{u}}$]
    {\includegraphics[width=0.3\textwidth]{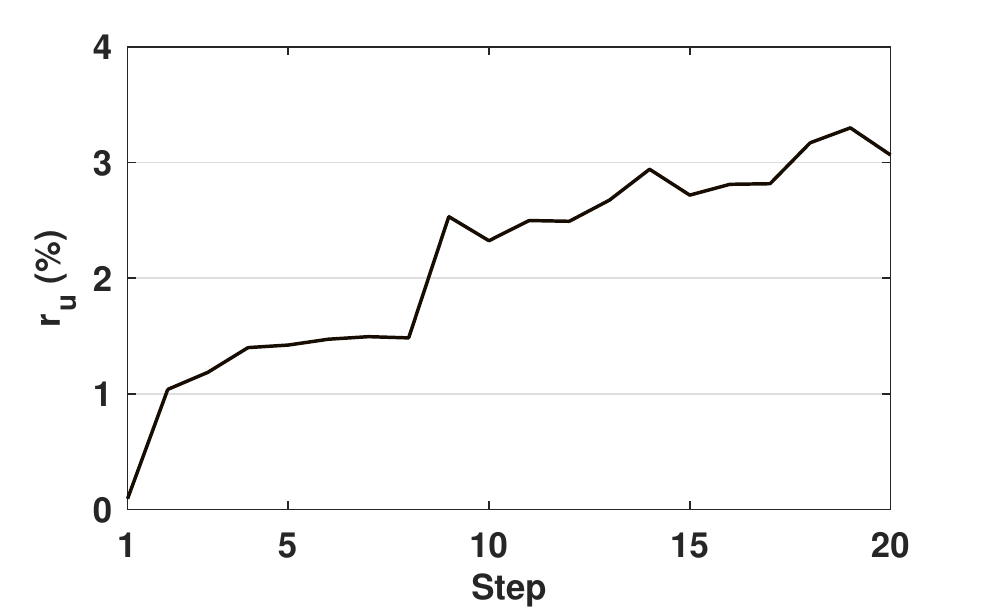}}
    \hspace{0.5cm}
    \subfigure[The number of active voxels]{\includegraphics[width=0.3\textwidth]{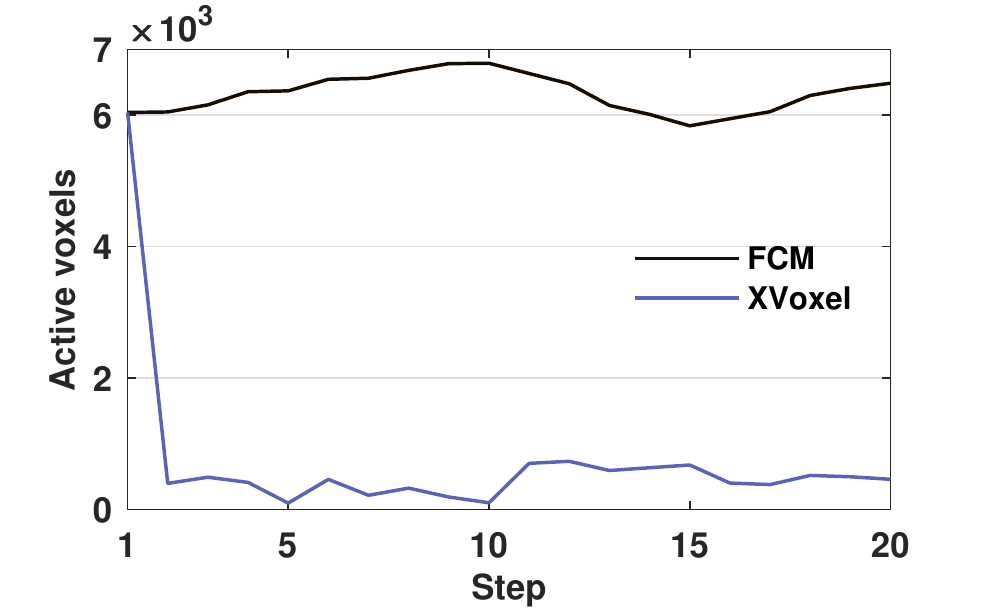}}
    \hspace{0.5cm}
    \subfigure[Timing]{\includegraphics[width=0.3\textwidth]{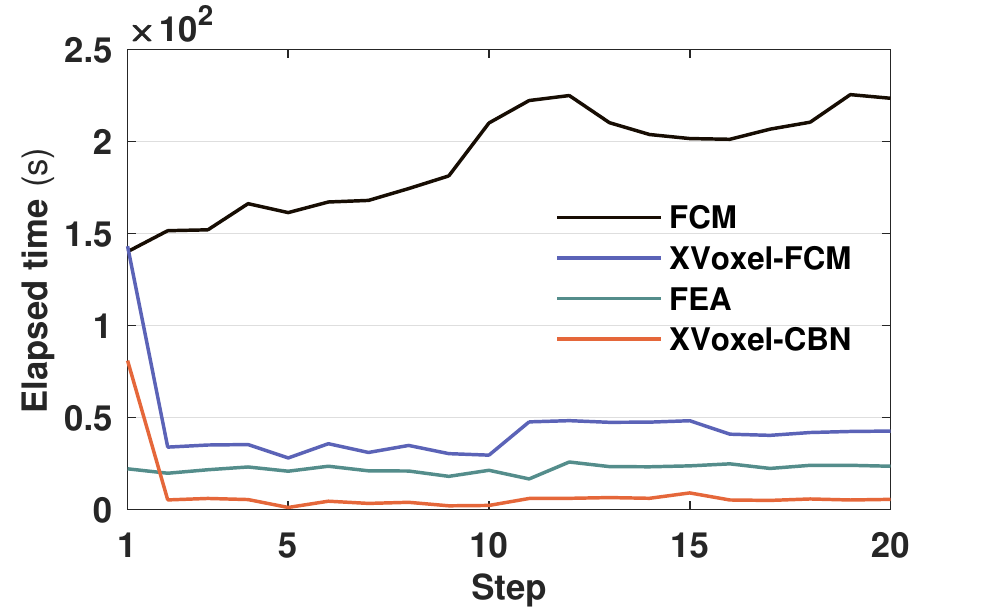}}
    \caption{Performance statistics of Example \#3 in Fig.\ref{fig-Cylinders_model_info}
    \rev{}{: (a) Displacement residual $r_u$ between XVoxel and FCM;
    (b) The number of active voxels by FCM and XVoxel;
    (c) Timing of four methods FEA, FCM, XVoxel-FCM  and XVoxel-CBN.}
    }
    \label{fig-Cylinders_statistics}
\end{figure*}

The proposed method's potentiality in handling drastic topology variations and varying loads was further tested on the complex pump model in Fig. \ref{fig-Cylinders_model_info}. The pump's bottom is fixed and its top and outer side are exerted by forces of 200$N$, 100$N$ respectively. The FEA has 216K tetrahedral elements while FCM (XVoxel) has 40K voxels. The model was edited by the following steps, during which both the model's topology and external loadings are varied: 
\begin{enumerate}
    \item[1.] Input an initial model consisting of different cylinders.
    \item[2.] Add a round corner feature $f_1$.
    \item[3.] Add feature $f_2$ which consists a cube and a cylinder.
    \item[4.] Add a negative feature $f_3$ as a union of four cylinders.
    \item[5.] Add a negative feature $f_4$.
    \item[6.] Add a feature $f_5$.
    \item[7.] Add a feature $f_6$ consisting of four spheres.
    \item[8.] Add a negative feature $f_7$.
    \item[9.] Add a feature $f_8$ consisting of a cube and a cylinder.
    \item[10.] Add a negative feature $f_9$.
    \item[11-15.] Modify design parameter $d_1$ from 76 to 56 at a step of -4.
    \item[16-20.] Modify design parameter $d_1$ from 50 to 70 at a step of 4.
\end{enumerate}

The resulting models during the modification were shown in Figs.~\ref{fig-Cylinders_results}(a)-(c), with the associated active voxels given in Fig.~\ref{fig-Cylinders_results}(b). Statistics of the relative errors, numbers of active voxels, timings were compared in Fig. \ref{fig-Cylinders_statistics}, which indicates XVoxel's high simulation accuracy and efficiency, as already confirmed in Examples \#1 and \#2.

We in particular observed from Fig. \ref{fig-Cylinders_statistics}(b) that the number of active voxels of XVoxel is only around 1/4 of FCM's, demonstrating XVoxel's strong ability in properly selecting active voxels regardless of the complex feature shape and feature operations. The nice property in turn resulted in a 4-time efficiency improvement of XVoxel-FCM in comparison to FCM; XVoxel-CBN even achieved a $50$ times efficiency improvement. Such efficiency is very useful in obtaining interactive simulation feedback in modifying designs.

\subsection{Example \#4: a bracket model for parametric design optimization}

\begin{figure*}[]
    \centering
    \subfigure[]{\includegraphics[width=0.25\textwidth]{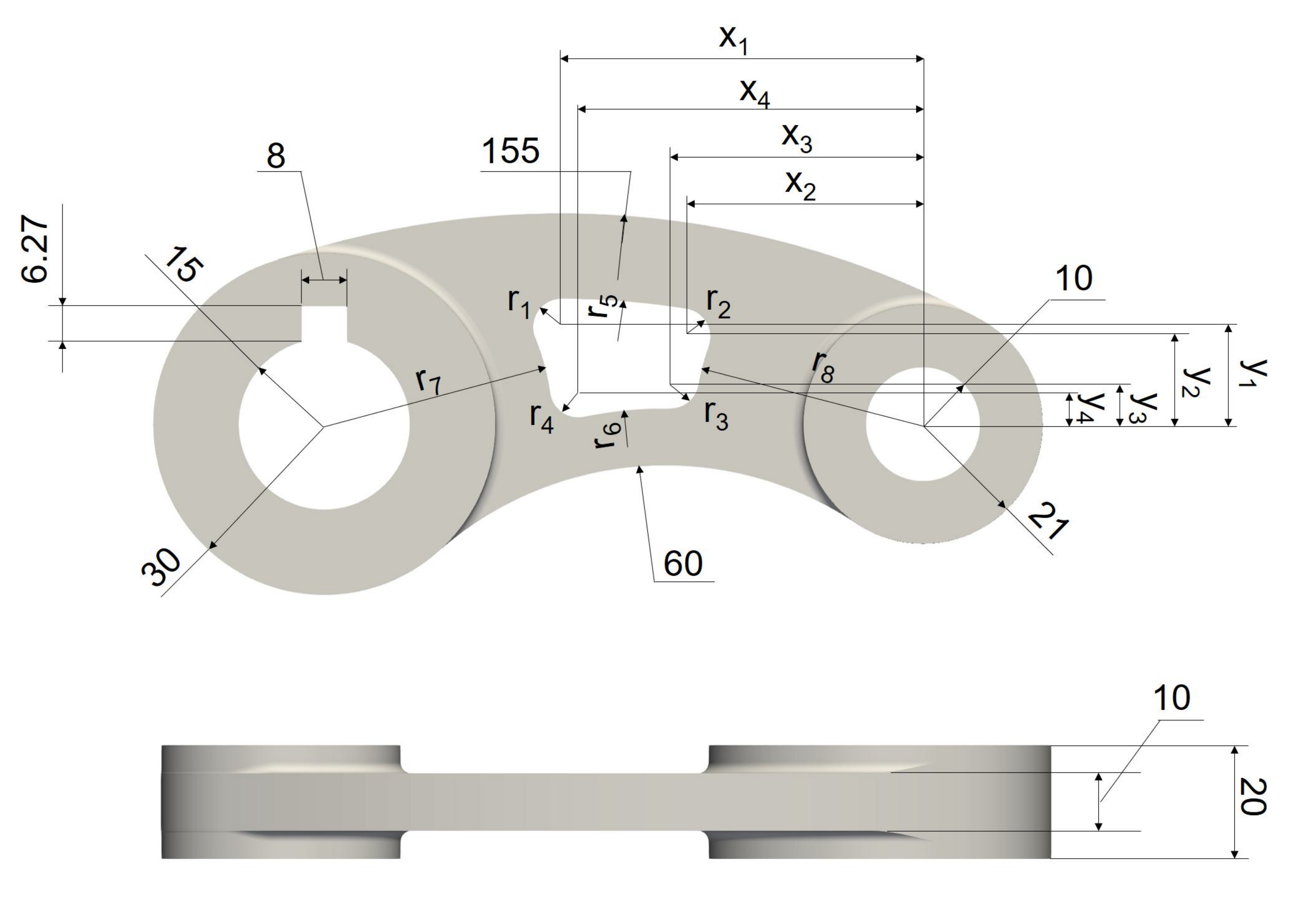}}
    \subfigure[]{\includegraphics[width=0.2\textwidth]{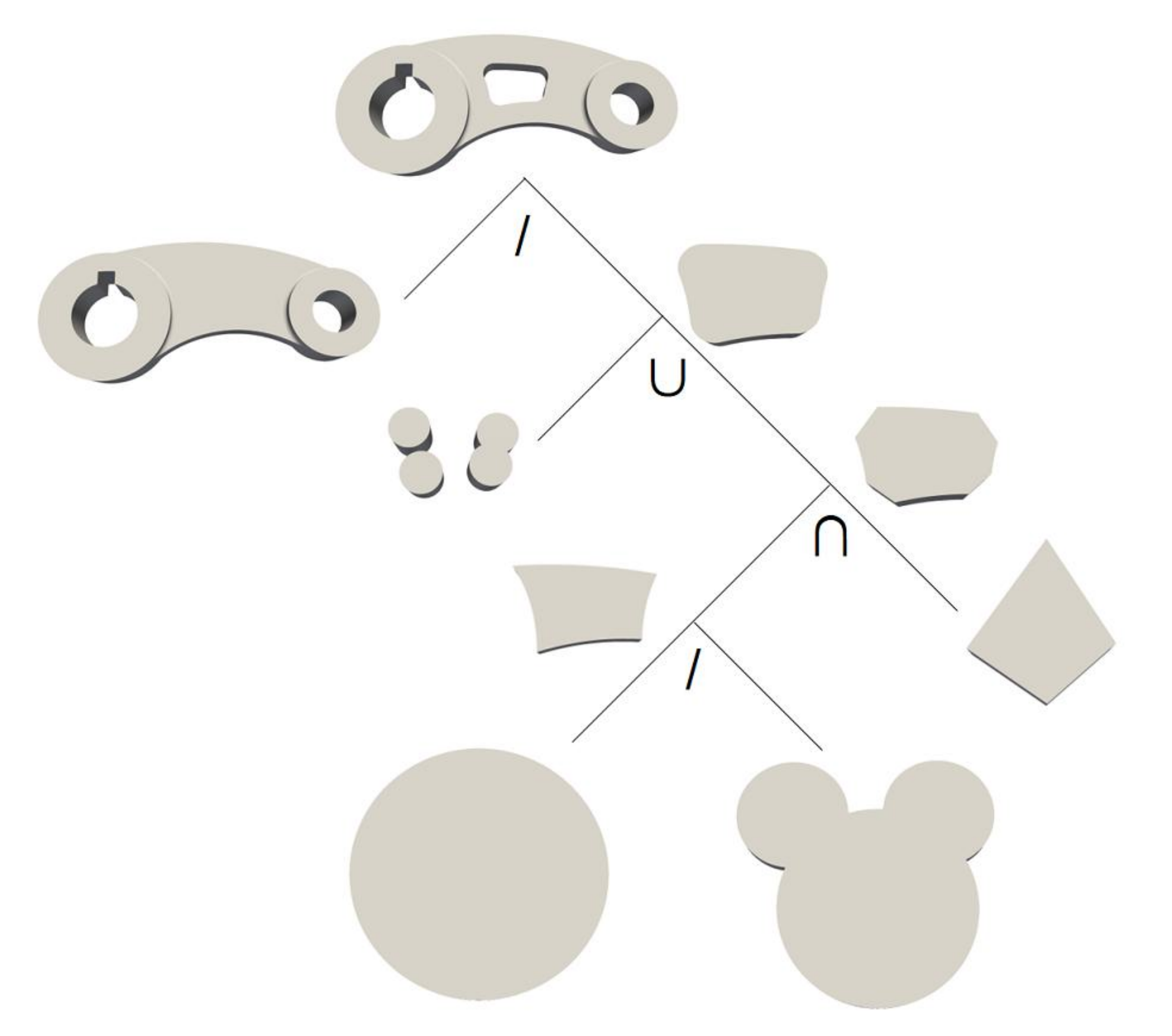}}
    \subfigure[]{\includegraphics[width=0.25\textwidth,trim=0 0 0 0,clip]{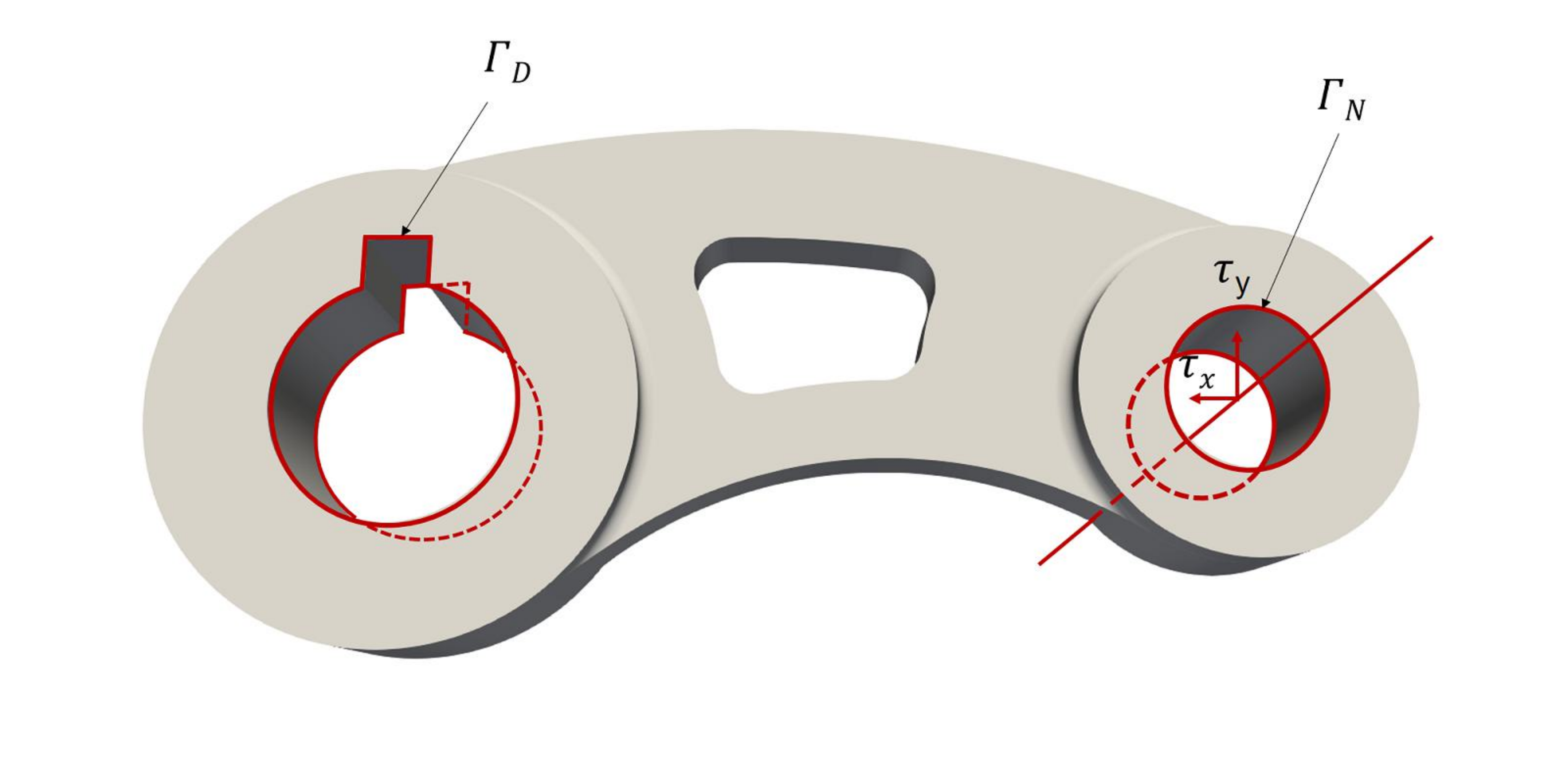}}
    \subfigure[]{\includegraphics[width=0.25\textwidth]{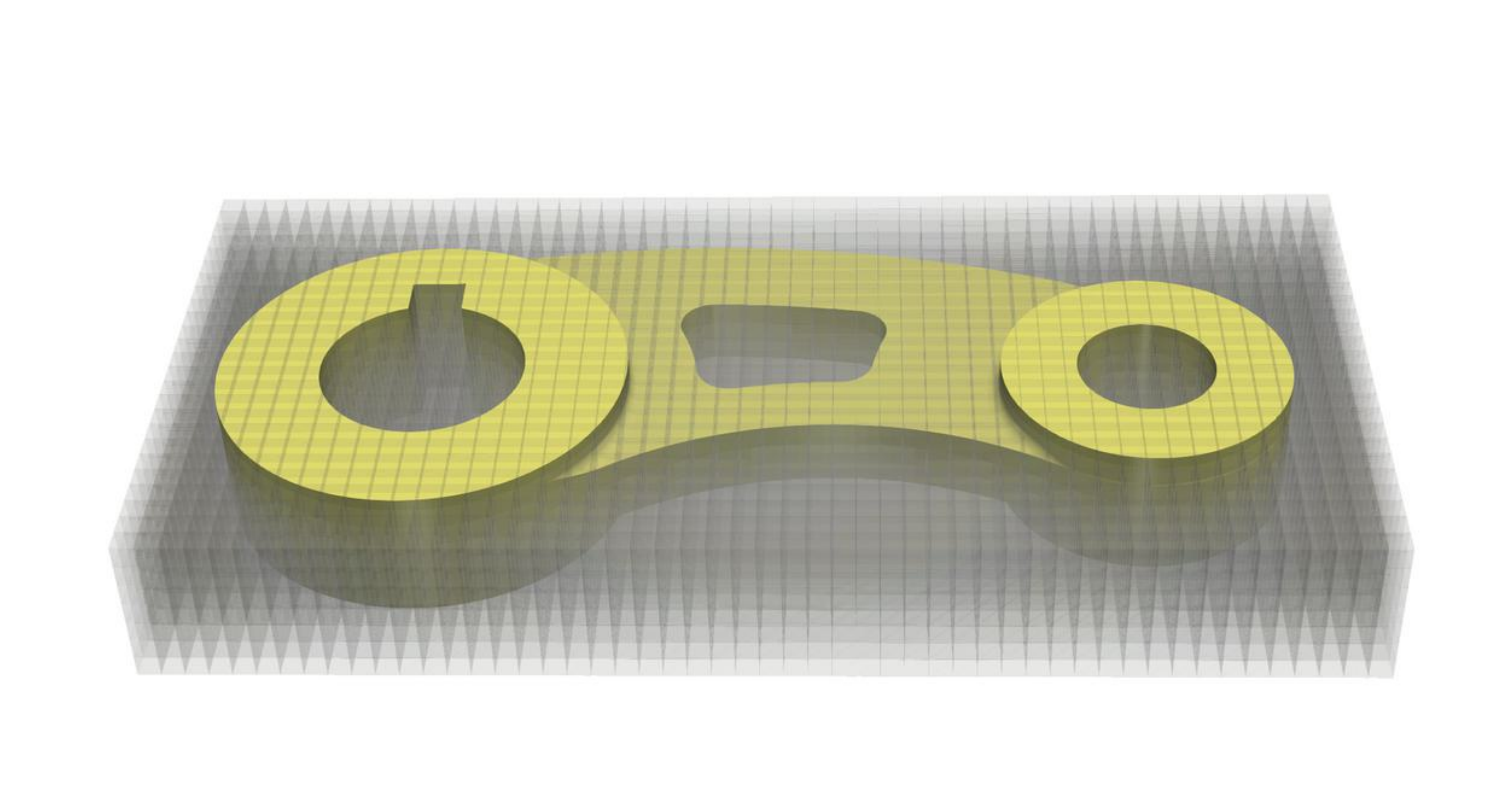}}
    \caption{
        Example \#4.
        (a) Parameters (mm) of the model\rev{,}{. The position and size of a groove are optimized by changing the position and radius of the tangential cylinders that form it,} where the design variables are \rev{}{} $x_i, y_i$ and $r_j$(i=1, 2, ..., 4; j=1, 2, ..., 8);
        (b) The CSG of the model;
        (c) Boundary conditions, where $\Gamma_D$ is fixed and $\tau_x=100N,\ \tau_y=200N$ are applied on $\Gamma_N$ as (sinusoidal) bearing loads; (d) FCM (XVoxel) mesh with $52\times25\times7$ voxels.
    }
    \label{fig-Bracket_model_info}
\end{figure*}

\begin{figure*}
    \centering
    \includegraphics[width=1\textwidth]{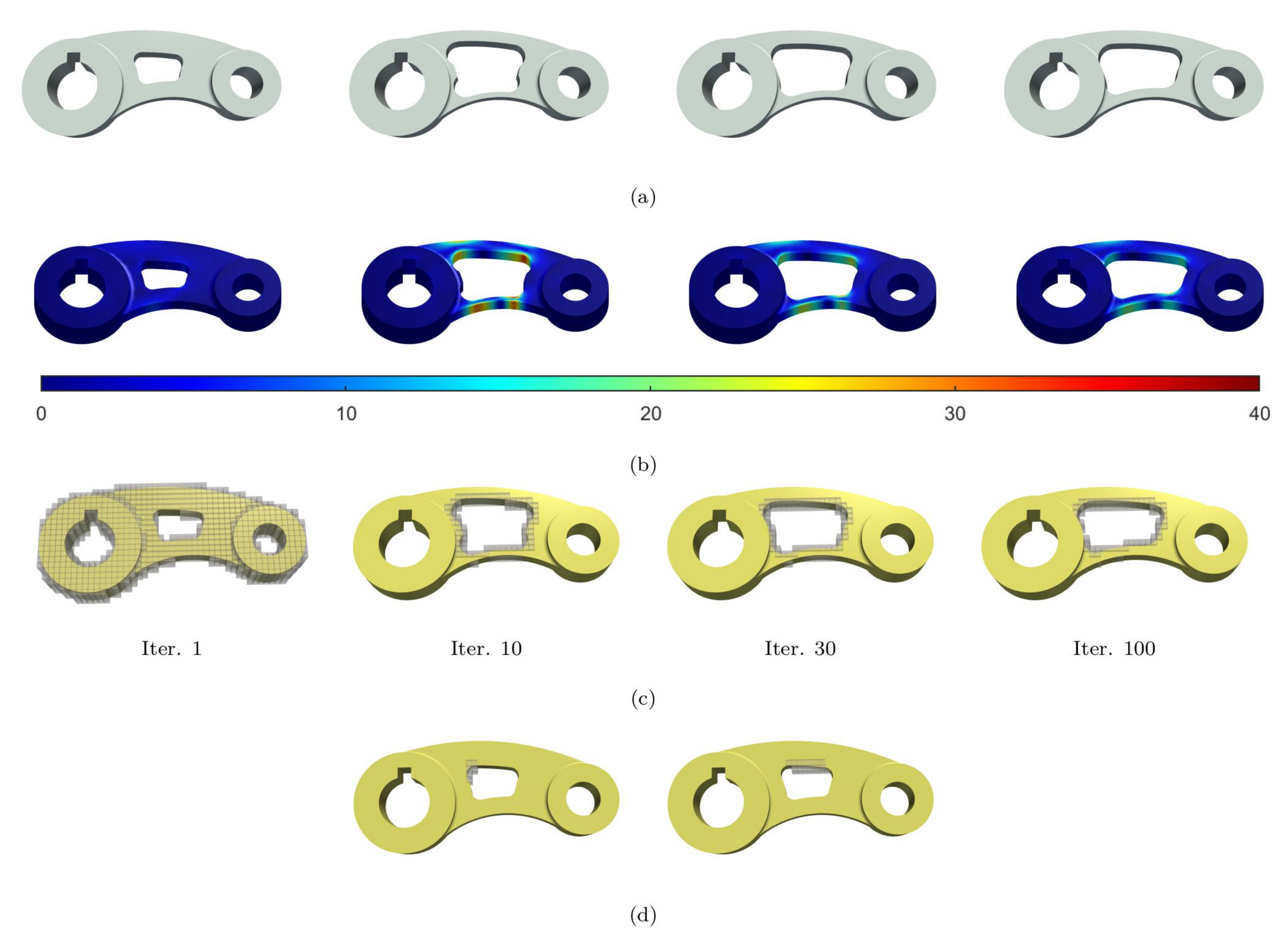}

    \caption{Results for a bracket model in optimization iterations of steps 1, 10, 30 and 100 (a) The model,
        (b) Von Mise stress (MPa) of FCM/XVoxel,
        (c) Active voxels (in grey) of XVoxel,
        (d) Active voxels (in grey) for calculating the sensitivities with respect to $x_1, y_1, r_1$ (left) and $r_5$ (right) in the first iteration.
    }
    \label{fig-Bracket_step_change}
\end{figure*}

\begin{figure*}
    \centering
    \subfigure[Convergence curves of compliance and volume]
    {\includegraphics[width=0.3\textwidth]{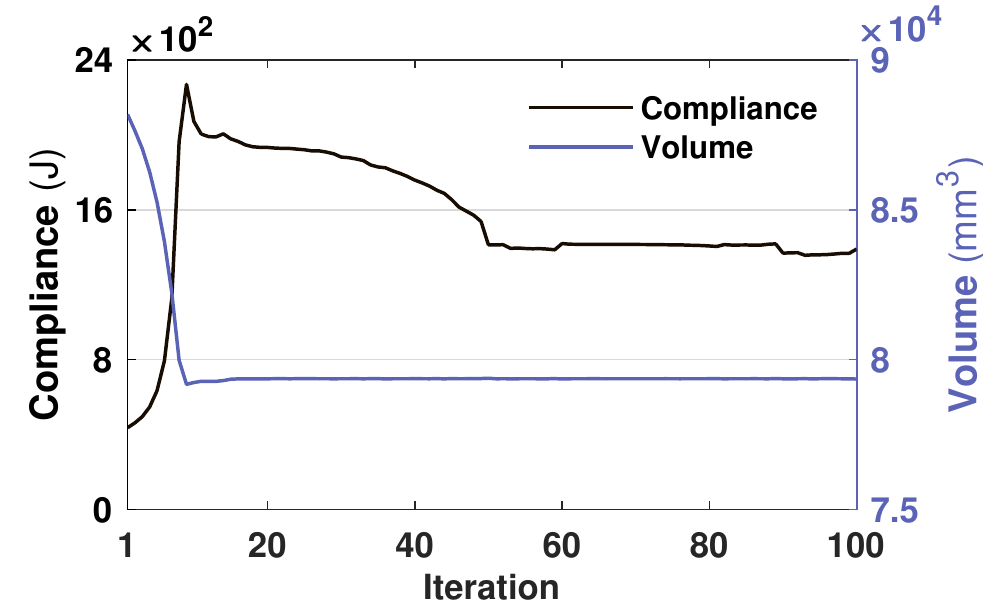}}
    \subfigure[The number of active voxels]
    {\includegraphics[width=0.3\textwidth]{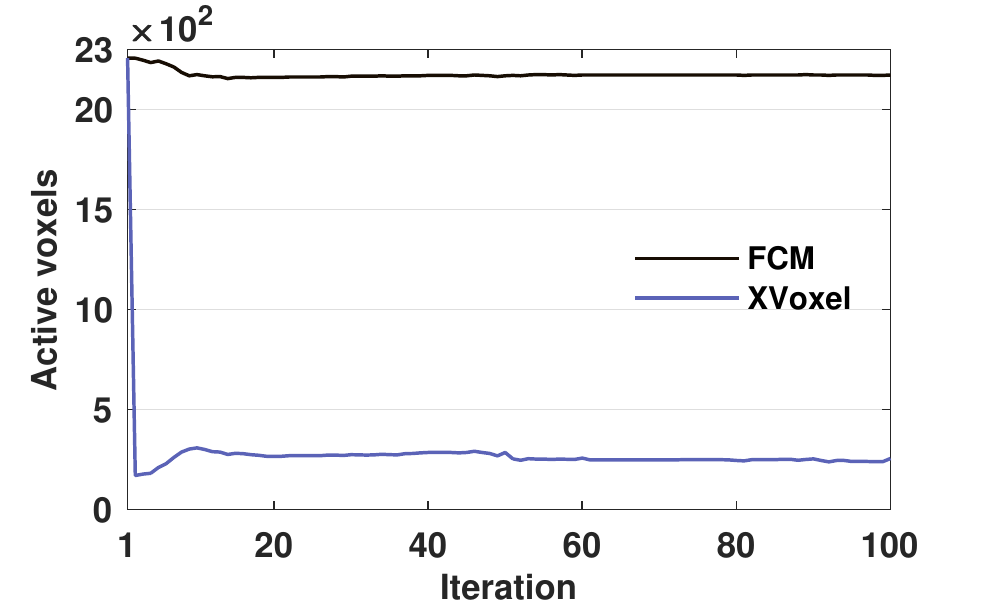}}
    \subfigure[Timing]
    {\includegraphics[width=0.3\textwidth]{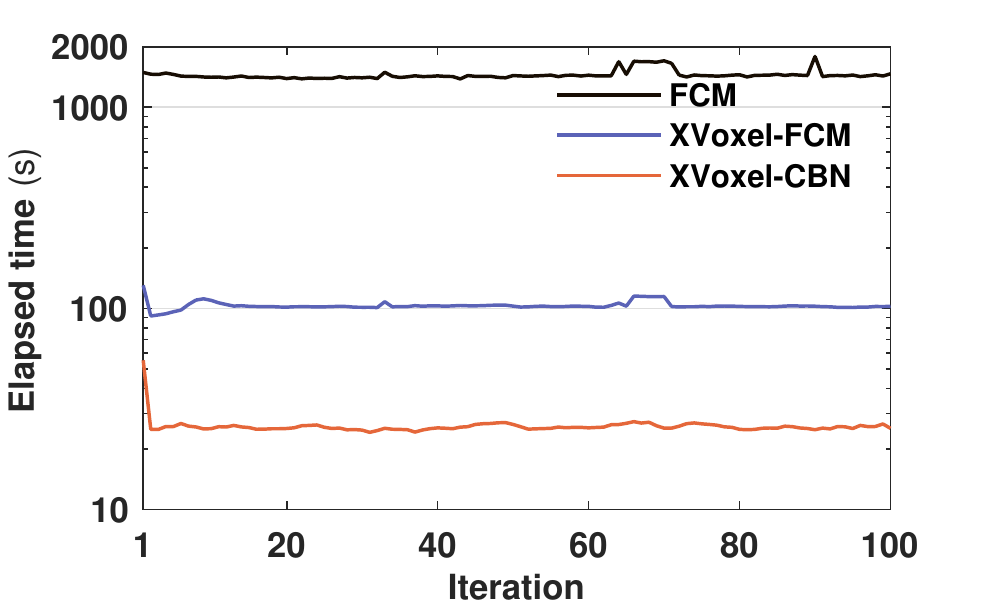}}
    \caption{Performance statistics of Example \#4 in Fig.~\ref{fig-Bracket_model_info}.
    \rev{}{: (a) Convergence curves of compliance and volume by XVoxel and FCM;
    (b) The number of active voxels by FCM and XVoxel;
    (c) Timing of three methods FCM, XVoxel-FCM  and XVoxel-CBN.}
    }
    \label{fig-Bracket_statistics}
\end{figure*}

The proposed method's ability in feature-based parametric design optimization was first tested on an asymmetric bracket model in Fig.~\ref{fig-Bracket_model_info}, which has a negative groove feature composed of several cylinders and prisms. The model was fixed on its left hole, and subject to an axial load of $100\sqrt{5}$N on its right hole. The model is discretized into 9.1K voxels for FCM (XVoxel) based simulation.

The design goal is to minimize the bracket's compliance under a volume ratio of 0.9 by varying the locations and sizes of the groove. The design variables are: circle centers $x_i, y_i$ and radii $r_i$ (i=1, 2, 3, 4) of the four corner circles, their inscribed circle radius $r_5$ and circumscribed circles' radii $r_6, r_7, r_8$ (in 2D plane). The associated geometric constraints are formulated as follows so as to produce a valid geometry:
\begin{equation*}
    \begin{aligned}
        \left\{\begin{array}{ll}
                   \bK\bu=\bF,                      &           \\
                   V\leq 0.9V_0,                    &           \\
                   ||(x_i,y_i)-O_5||+r_i=r_5,       & i=1,2     \\
                   ||(x_i,y_i)-O_6||-r_i=r_6,       & i=3,4     \\
                   ||(x_i,y_i)-O_7||-r_i=r_7,       & i=1,4     \\
                   ||(x_i,y_i)-O_8||-r_i=r_8,       & i=2,3     \\
                   x_i^{min}\leq x_i\leq x_i^{max}, & i=1,...,4 \\
                   y_i^{min}\leq y_i\leq y_i^{max}, & i=1,...,4 \\
                   r_j^{min}\leq r_j\leq r_j^{max}, & j=1,...,8 \\
               \end{array}\right.
    \end{aligned}
\end{equation*}
where $V_0$ is the volume of the original model, $O_i$'s are the 2D circle centers' coordinates, and ranges of $x_i,y_i,r_j$ were set so that the features would not move out of the bracket. 

Some intermediate structures during optimization were shown in Fig. \ref{fig-Bracket_step_change}(a), where the groove gradually enlarged its size to meet the volume constraint while moving to the left side for performance improvement. Stress distributions were also plotted in Fig. \ref{fig-Bracket_step_change}(b), and the active voxels were shown in gray in Fig. \ref{fig-Bracket_step_change}(c).

The optimization was stopped after $100$ iterations, where the features and their relative constraints were all maintained during the optimization; the convergence curve was plotted in Fig.~\ref{fig-Bracket_statistics}(a). During the optimization, XVoxel only had approximately $1/8$ active voxels of FCM, with a much-improved efficiency; see also Figs.~\ref{fig-Bracket_statistics}(b) and (c). XVoxel-FCM and XVoxel-CBN respectively achieved around $13 \times$ and $50 \times$ efficiency improvements compared to FCM. By maintaining the feature lists during optimization, the simulation was greatly accelerated by local recomputations for active voxels.

\rev{}{As can be seen from Fig.~\ref{fig-Bracket_statistics}(a), the compliance curve did not go steadily, but first went up quickly to the peak, then went down. This is because compliance is highly sensitive to volume changes, and the volume factor dominates the optimization before reaching the specified volume limit. That is, the optimization will quickly reduce the volume toward the specified volume limit at the beginning (see the volume curve in the same figure). After getting the peak (at the $10$th iteration), the optimization algorithm (i.e., the globally convergent method of moving asymptotes) reduces compliance effectively while keeping the volume above the limit. This process corresponds to the going down phase in Fig.~\ref{fig-Bracket_statistics}(a).}

In conducting the optimization, the sensitivities for this example were derived using finite differences. Without proper handling, the computation would be very expensive as it requires a complete FE recomputation for all 16 design parameters.
Instead, XVoxel requires much fewer active voxels for the finite difference computations, as only one feature parameter was varied in each finite difference process; see also Fig. \ref{fig-Bracket_step_change}(d) for an illustration.

\subsection{Example \#5: a bearing bracket model for parametric design optimization with varied topology}

\begin{figure*}[]
    \centering
    \subfigure[]{\includegraphics[width=0.275\textwidth]{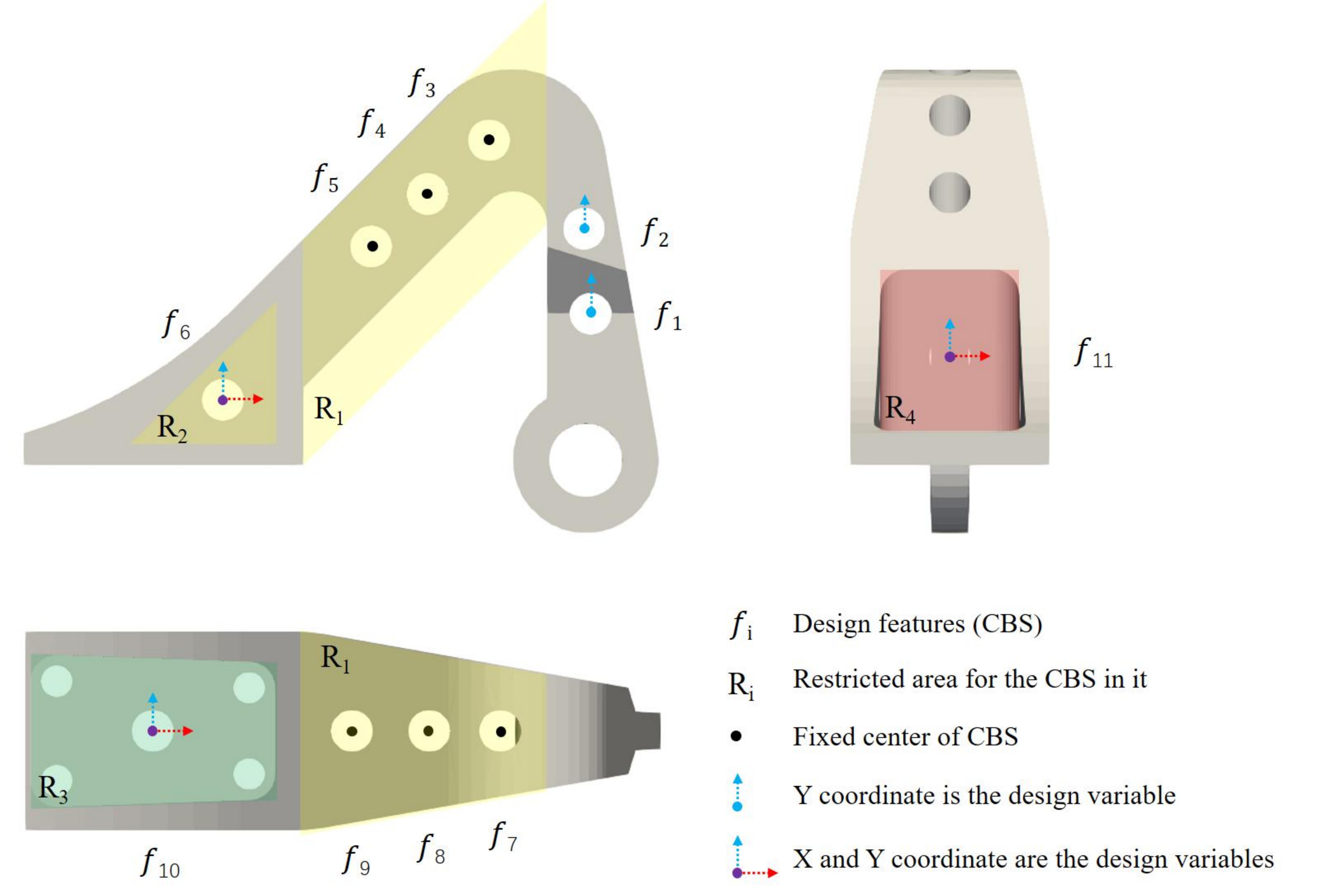}}
    \subfigure[]{\includegraphics[width=0.2\textwidth]{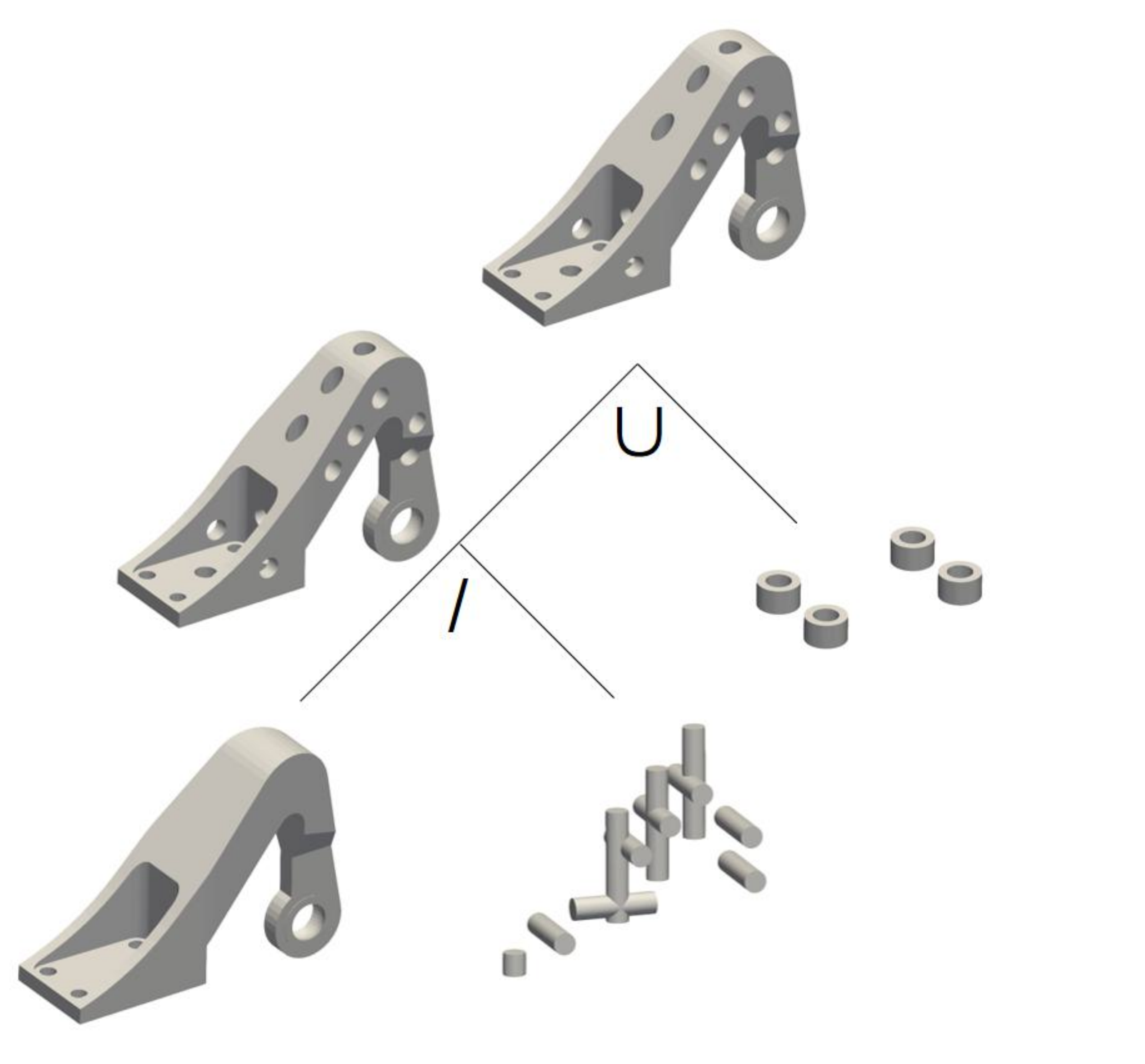}}
    \subfigure[]{\includegraphics[width=0.225\textwidth,trim=0 0 0 0,clip]{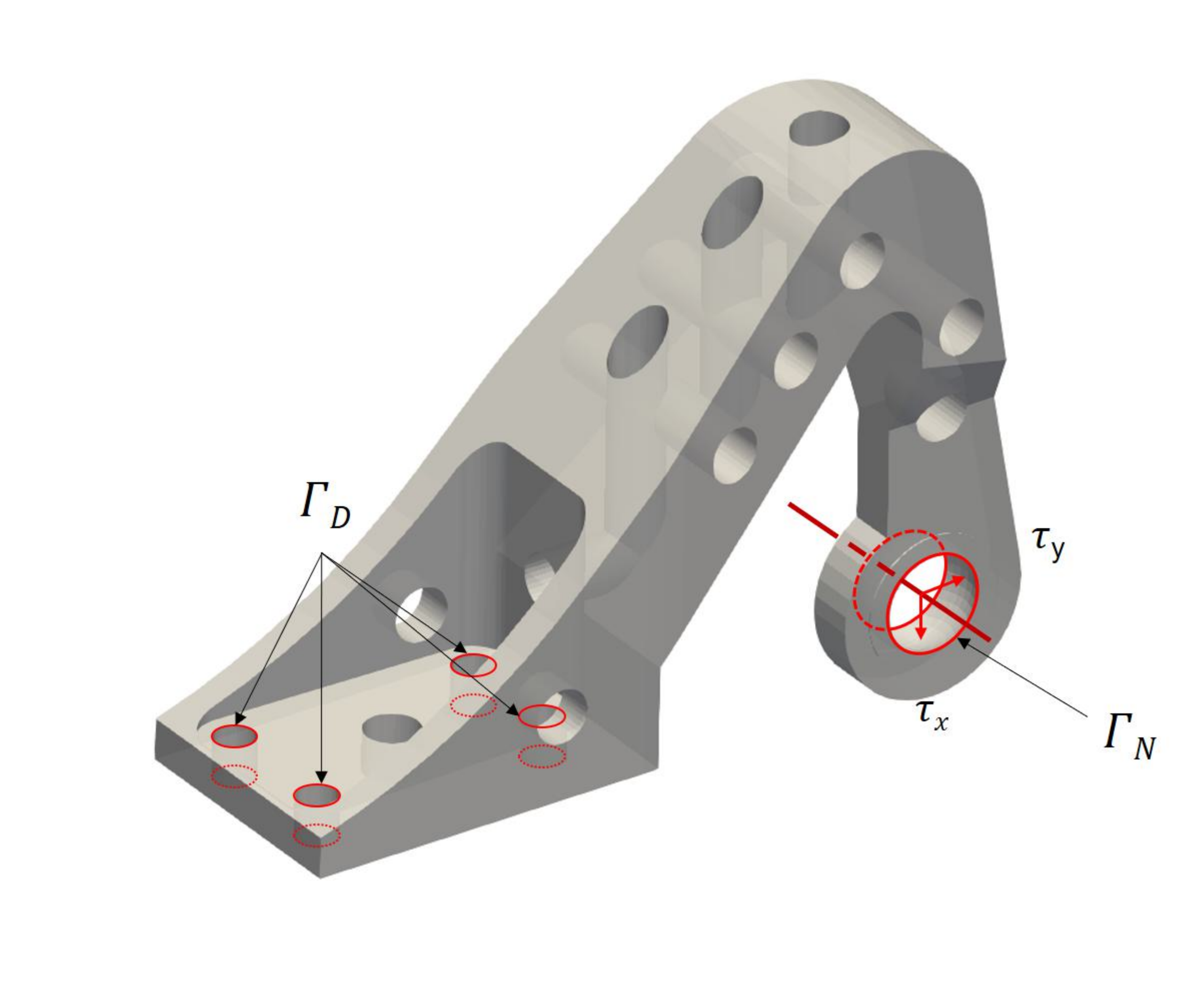}}
    \subfigure[]{\includegraphics[width=0.225\textwidth,trim=300 75 300 0,clip]{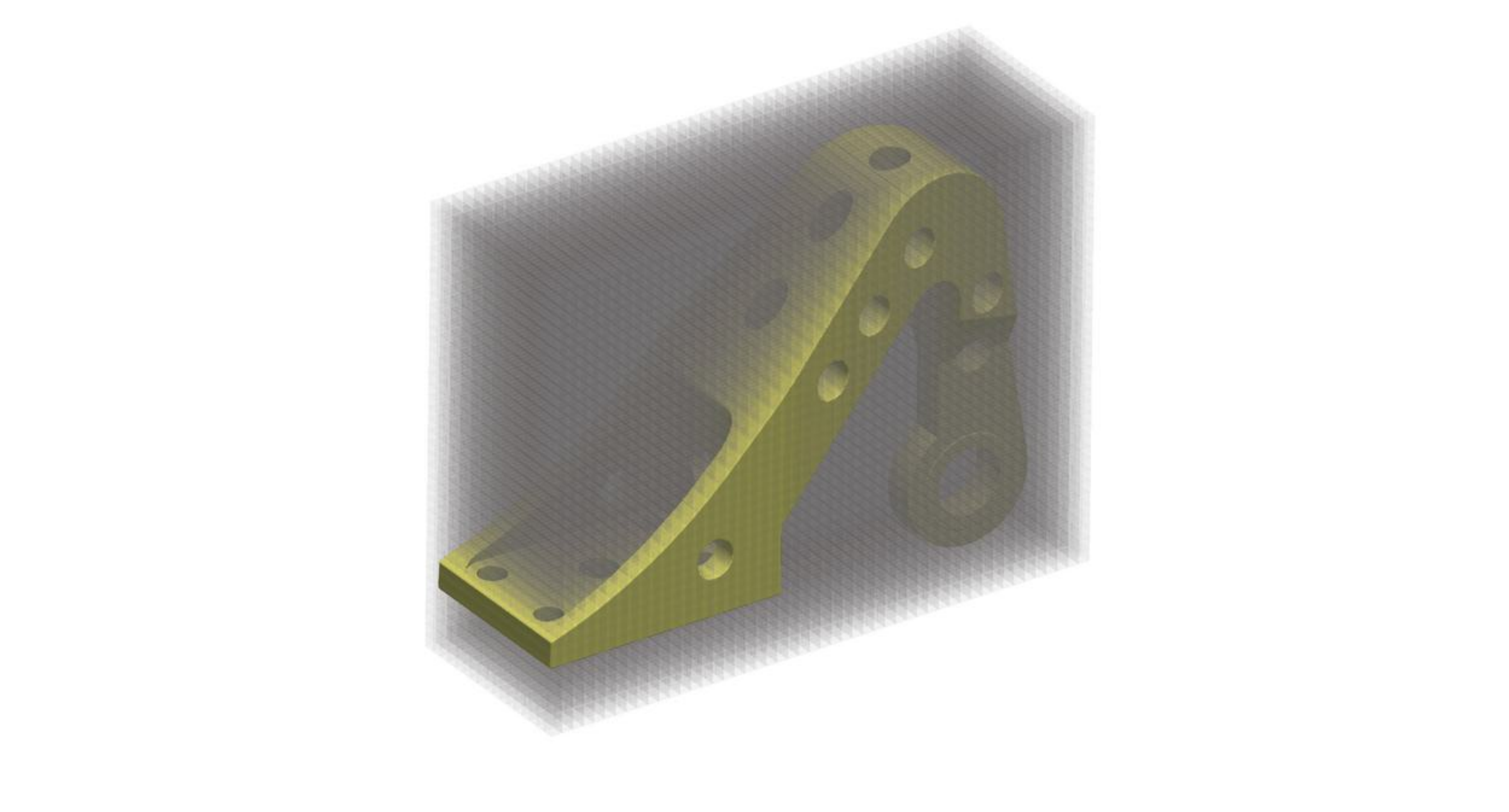}}
    \caption{
        Example \#5.
        (a) The \rev{}{CBS (Closed B-Spline)} design features  $f_i,\ i=1,...,11$ and restricted area $R_1$ \rev{}{(in yellow)} for $f_j,\ j=3,4,5,7,8,9$, $R_2$ \rev{}{(in yellow)} for $f_6$, $R_3$ \rev{}{(in green)} for $f_{10}$ and $R_4$ \rev{}{(in red)} for $f_{11}$. The x-coordinate of $f_i,\ i=1,2$ and x-, y- coordinates of $f_j,\ i=3,4,5,7,8,9$ are fixed. 
        \rev{}{There are $250$ design variables in total, and each feature contains $22$ control radii and unfixed x-, y-coordinates};
        (b) The CSG of the model;
        (c) Boundary conditions, where $\Gamma_D$ is fixed and $\tau_x=100N$ and $\tau_y=100N$ are exerted on $\Gamma_N$ as (sinusoidal) bearing loads;
        (d) FCM (XVoxel) mesh with $50\times17\times39$ voxles.
    }
    \label{fig-BearingBracket_model_info}
\end{figure*}

\begin{figure*}[]
    \centering
    \includegraphics[width=1\textwidth]{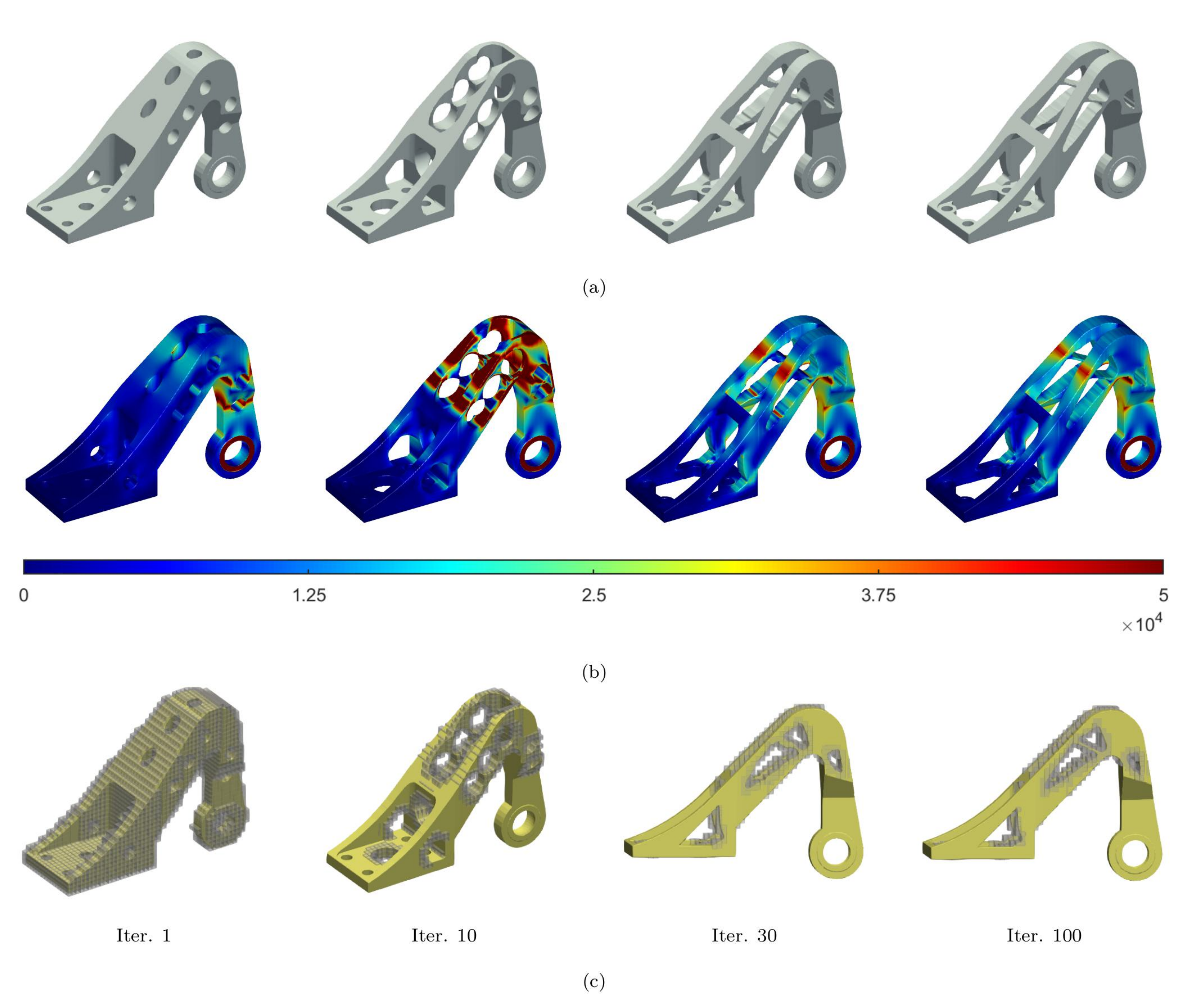}
    \caption{Results for a bearing bracket model in optimization iterations of steps 1, 10, 30 and 100: (a) the models, (b) Von Mise stress (MPa) of FCM/XVoxel, (c) active voxels (in grey) of XVoxel.}
    \label{fig-BearingBracket_step_change}
\end{figure*}

\begin{figure*}[]
    \centering
    \subfigure[Convergence curves of compliance and volume]
    {\includegraphics[width=0.3\textwidth]{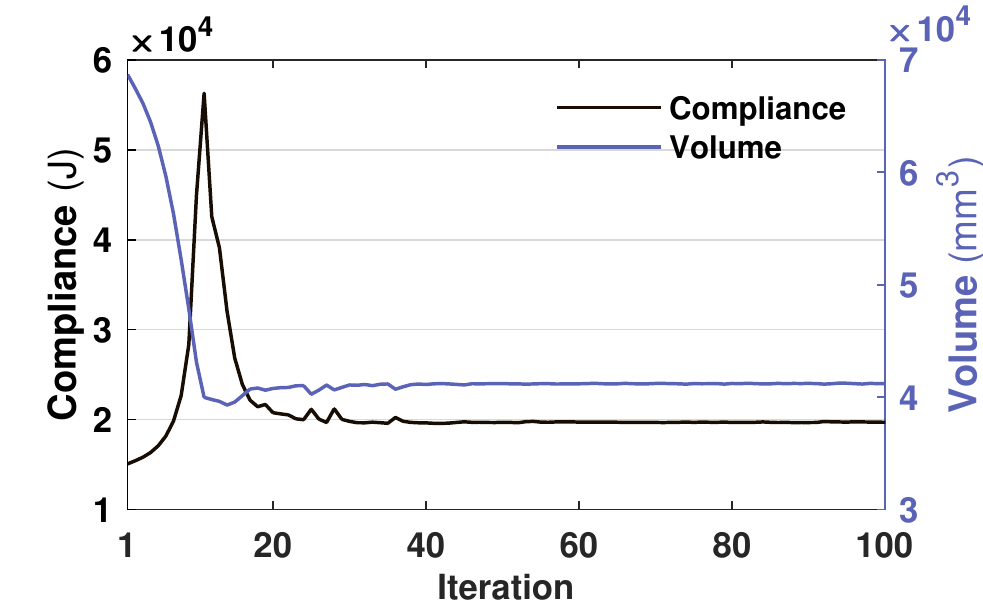}}
    \hspace*{0.5cm}
    \subfigure[The number of active voxels]
    {\includegraphics[width=0.3\textwidth]{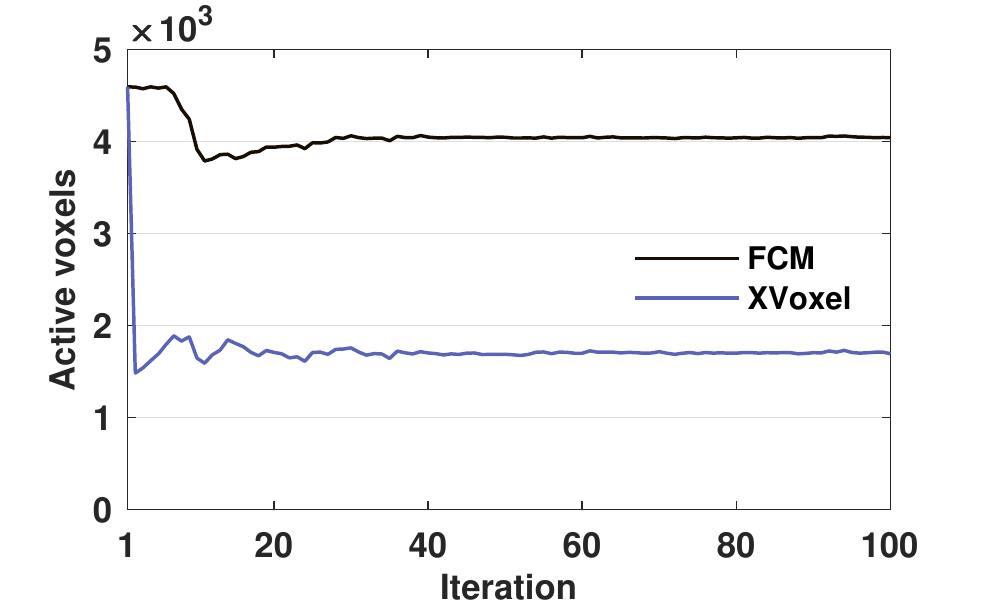}}
    \hspace*{0.5cm}
    \subfigure[Timing]
    {\includegraphics[width=0.3\textwidth]{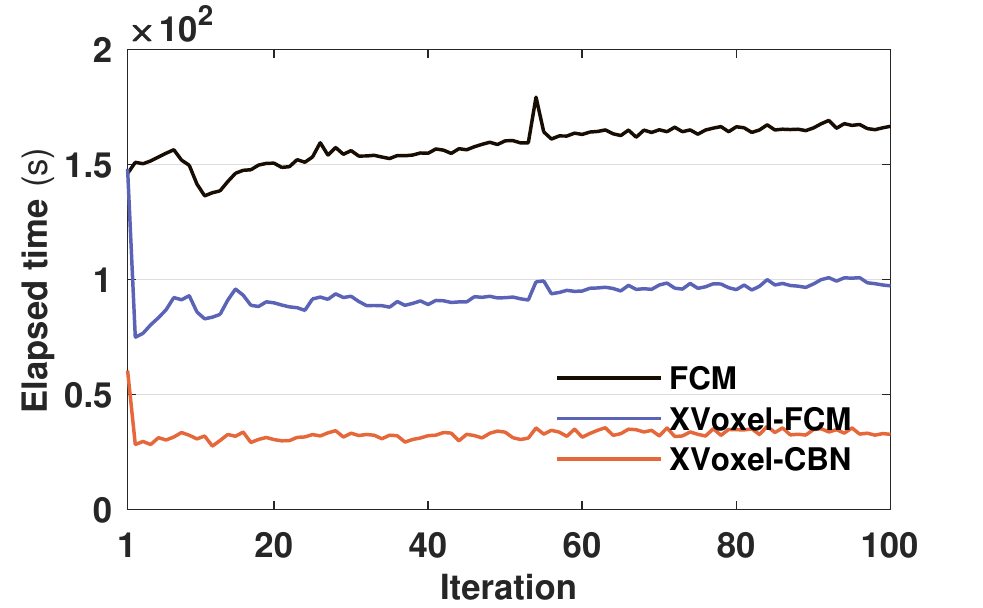}}
    \caption{Performance statistics of Example \#5 in Fig.~\ref{fig-BearingBracket_model_info}
    \rev{}{: (a) Convergence curves of compliance and volume by XVoxel and FCM;
    (b) The number of active voxels by FCM and XVoxel;
    (c) Timing of three methods FCM, XVoxel-FCM  and XVoxel-CBN.}
    }    
    \label{fig-BearingBracket_statistics}
\end{figure*}
The proposed method's robustness in handling complex parametric design optimization with varied structural topology was tested on a bearing bracket model in Fig. \ref{fig-BearingBracket_model_info}, where the design model, CSG tree, boundary conditions, discrete voxels were respectively shown in Figs.~\ref {fig-BearingBracket_model_info}(a), (b), (c), (d). The FCM (XVoxel) has 33K voxels, and the volume ratio constraint was set to be $0.6$.

The model has 11 CBS (Closed B-Spline) negative features, each with 24 control radii and 2 position coordinates, where the first and last radii of each feature are driven parameters for higher-order geometric continuity. In order to maintain the features to produce a valid structure of complex topology, features $f_1$ and $f_2$ were restricted to move only along the y-axis, features $f_3, f_4, f_5, f_7, f_8, f_9$ were fixed, and features $f_6, f_{10}$ and $f_{11}$ only moved along a prescribed plane; see also Fig. \ref{fig-BearingBracket_model_info}(a). Altogether, there are in total $(24-2+2)\times 11-14=250$ design variables.

The intermediate results of the examples are given in Fig.~\ref{fig-BearingBracket_step_change} and performance statistics of the example in Fig.~\ref{fig-BearingBracket_statistics}. As can be seen from Fig.~\ref{fig-BearingBracket_step_change}, during the optimization, the CBS features gradually expanded to meet the volume constraint, resulting in drastic topology variations in the arm and base part of the bearing bracket, before reaching convergence after 35 iterations. The XVoxel-based optimization successfully handled the topological changes.

The XVoxel model has around half active voxels of FCM, and XVoxel-FCM is about $1.3 \times$ faster than FCM in each iteration. The speedup is much smaller in comparison with Example \#4, as the CBS features were distributed more broadly within the bracket model, and therefore resulted in more active voxels. Nevertheless, XVoxel-CBN has gained about $5\times$ efficiency improvement. Again, the proposed method remains robust in the feature-based design optimization framework in handling such a large-scale model with a large number of design variables and drastic topology changes.


\subsection{Example \#6: bearing plates containing varying cylindrical supports}

\begin{figure*}[]
    \centering
    \subfigure[]{\includegraphics[width=0.35\textwidth,trim=120 0 0 0,clip]{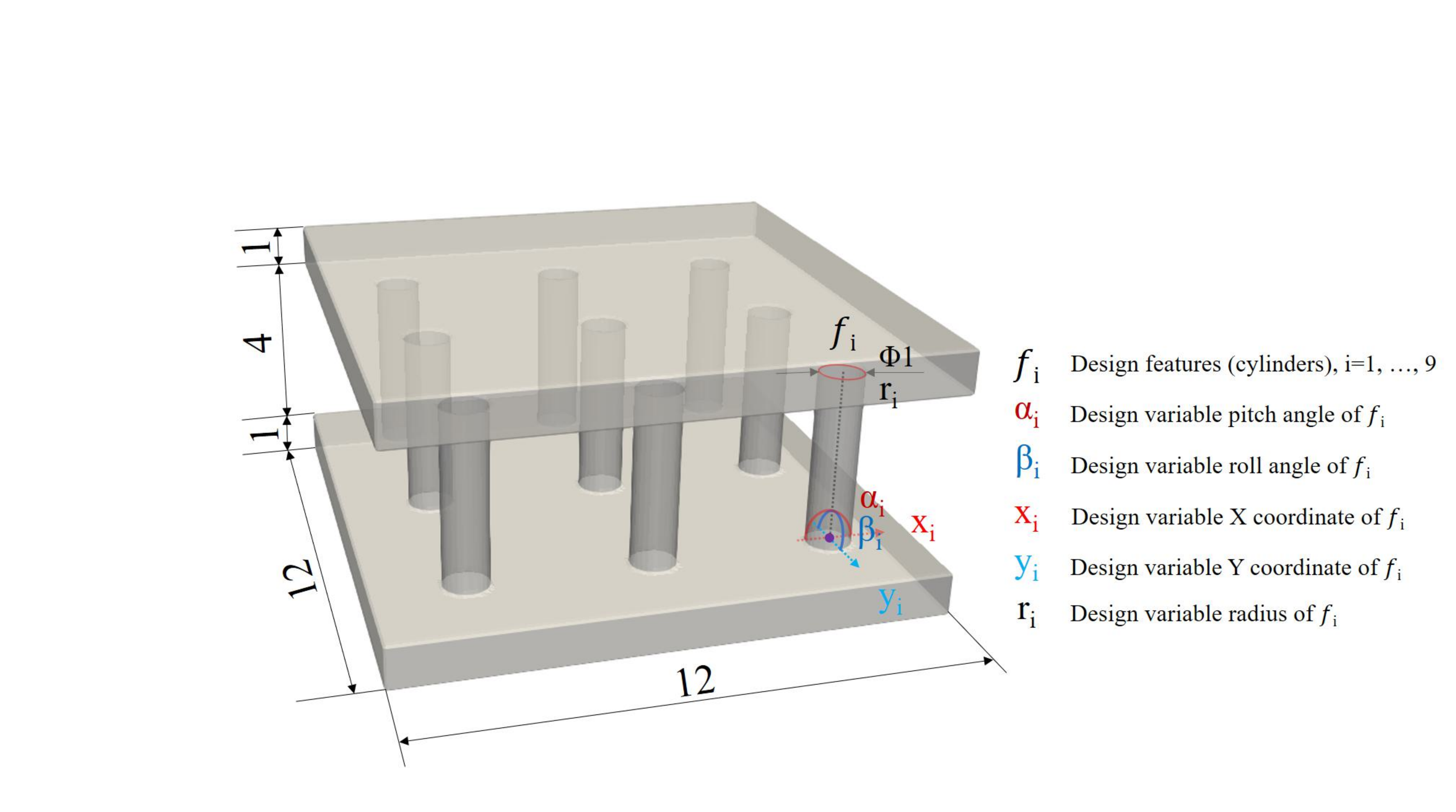}}
    \subfigure[]{\includegraphics[width=0.25\textwidth]{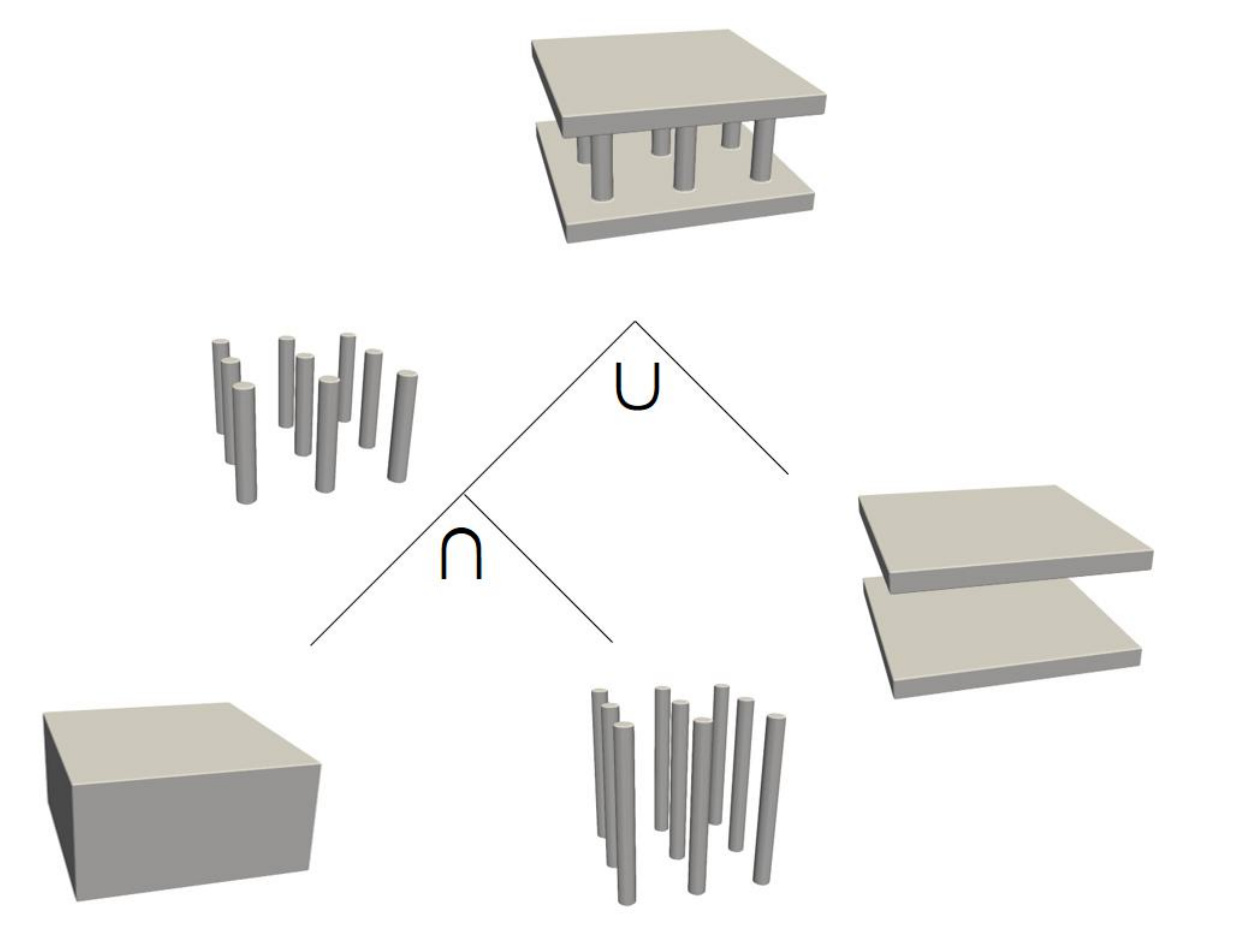}}
    \subfigure[]{\includegraphics[width=0.25\textwidth,trim=0 0 0 0,clip]{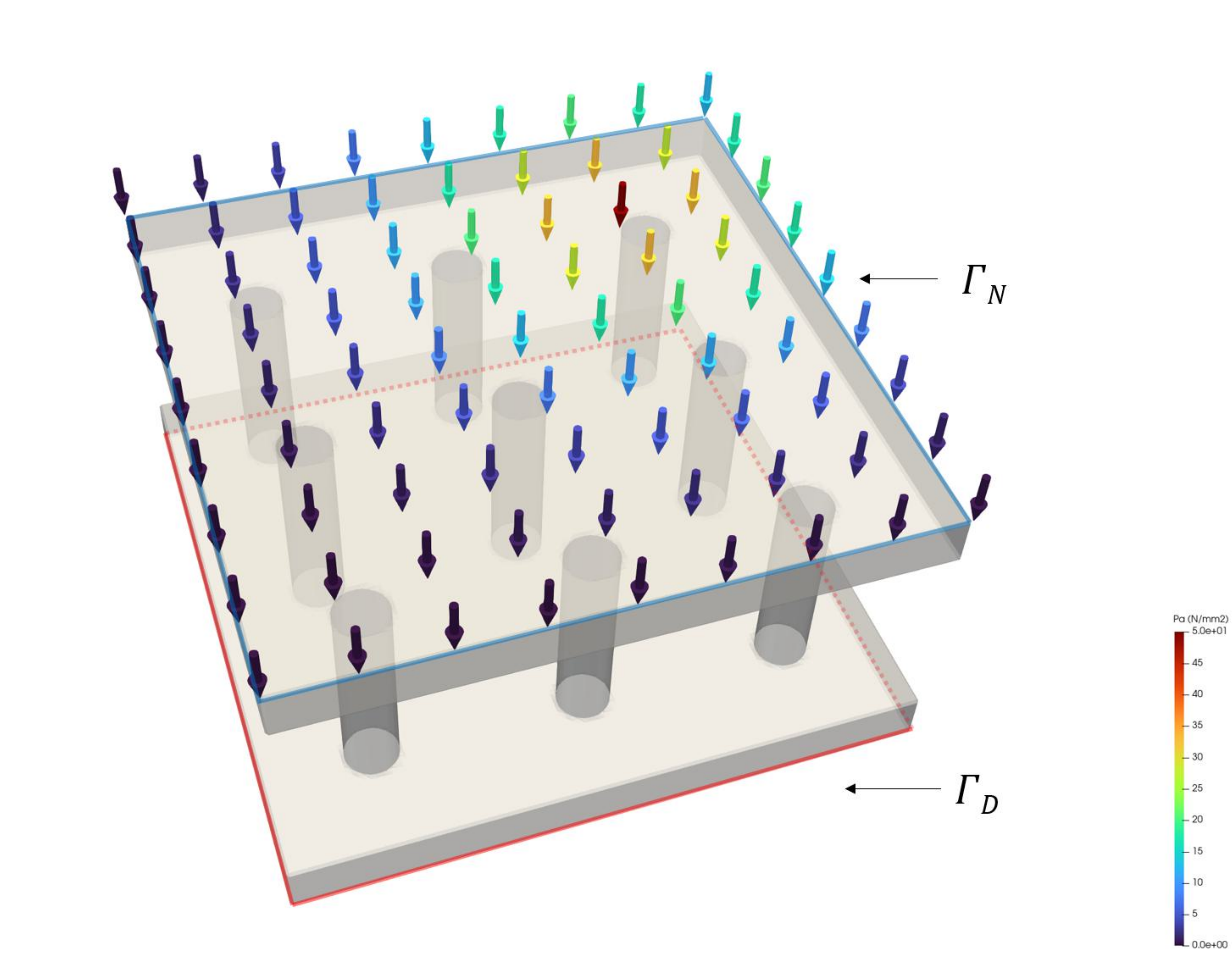}}
    \caption{
        Example \#6.
        (a) Parameters (mm) of the model, where the design variables are $x_i, y_i, \alpha_i, \beta_i$ and $r_i$ for $i=1, 2, ...,  9$;
        (b) The CSG of the model; 
        (c) Boundary conditions, where $\Gamma_D$ is fixed and $\Gamma_N$ is exerted by 
        a radially decayed force field 
        $F(x,y)=F_{c}\cos{(\sqrt{(x-x_{c})^2+(y-y_{c})^2}/(9\sqrt{2})\cdot\frac{\pi}{2})}^8$, where $F_{c}=50N$, $x_{c}=x_{c}=9mm, 0\leq x,y \leq 12$.
    }
    \label{fig-BearingCylinders_model_info}
\end{figure*}

\begin{figure*}[]
    \centering
    \begin{minipage}[]{1\linewidth}
    {
        \centering
        \includegraphics[width=0.3\textwidth]{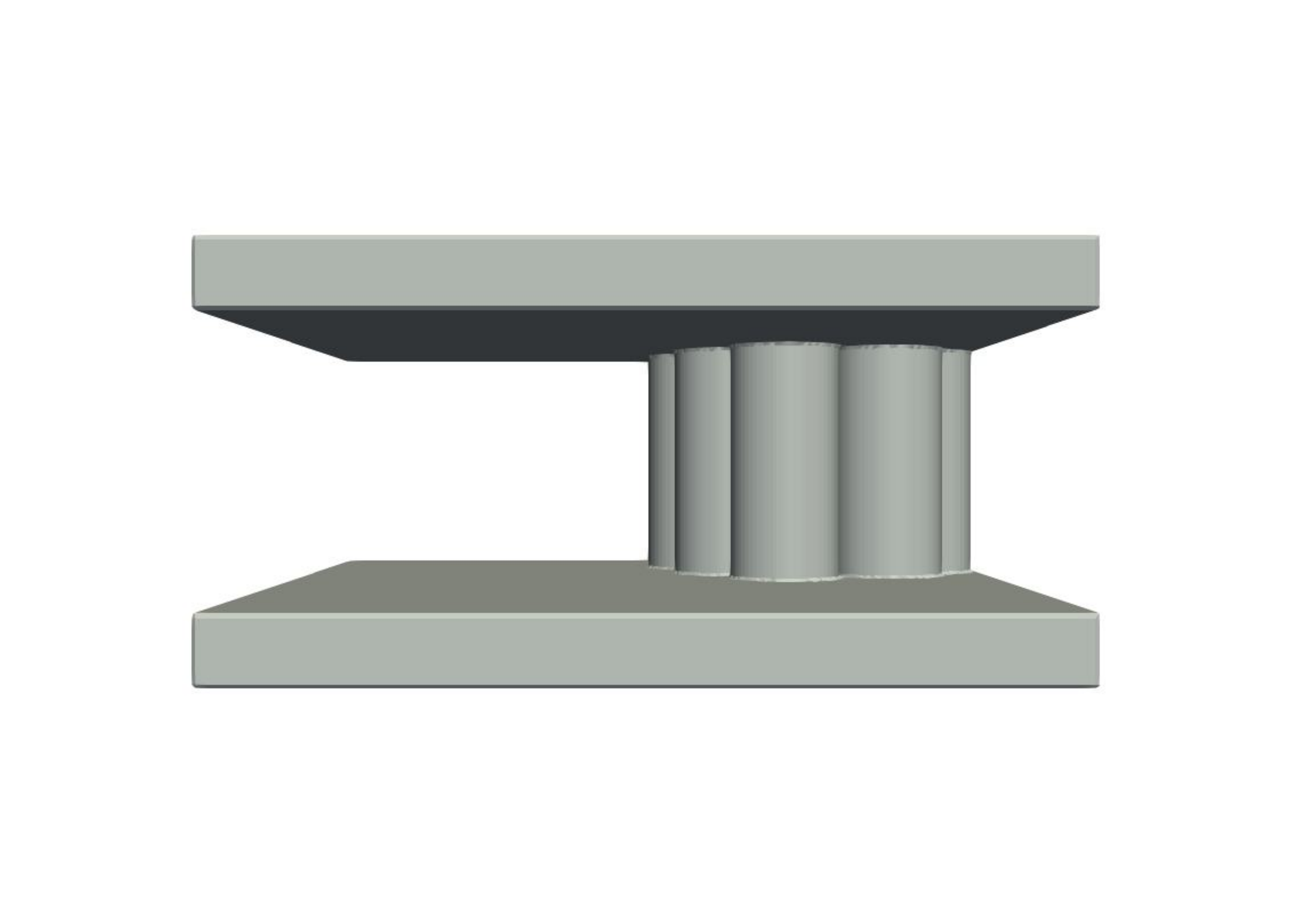}
                 \hspace*{0.5cm}
        \includegraphics[width=0.3\textwidth]{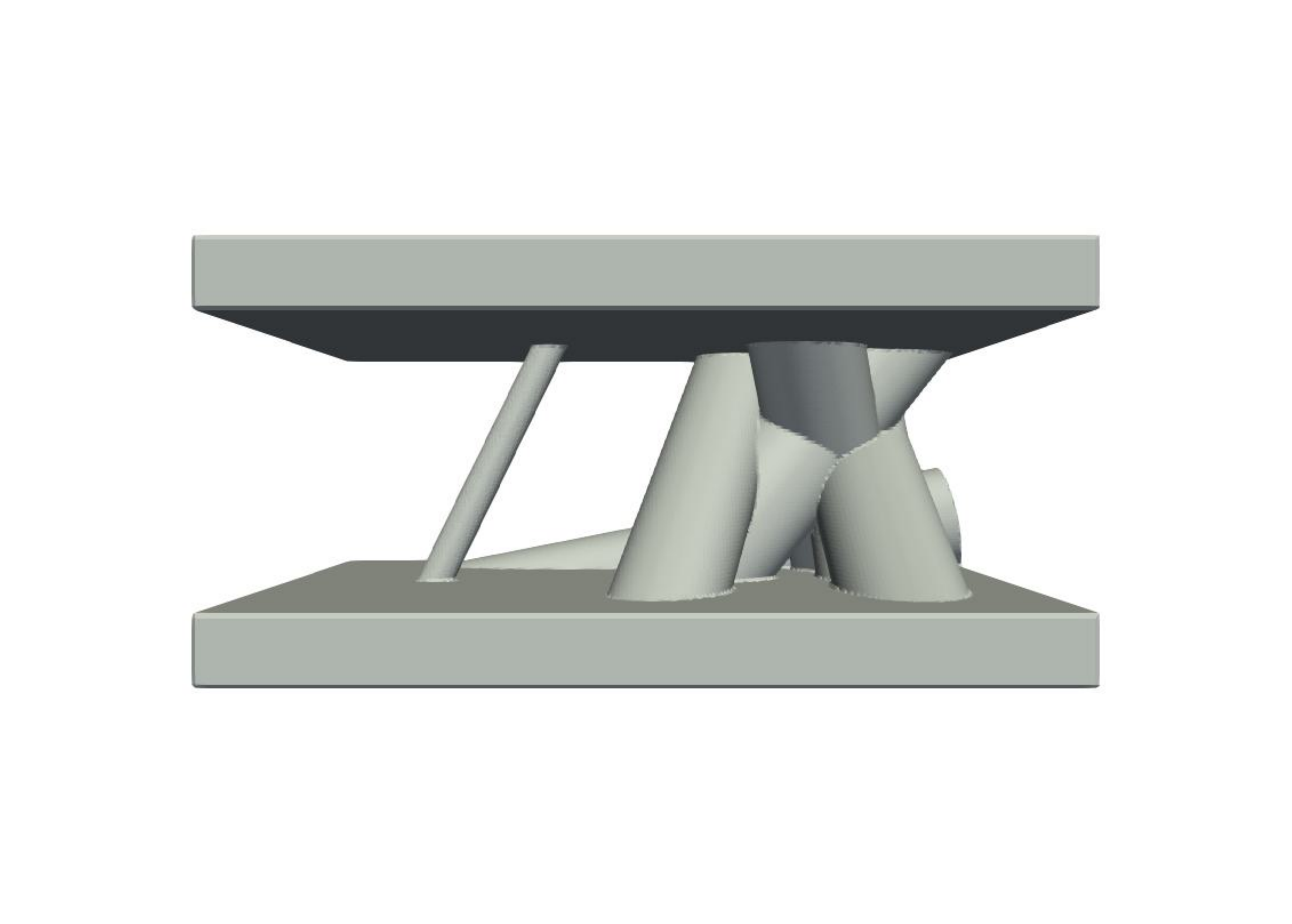}
                 \hspace*{0.5cm}
        \includegraphics[width=0.3\textwidth]{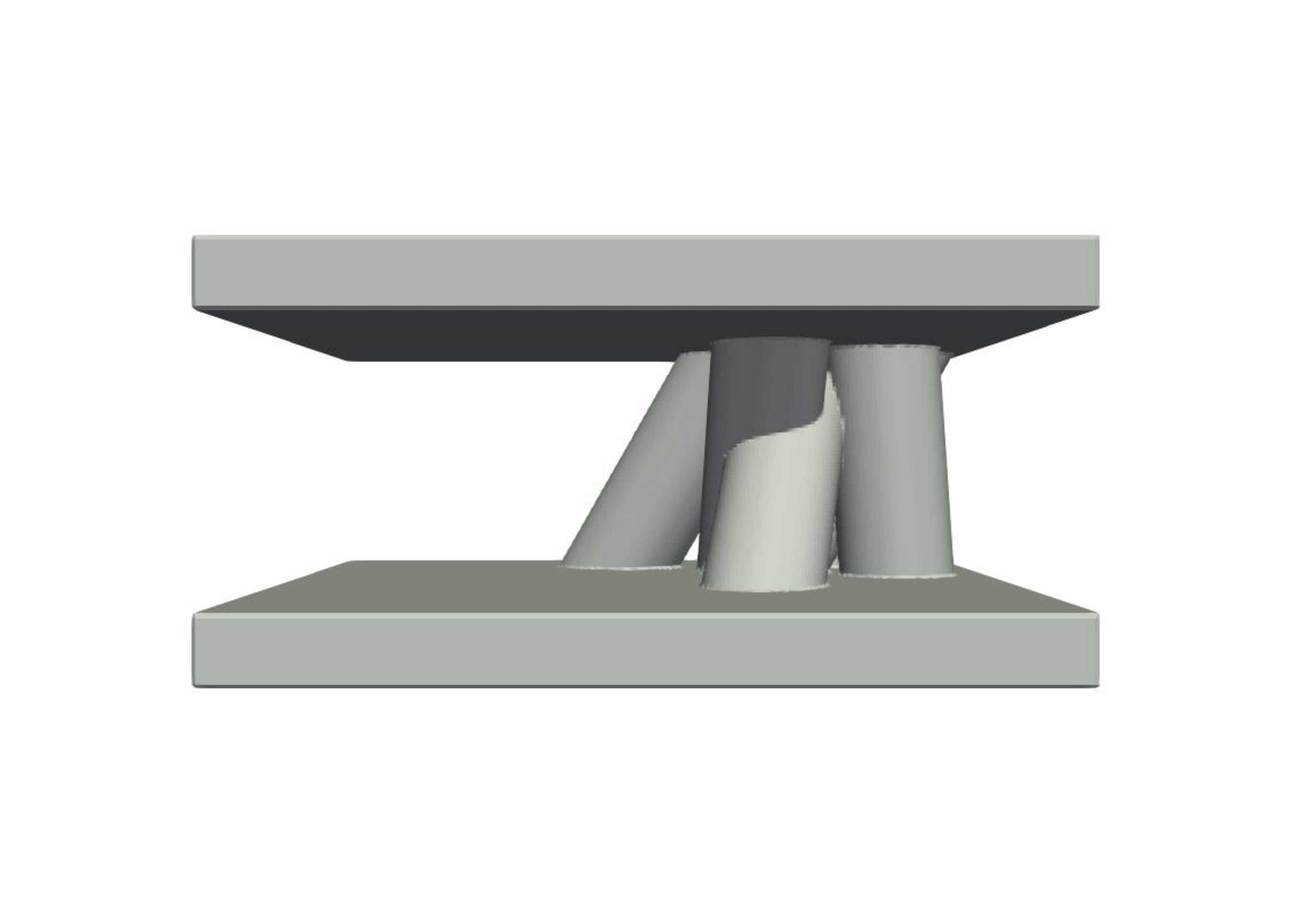}
    }
    \end{minipage}

    \begin{minipage}[]{1\linewidth}
    {
        \centering
        \subfigure[]
        {\includegraphics[width=0.3\textwidth]{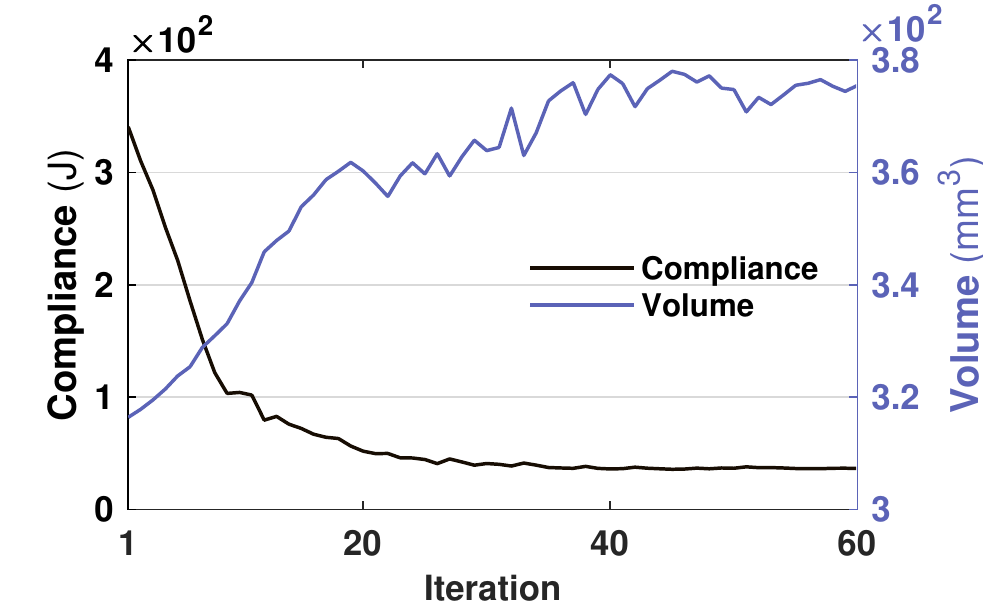}}
         \hspace*{0.5cm}
        \subfigure[]
        {\includegraphics[width=0.3\textwidth]{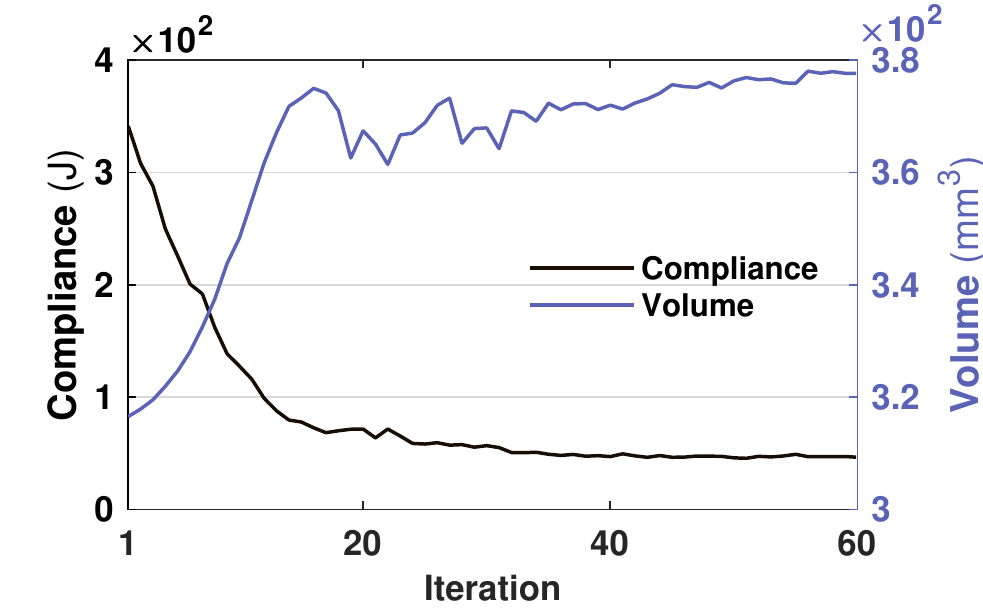}}
        \hspace*{0.5cm}
        \subfigure[]
        {\includegraphics[width=0.3\textwidth]{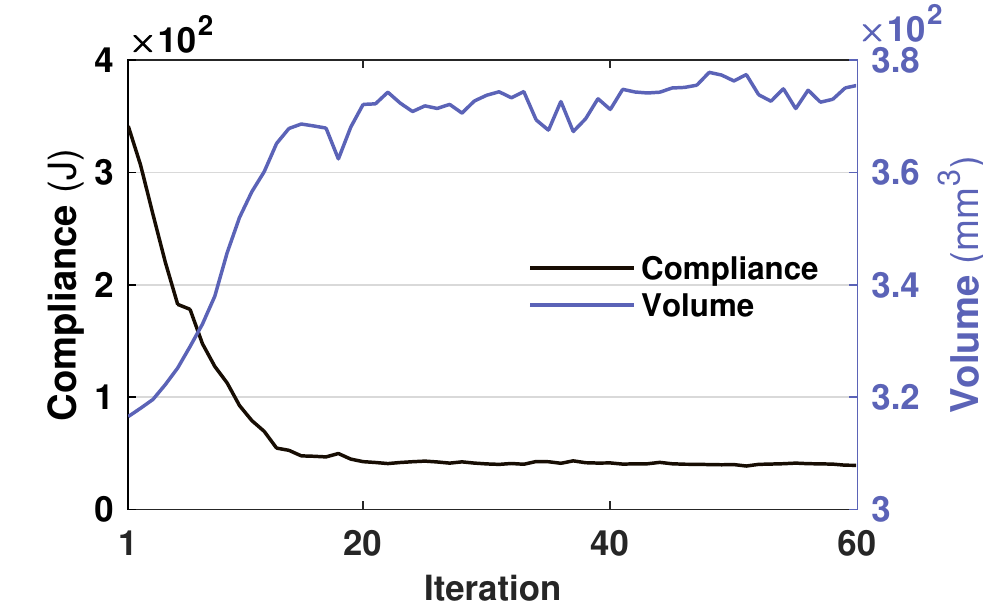}}
    }
    \end{minipage}
    \caption{
            Optimized models and convergence curves of structural compliance and cylindrical volumes: 
            (a) $x_i, y_i$ and $r_i$ as design variables;
            (b) $\alpha_i, \beta_i$ and $r_i$ as design variables;
            (c) all $5$ design variables; 
            in Fig.~\ref{fig-BearingCylinders_model_info}.}
    \label{fig-BearingCylinders_optimazation}
\end{figure*}

\begin{figure*}[]
    \centering
    \subfigure[loop 14]{\includegraphics[width=0.3\textwidth,trim=0 0 0 0,clip]{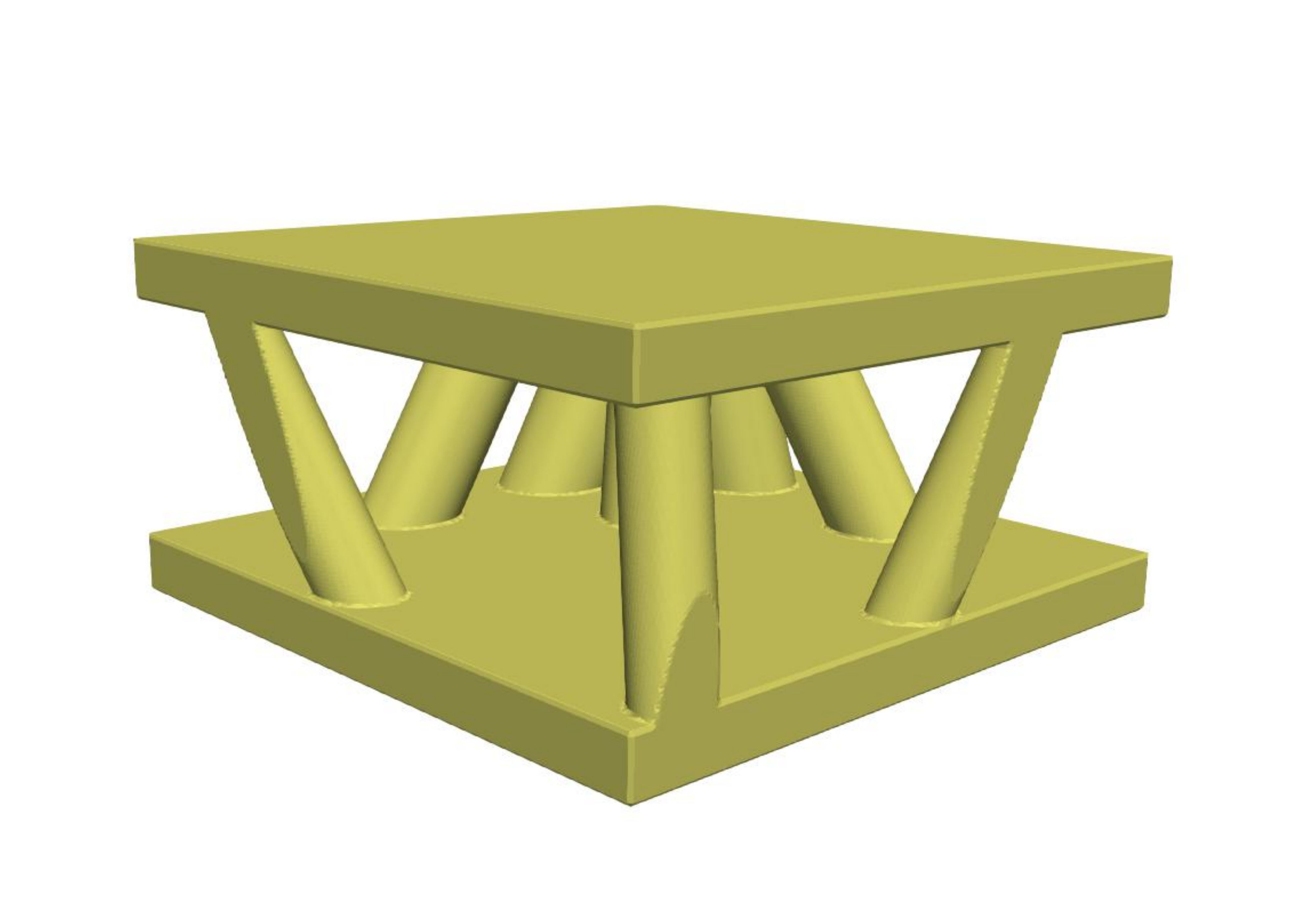}}
    \subfigure[loop 17]{\includegraphics[width=0.3\textwidth]{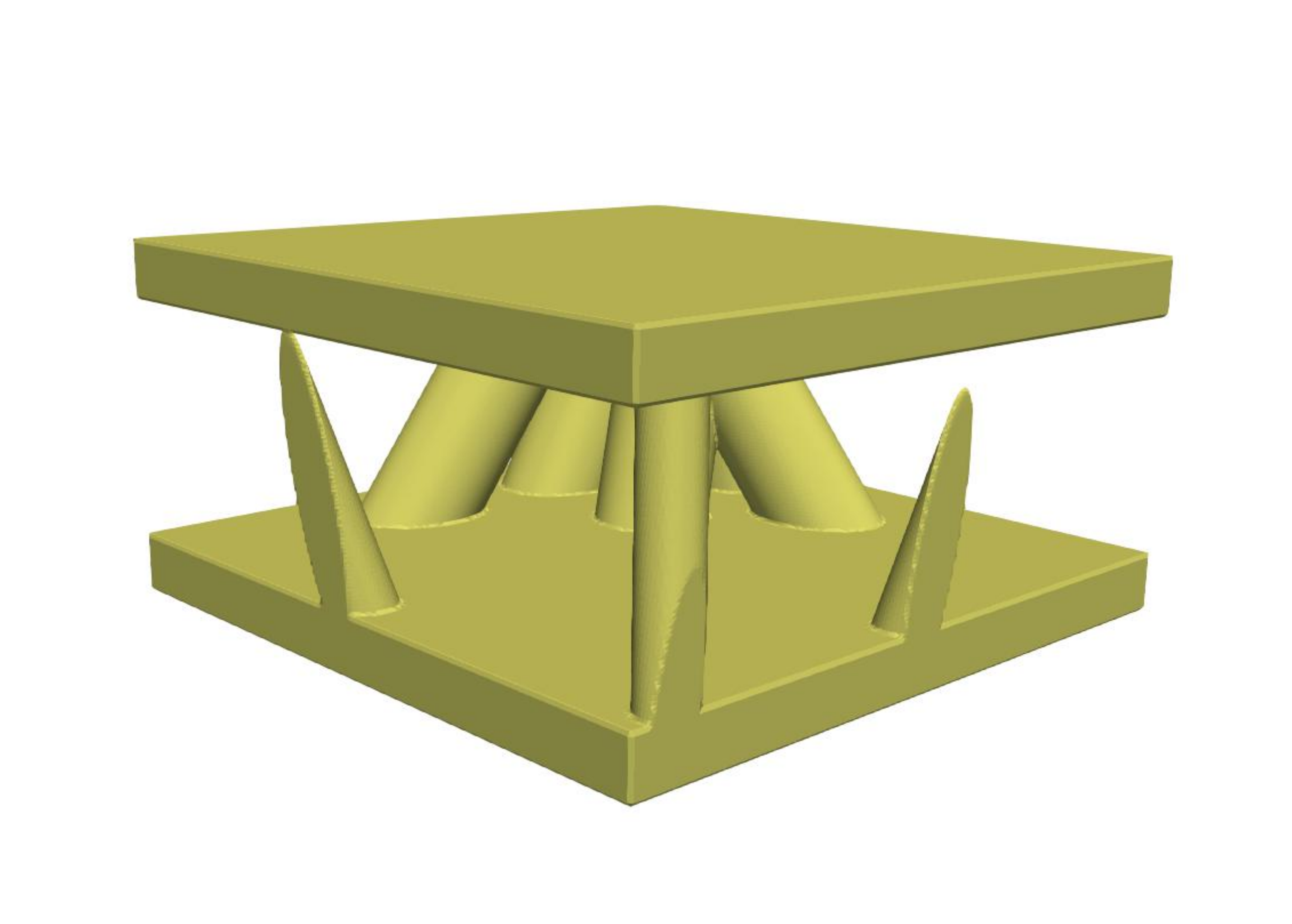}}
    \subfigure[loop 20]{\includegraphics[width=0.3\textwidth,trim=0 0 0 0,clip]{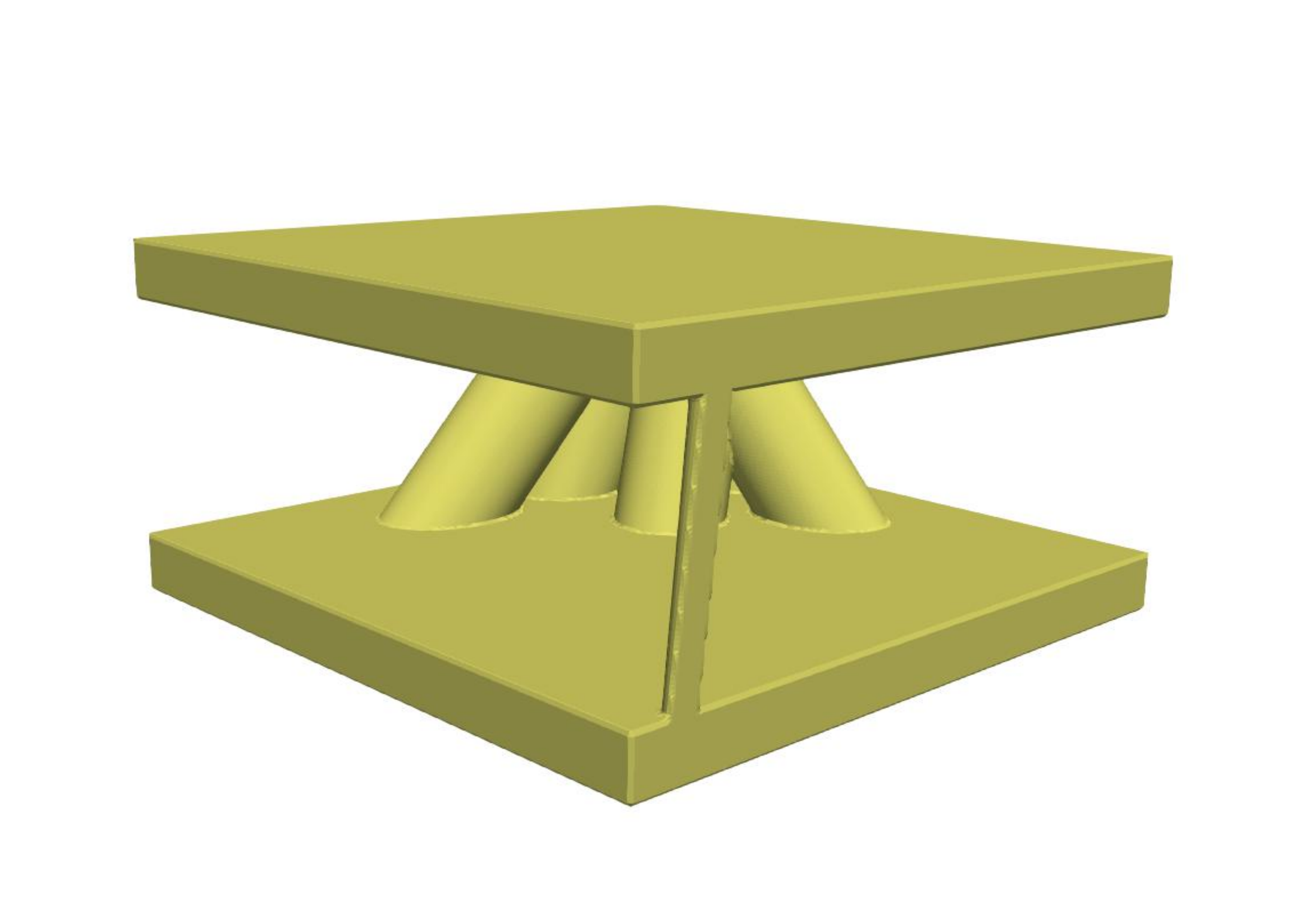}}
    \caption{The disappearance of features during optimization in Fig.~\ref{fig-BearingBracket_step_change}(c). %
    }
    \label{fig-BearingCylinders_Vanishing}
\end{figure*}

\begin{table*}[htbp]
\rev{}{
    \centering
    \caption{\rev{}{Performance comparison of FCM, XVoxel-FCM and XVoxel-CBN on the tested numerical examples. The table shows the size of mesh, the number
of DOFs, the total time and the total acceleration ratio.}
    }
\resizebox{1\textwidth}{!}{
\begin{tabular}{lllllllll}
\hline
\multicolumn{1}{c}{\multirow{2}{*}{Example}} & \multicolumn{1}{c}{\multirow{2}{*}{Mesh Size}} & \multicolumn{2}{c}{DOFs}     & \multicolumn{3}{c}{Timings (total Steps/Iterations)}   & \multicolumn{2}{c}{Total Acceleration Ratio ( Based on FCM )} \\
\multicolumn{1}{c}{}                         & \multicolumn{1}{c}{}                           & FCM(XVoxel-FCM) & XVoxel-CBN & FCM       & XVoxel-FCM & XVoxel-CBN & XVoxel-FCM                  & XVoxel-CBN                 \\ \hline
\#1                                          & 675                                            & 63,480          & 42,280     & 116.0     & 117.2      & 12.8       & 0.990                       & 9.06                      \\
\#2                                          & 4,004                                          & 108,135         & 255,024    & 259.9     & 84.5       & 21.6       & 3.08                       & 12.0                     \\
\#3                                          & 39,775                                         & 998,325         & 2,365,080  & 3,801.4   & 883.6      & 174.6      & 4.30                       & 21.8                     \\
\#4                                          & 9,100                                          & 240,975         & 250,638    & 145,022.8 & 10,321.3   & 2,597.2    & 14.1                      & 55.8                     \\
\#5                                          & 33,150                                         & 279,265         & 282,006    & 15,849.0  & 9,358.5    & 3,278.5    & 1.69                       & 4.83                      \\ \hline
\end{tabular}
}
\label{tb-acceleration}
}
\end{table*}

\rev{}{In the end, we further test the approach's ability in handing features of varied locations, sizes or orientations using the example in Fig.~\ref{fig-BearingCylinders_model_info}. The approach's potentiality is also tested in removing unnecessary features in the construction history during the optimization. The model has  
two plates supported by  9  cylinders in the middle. Each cylindrical feature contains $5$ design variables, including position coordinates $x_i, y_i$, directions $\alpha_i, \beta_i$ and a radius $r_i$ for $i=1,\ldots,9$. The upper surface of the top plate is exerted by a radiant force F defined as $F(x,y)=F_{c}\cos{(\sqrt{(x-x_{c})^2+(y-y_{c})^2}/(9\sqrt{2})\cdot\frac{\pi}{2})}^8$, where $F_{c}=50N$, $x_{c}=x_{c}=9mm, 0\leq x,y \leq 12$ while the lower surface of the bottom plate is fixed, as shown in Fig. \ref{fig-BearingCylinders_model_info}(c). Three different tests are conducted at a volume fraction of $1.2$ times the original volume of the cylinders by optimizing  $x_i, y_i$ and $r_i$;$\alpha_i, \beta_i$ and $r_i$; all the variables. }

\rev{}{The optimized structures and convergence curves are shown in Fig. \ref{fig-BearingCylinders_optimazation}. All the cases lead to reliable convergence and produce structures respectively of compliance  $39.3$, $47.1$, and $36.0$. It can be observed that the full variable optimization has the best convergence rate.}
\rev{}{Meanwhile, note in Fig. \ref{fig-BearingCylinders_Vanishing}(c) that three design features are able to run out of the design area during optimization, demonstrating the approach's ability in removing unnecessary features  automatically. In this case, we can conveniently remove the features from  the feature list. The variations of the features are also shown in Fig.~\ref{fig-BearingCylinders_Vanishing}. 
}

\subsection{\rev{}{Discussion and limitations}}\label{sec:discussion}
\rev{}{As one may have noticed from the above examples, although the final results of our XVoxel method can admit topology changes, their overall shapes still follow a similar structural pattern to those of the initial designs before optimization. This is because the method is designated for feature-based CAD and parametric optimization, where feature semantics (and therefore the overall shapes) often need to be respected. For example, the boundary representation of the final results is consistently composed of smooth parametric surfaces; there is no way the boundary takes discrete, free shapes (which is the case for SIMP or the voxel density method \cite{bendsoe2003Topology}).}

\rev{}{For this reason, our method works better for situations like fine or semi-fine design tuning. Dimensional variations and topology changes can be large (as demonstrated by the example in Fig.~\ref{fig-BearingBracket_step_change}) but cannot be radically different. If design changes of this sort are desired, other methods, e.g., the voxel density method, should be used. For the same reason, our method needs to take as input an initial design that does not deviate too much from the final result. This is different from methods like the voxel density method. Their input can even be a block without bearing any similarity to the final result.}

\section{Conclusions}\label{sec:conclusion}
An XVoxel-based parametric design optimization method for feature models has been presented in this paper. The proposed method combines the local regularity of voxel models and the global semantic information of feature models to facilitate the automatic linking between CAD and CAE. By further integrating XVoxels, FCM and CBN, design modifications and simulation updates can be looped in an efficient and robust manner, without involving labor-intensive conversion between CAD models and CAE models. The effectiveness of the proposed method has been validated by various numerical examples with complex topology variations and varying loads. \rev{}{And a computational efficiency improvement of up to 55.8 times the existing FCM method has been achieved, see Table \ref{tb-acceleration}.}

\rev{}{We consider the proposed XVoxel method as an alternative attempt toward the long-standing research objective of a unified model representation scheme that can completely, compactly, and associatively represent the contents of both CAD and CAE models. The method builds itself upon a new concept called semantic voxels to provide the advantage of avoiding B-rep model simplification and mesh generation. These two procedures could present a particular challenge for existing methods; for example, the quad/hex meshing required by IGA is never easy if the geometry is complex. For this reason, a typical use case where XVoxel is preferable over the others is when the design to be optimized is given as a feature model and its overall shape is complex.} 

A couple of interesting improvement directions for the XVoxel method are noted here. As already noted in Sections~\ref{sec:analysis-geometry} and \ref{sec:SimFeature}, XVoxel models can much reduce the dependence of simulation on model simplification and can be directly used to guide the simulation-suitable model simplification process. Nevertheless, the method, in its current form, is still not able to handle dimension reduction, which is the other important step in model idealization~\cite{bb-ARMSTRONG1994573,2000Automated,chong2004automatic}. Extending the method to including dimension reduction is among the research studies to be carried out in our group. Another interesting improvement direction is that the proposed method has only been implemented in Matlab for proof of concept. Its further implementation on basis of commercial/opensource feature modelers, e.g., Open CASCADE, and then release as an open-source plugin is of great interest to our future research work.


\rev{}{In industrial design optimization, innovative designs often require heavy optimization within a large design space but time resources are limited. This entails the use of dimensionality reduction techniques on the design space during optimization. The proposed XVoxel method has a good potential to integrate with dimensionality reduction methods, e.g., parametric model embedding~\cite{serani2023parametric}, due to its generality on the input model. Such an interesting integration is among our future work. Another improvement direction lies in the efficiency of the proposed method. Currently, our use of octrees in finding Gaussian points leads to a time-consuming simulation. If augmented with some adaptive meshing method (e.g., \cite{qian2012automatic}), the proposed method can be much accelerated.}

\rev{}{It is also worth noting that a model may correspond to multiple construction ways using Boolean operations. For different construction ways, their design variation spaces may be different, so do the corresponding optimization processes. Therefore, the construction way needs to be carefully thought out before using our method. The proposed method, in its current form, only focuses on parametric optimization on given models, and it cannot automatically find an appropriate construction way. Such an automatic selection mechanism is of great interest to future work.
In addition, If a B-Rep model is provided, a Boundary-to-CSG conversion procedure (e.g., the method presented in \cite{shapiro1993separation}) is necessary for the proposed method to work.
Another limitation of the proposed method is that the optimization result is affected by the size of the cells used in the simulation, a consequence of using FCM for simulation (see \cite{duster2008finite} for a detailed discussion). Currently, there is no principled way to choose the best cell size, and the usual solution is using empirical tuning to find a good cell size.
}

\section*{Acknowledgements}
This work has been partially supported by the National Key Research and development Program of China (No. 2020YFC2201303), the National Natural Science Foundation of China (No. 62102355, 61872320), the Natural Science Foundation of Zhejiang Province (No. LQ22F020012), and Key Research and development Program of Zhejiang Province (No. 2022C01025).

\bibliographystyle{elsarticle-num}
\bibliography{XVoxel}
\end{document}